\documentclass[aps,twocolumn,floatfix]{revtex4}

\usepackage{custom_defn}
\usepackage{graphicx} 
\usepackage{dcolumn} 
\usepackage{amssymb}
\usepackage{amsmath}
\usepackage{bm}
\usepackage[dvips]{color}
\usepackage{times}
\usepackage{graphicx}
\usepackage{ulem}
\usepackage{subfigure}
\usepackage{tabularx}
\usepackage{feynmp} \unitlength=1mm
\usepackage{slashed}

\usepackage{color}   
\usepackage[colorlinks=true
,urlcolor=blue
,anchorcolor=blue
,citecolor=red
,filecolor=blue
,linkcolor=blue
,menucolor=blue
,pagecolor=blue
,linktocpage=true
,pdfproducer=medialab
]{hyperref} 

\allowdisplaybreaks 

\newcommand{\bqa}{\begin{eqnarray}} 
\newcommand{\eqa}{\end{eqnarray}}
\newcommand{\nn}{\nonumber \\}

\definecolor{new_color}{RGB}{50,155,0}

\newcommand{\msr}[1]{\frac{d#1}{2\pi}}
\newcommand{\step}[1]{\Theta\lt(#1\rt)}
\newcommand{\td}{\tilde} 
\newcommand{\what}{\widehat}
\newcommand{\om}{\omega}

\begin{document}
\title{ Chiral non-Fermi Liquids}

\author{Shouvik Sur$^{1}$ and Sung-Sik Lee$^{1,2}$\\
\vspace{0.3cm}
{\normalsize{$^1$Department of Physics $\&$ Astronomy, McMaster University,}}\\
{\normalsize{1280 Main St. W., Hamilton ON L8S 4M1, Canada}}
\vspace{0.2cm}\\
{\normalsize{$^2$Perimeter Institute for Theoretical Physics,}}\\
{\normalsize{31 Caroline St. N., Waterloo ON N2L 2Y5, Canada}}
}

\date{\today}

\begin{abstract}

A non-Fermi liquid state 
without time-reversal and parity symmetries arises 
when a chiral Fermi surface is coupled 
with a soft collective mode
in two space dimensions.
The full Fermi surface is described by
a direct sum of chiral patch theories, 
which are decoupled from each other in the low energy limit.
Each patch includes low energy excitations 
near a set of points on the Fermi surface
with a common tangent vector.
General patch theories are classified 
by the local shape of the Fermi surface, 
the dispersion of the critical boson,
and the symmetry group, 
which form the data for distinct universality classes.
We prove that a large class of chiral non-Fermi liquid states exist
as stable critical states of matter.
For this, we use a renormalization group scheme 
where low energy excitations of the Fermi surface are 
interpreted as a collection of $(1+1)$-dimensional chiral fermions 
with a continuous flavor labeling the momentum along the 
Fermi surface. 
Due to chirality, 
the Wilsonian effective action 
is strictly UV finite.
This allows one to extract the exact scaling exponents
although the theories flow to 
strongly interacting field theories at low energies.
In general, the low energy effective theory 
of the full Fermi surface includes
patch theories 
of more than one universality classes.
As a result, physical responses
include multiple universal components
at low temperatures.
We also point out that, 
in quantum field theories with extended Fermi surface,
a non-commutative structure naturally emerges
between a coordinate and a momentum which are orthogonal
to each other.
We show that the invalidity of patch description
for Fermi liquid states is tied with
the presence of UV/IR mixing associated with the emergent
non-commutativity.
On the other hand,
UV/IR mixing is suppressed in
non-Fermi liquid states due to UV insensitivity,
and the patch description is valid.
\end{abstract}

\maketitle

\tableofcontents

\section{Introduction}

Landau Fermi liquid theory 
is the low energy effective theory for 
conventional metals\cite{landau}.
In Fermi liquids, 
kinematic constraints 
suppress non-forward scatterings 
caused by short-range interactions
except for the pairing channel\cite{shankar, polch-1}.
In the absence of  both time reversal and parity
symmetries,
even the pairing interactions
are suppressed, and 
Fermi liquid states can exist 
as a stable phase of matter at zero temperature.
In Fermi liquid states, 
shape of  the Fermi surface is 
a good `quantum number' at low energies,
and many-body eigenstates can be constructed
by filling single-particle states
even in the presence of  interactions.

The Fermi liquid theory breaks down in metals 
where a soft collective excitation
mediates a singular interaction between fermions
at quantum critical points
or in quantum critical phases\cite{holstein, reizer,PLEE89,Nagaosa92,
 halperin, altshuler, polch-2, YBkim, abanov,
motrunich1, lohneysen, coleman, senthil_mott, podolsky,
oganesyan, metzner, dellanna, kee, lawler, rech, wolfe,
maslov, quintanilla, yamase1, yamase2, halboth, jakubczyk,
zacharias, EAkim, huh, motrunich2, slee-2, plee,slee,metlitski,jiang}.
As a result of strong mixing between 
particle-hole excitations of the Fermi surface
and the critical collective mode, 
quantum fluctuations of Fermi surface 
remain important at low energies,
and the many-body ground state
is no longer described by a serene Fermi sea.

Recently, different perturbative schemes have been developed
in order to tame quantum fluctuations of Fermi sea
and gain a controlled access to non-Fermi liquid states.
One can continuously tune the energy dispersion
of the soft collective mode\cite{nayak,mross},
or the co-dimension of  the Fermi
surface\cite{dalidovich}
to obtain perturbative non-Fermi liquid fixed points. 
Non-Fermi liquid behaviors in an intermediate scale
are also obtained from an alternative
scheme\cite{fitzpatrick}.
Having established that there exist 
perturbative non-Fermi liquid fixed points,
it is of interest to find examples of strongly
interacting non-Fermi liquid states 
that can be understood beyond the perturbative limits.

It is an outstanding theoretical problem
to understand the non-perturbative nature of 
wildly fluctuating
Fermi surfaces in non-Fermi liquid states,
especially in two space dimensions
where strong quantum fluctuations 
persist down to arbitrarily low energies.
Compared to the critical systems
which have only discrete gapless points in momentum space,
field theories for Fermi surface
are more challenging due to the 
extra softness of the Fermi surface
associated with the presence of an infinitely
many gapless modes\cite{slee,metlitski}.

In this paper, we study a class of 
$(2+1)$-dimensional chiral non-Fermi liquid states
without time-reversal and parity invariance. 
The metallic state in two space dimensions
is chiral in the sense that 
one of the components of the Fermi velocity
has a fixed sign.
Although there are indications that 
the chiral non-Fermi liquid states are stable,
so far there has been no rigorous proof of the stability
due to a lack of control over the strongly interacting  
field theories\cite{slee}.
In this work, we systematically exploit 
extra kinematic constraints
imposed by the chiral nature of the theory to
show that a large class of chiral non-Fermi liquid states 
indeed do exist as stable critical states of matter.
For this we use the patch description 
where the full Fermi surface is decomposed into 
local patches in momentum space.
General chiral patch theories
are characterized by  
the geometric data of the local Fermi surface
and the dispersion of the critical boson.
Thanks to chirality,
exact critical exponents 
can be obtained.

The chiral non-Fermi liquids are the two-dimensional
analogs of the chiral Luttinger liquids\cite{CLL}. 
In both cases,
the stability is guaranteed by 
the absence of back scatterings.
Furthermore, 
critical exponents are protected 
by chirality, and 
can be computed exactly
based on kinematic considerations.

The paper is organized in the following way.
In Sec. \ref{sec:exp}, we motivate the low energy effective theory
for chiral non-Fermi liquids 
from a set-up that can be potentially realized in experiments.
It is based on the chiral metal\cite{balents1},
where a two-dimensional chiral Fermi surface
arises on the surface of
a stack of  quantum Hall layers 
in the presence of inter-layer tunnelings.
A flavor degree of freedom is introduced by
bringing two such stacks with opposite magnetic fields 
close to each other
as is shown in Fig. \ref{f.qh1}.
In the absence of tunneling between the two stacks, 
there is a flavor symmetry associated with 
a rotation in the space of flavor.
The flavor symmetry can be spontaneously broken
by an interaction between electrons.
The quantum critical point associated with the
symmetry breaking is described 
by a chiral non-Fermi liquid state.
In Sec. \ref{sec:model}, we construct 
a low-energy effective theory for 
chiral non-Fermi liquid states
using the local patch description.
First, we focus on the most generic 
patches with nonzero quadratic curvatures
of the Fermi surface.
Near one of the generic points, 
the local dispersion of fermion
is given by $\epsilon_k = k_x + \gamma_2 k_y^2$,
where $\vec k = (k_x, k_y)$ is the momentum away from a 
point on the Fermi surface.
Higher order curvatures become important
at isolated points on the Fermi surface
where the quadratic curvature $\gamma_2$ vanishes
as is shown in Fig. \ref{fig:inflection}.
Theories for more general shapes of Fermi surface
with the dispersion of the form,
$\epsilon_k = k_x + \gamma_u k_y^u$
with $u>2$ will be discussed in Sec. \ref{sec:general}.
After the patch theory is introduced,
the two-dimensional Fermi surface 
is mapped into a collection of one-dimensional chiral 
fermions carrying a continuous flavor 
which corresponds to the momentum along the Fermi surface.
In the following sections, 
we develop a renormalization group scheme 
based on the one-dimensional picture.
Although the theory is superficially written 
as an one-dimensional theory,
it remembers the two-dimensional nature of the underlying theory
through a kinematic constraint between the 
two momentum components.
In particular, the momentum along the Fermi surface (the continuous flavor)
and the coordinate perpendicular to the Fermi surface
obey an emergent non-commutative relation.
The physical origin of the non-commutativity
and its consequences are discussed in Sec. \ref{sec: non-comm}
and Appendix \ref{app: 1}.
In Sec. \ref{sec:regularization}, 
the regularization scheme and 
the renormalization group (RG) prescription
for the field theory are introduced.
Because the dynamical critical exponent is not known a
priori, we choose a prescription which is compatible
with any dynamical critical exponent.
Namely, a cut-off 
is imposed only along the spatial direction 
perpendicular to the Fermi surface, but not in time.
Because of this rather unusual choice of regularization scheme,
the Wilsonian effective action obtained by integrating out
short-distance (in space but not in time) modes 
includes terms that are non-local in time, although 
locality is maintained along the spatial direction.
In Sec. \ref{sec:con}, 
which constitutes the central part of the paper,
we show the stability of the chiral path theories.
In Sec. \ref{sec:UV_SC},
we consider the general form of the Wilsonian
effective action allowed by scaling analysis.
In the scaling under which
the interaction remains invariant, 
a part of the kinetic term is irrelevant,
and it introduces a UV cut-off scale to the theory.
If the Wilsonian effective action were dependent on 
the UV cut-off scales in a singular way,
the scaling dimensions would deviate from the `bare' ones.
It turns out that the present theory is UV finite 
thanks to the chiral nature of the theory
as is proved in Sec. \ref{sec: UV_finite}.
As a result, one can take the limit of infinite UV cut-off scales.
The UV finiteness and the absence of IR scale at the critical point 
implies that the theory should remain critical 
at low energies 
provided that the theory is IR finite
as is explained in Sec. \ref{sec: IR_finite}.
In Sec. \ref{sec: EC}, 
it is shown that the theory is indeed IR finite
through an explicit computation of 
the Wilsonian effective action.
As a result, one can show that the exact scaling
dimension is given by the bare scaling under which
the interaction is kept invariant.
In Sec. \ref{sec: WQ}, we emphasize
the difference between the
Wilsonian effective action and the full quantum effective action,
which allows one to compute the Wilsonian effective action
of the chiral patch theory perturbatively 
in the limit where the external momenta are smaller
than the running cut-off scale
while the full quantum effective action 
can not be computed perturbatively\cite{slee}.
In Sec. \ref{sec:RG}, the exact beta functions are derived,
from which the dynamical critical exponent
and the scaling dimension of the fermionic field 
are obtained, 
which coincide with the ones obtained
from the general scaling analysis in Sec. \ref{sec:UV_SC}.
In Sec. \ref{sec:general}, 
we turn to general patch theories 
with cubic or higher order local curvatures
of the Fermi surface.
In particular, inflection points on the Fermi surface
are described by the patch theory
with the local dispersion, 
$\epsilon_k = k_x + \gamma_3 k_y^3$.
The results obtained 
in Sec. \ref{sec:regularization} - Sec. \ref{sec:RG} 
are extended to the general cases.
From this, 
it is shown that the general chiral patch theories are also stable, 
and the exact dynamical critical exponents are obtained.
In Sec. \ref{sec:TR}, 
we discuss the thermodynamic response of the system.
The full Fermi surface is composed of 
local patch theories, 
some of which belong to different universality classes.
As a result, physical response functions 
of the system possess multiple universal components.
In Sec. \ref{sec:summary}, 
we close with a summary and discussions.

\section{A potential experimental realization}
\label{sec:exp}

\begin{figure}[!ht]
 \centering
\subfigure[]{\includegraphics[scale=0.35]{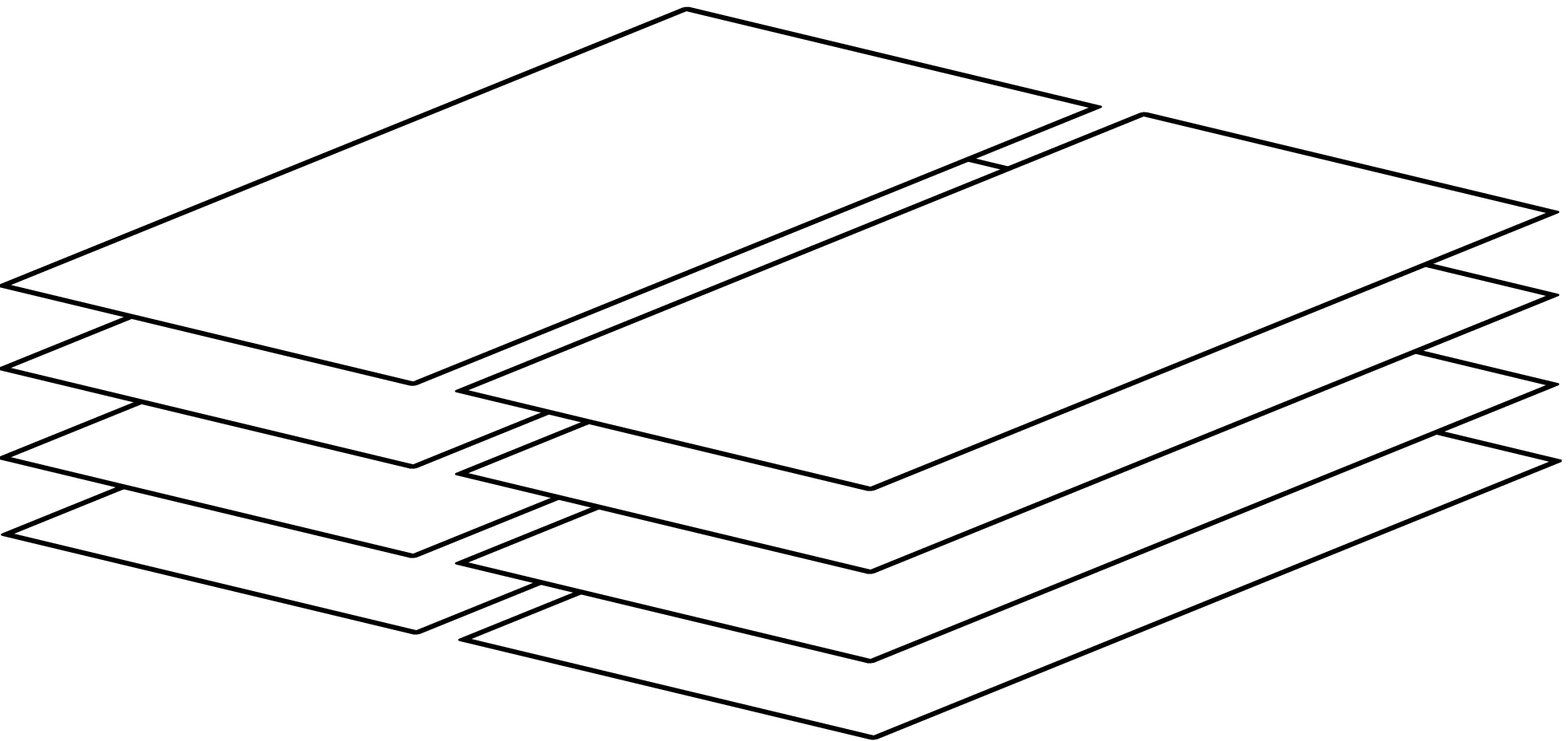} 
\label{f.qh1}} \\
\subfigure[]{\includegraphics[scale=0.25]{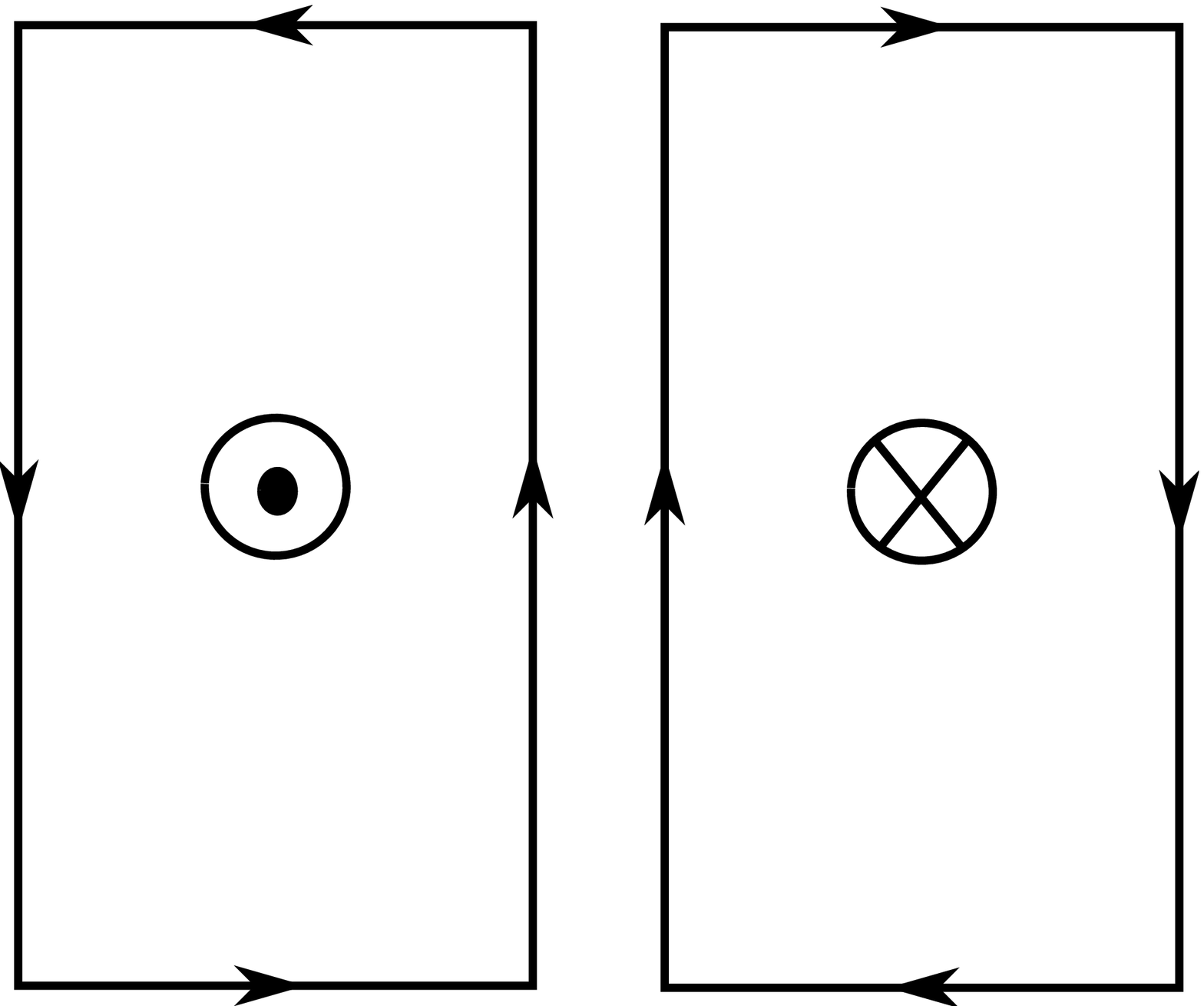} 
\label{f.qh2}}
\caption{
(a) Two stacks of integer quantum Hall layers with $\nu=1$.
(b) When the directions of magnetic field are opposite in
the two stacks,
the edge modes have same chirality near the region where 
two stacks are close to each other.
The interface of the two stacks is described
by a two-dimensional chiral metal with two flavors.
}
\end{figure}

To motivate the field theory for chiral non-Fermi liquids, 
we consider two stacks of $\nu=1$ integer quantum Hall
 layers  
as is shown in Fig. \ref{f.qh1}.
Each layer supports one-dimensional chiral edge mode.
If magnetic field is applied in opposite directions in
the two stacks, the edge modes  
have the same chirality 
in the region where the two stacks face each other
as is shown in Fig. \ref{f.qh2}.

In the presence of tunneling between the layers within each
stack, the chiral edge modes disperse along the 
direction perpendicular to the quantum Hall layers,
and form a two-dimensional chiral Fermi 
surface\cite{balents1}.
The low energy modes near the chiral Fermi surface 
can be described by two fermionic fields $\psi_i$,
where the flavor $i=1,2$ labels different stacks. 
We assume that electron spin is fully polarized,
and consider spinless fermions.
In the absence of  tunneling between the two stacks,
the quadratic action has $SU(2)$ flavor symmetry. 
Suppose that there is a short-range density-density
interaction that respects the flavor symmetry,
\bqa
H_{int} = V  ( c_{1}^* c_1 + c_2^* c_2 )^2,
\eqa
with $V>0$. 
Here $c_i$ represents the microscopic electron operator
in the $i$-th stack. It can be written as a
superposition of low energy fields which include not only
the gapless chiral mode
but also other gapped non-chiral modes which reside near
the edge.
By using the identity 
$\vec{\sig}_{ij} \cdot \vec{\sig}_{kl} = 2(\dl_{il} \dl_{jk} - \half \dl_{ij}
\dl_{kl})$ the interaction can be expressed as 
\bqa
H_{int} = -\frac{V}{3} (c_i^* \vec \sigma_{ij} c_j) 
\cdot 
(c_k^* \vec \sigma_{kl} c_l ).
\eqa
When $V$ is large enough, $H_{int}$ can lead
to a condensation in the particle-hole channel, 
$\vec \phi \sim < c_i^* \vec \sigma_{ij} c_j >$.
The exciton condensation spontaneously
breaks the flavor symmetry,
inducing a charge imbalance 
(coherent tunneling) between the stacks 
when $\vec \phi$ points along (perpendicular to)
the $\hat{z}$ direction.
Once $\langle \vec \phi \rangle$ becomes nonzero,
the system becomes a chiral Fermi liquid
with a reduced symmetry.
If the phase transition is continuous,
the corresponding quantum critical point is 
described by a chiral metal 
which is coupled with a critical boson.
If the $SU(2)$ symmetry is explicitly broken 
by a small energy scale, there will be no
sharp phase transition.
Nonetheless, critical behaviors will show up 
at temperatures larger than the energy scale.

\section{Patch description for chiral non-Fermi liquids} 
\label{sec:model}

\begin{figure}[!ht]
     \centering
     \includegraphics[scale=0.5]{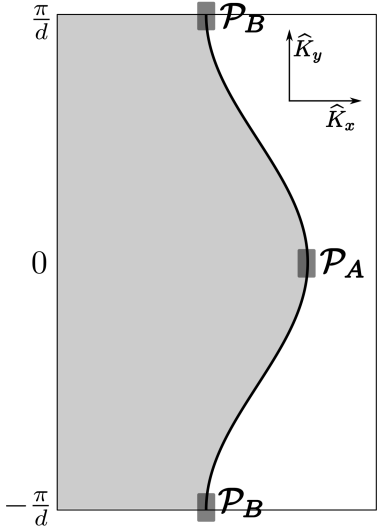}
     \caption{
The chiral Fermi sea. 
Generically, there are two points on the Fermi surface
which have a common tangent vector. 
For example, the points $\mc{P}_A$ and $\mc{P}_B$
have the same tangent vectors and they 
remain strongly coupled to each other  in the low energy
limit.
A minimal patch theory should include 
all the points with the common tangent vector.
}      \label{fig: chiral_FS}
\end{figure}

In this section, we construct a minimal theory 
for the quantum phase transition described in the previous 
section. If there are only nearest neighbor hopping 
between chiral edge modes within each stack,
the kinetic energy of the chiral metal in each stack can be 
written as
\bqa
\varepsilon_K = K_x -2 t \cos( K_y d ),
\label{ek}
\eqa
where 
$K_x$ is the momentum along the edge,
$K_y$ is the momentum perpendicular to quantum Hall layers,
and 
$t$ ($d$) is the hopping matrix element (distance) 
between nearest layers within each stack.
The velocity along the edge is normalized to be one.
At low energies, regions near the Fermi surface
with different tangent vectors
are decoupled from each other 
because of the kinematic separation\cite{polch-2}.
As a result, one can focus on patches with 
a common tangent vector at low energies.
Except for the inflection points at
$(K_x = 0, K_y = \pm \pi/2d)$,
there are two points at which the tangent vectors of the
Fermi surface 
are parallel to each other.
Here we focus on the generic patches away from the
inflection points.
Patch theories for the inflection points 
will be discussed in Sec. \ref{sec:general}.

At the generic points, the local Fermi surface is parabolic.
As a concrete example 
for patches with quadratic curvatures, 
we consider the low-energy modes near 
$\mc{P}_A \equiv (K_x = 2t, K_y =0)$ and $\mc{P}_B \equiv 
(K_x = -2t, K_y = \pi/d)$ 
which have parallel tangent vectors,
as is shown in  Fig. \ref{fig: chiral_FS}.
They are described by the quadratic action,
\eqn{
S_2 &=  \sum_{j=1}^2 \sum_{s=\pm} 
\int \frac{d\omega ~d^2 \vec{k}}{(2\pi)^3} \nn
& \quad \times \Bigl( i \omega + k_x + s \gamma k_y^2
\Bigr)\td \psi_{s, j}^*(\omega, \vec{k}) \td 
\psi_{s,j}(\omega, \vec{k}), 
}
where $\td \psi_{+,j}$ ($\td \psi_{-,j}$) 
represents the low energy excitations near $\mc{P}_A$ 
($\mc{P}_B$)  in the $j$-th stack with $j=1,2$,
$(k_x, k_y)$ refers to the deviation of the momentum 
of $\td \psi_{+,j}$ ($\td \psi_{-,j}$) 
from $\mc{P}_A$ 
($\mc{P}_B$),
and $\gamma = t d^2$ is the absolute value of the local
curvature.
It is noted that the curvatures at the two points are equal
in magnitude
but opposite in sign.

Now, we consider a general theory 
where the fermions with $N$ flavors 
are coupled with a boson, 
\eqn{
S &=   \int \frac{d\omega ~d^2 \vec{k}}{(2\pi)^3} ~ \lt(i
\eta \omega +  k_x + s ~ \gamma k_y^2 \rt) \td
\psi_{s,j}^*(\omega,\vec{k}) \td \psi_{s,j}(\omega,\vec{k})
\nn
& + \frac{1}{2} \int \frac{d\nu ~d^2 \vec{q}}{(2\pi)^3} ~
( \nu^2 + | \vec q|^2 + \mu^2 ) 
~\phi_{\alpha}(-\nu,-\vec{q} ) \phi_{\alpha}(\nu,\vec{q}) 
\nn
& + g \int \frac{d\om ~d^2 \vec{k}}{(2\pi)^3} ~
\frac{d\nu ~d^2 \vec{q}}{(2\pi)^3} ~ \nn
& \quad \times \phi_{\alpha}(\nu,\vec{q}) ~
\td \psi_{s,i}^*(\om + \nu, \vec{k}+ \vec{q}) ~
\td T^{\alpha}_{ij} ~ \td  \psi_{s,j}(\om,\vec{k}).
 \label{eq: general_action_1}
}
The summation over $s=\pm$ and $i, j=1,2,..,N$ are assumed 
in the expression.
The example discussed in the previous section corresponds
to the case with $N=2$.
In the following, we assume general $N$ 
which are not necessarily large.
In general, the low energy physics can be 
governed by an effective theory
with a dynamical critical exponent $z > 1$,
in which case the term that is linear in frequency
in the fermion kinetic term is irrelevant.
Therefore, we keep a `coupling'
$\eta$ for the dynamical term.
$\phi_\alpha$ is a boson field, and
$\td T^{\alpha}_{ij}$ is a matrix that characterizes
the flavor quantum number carried by the boson.
Here we consider the $SU(N)$ flavor group, 
where $\td T^{\alpha}_{ij}$'s are traceless Hermitian
matrices
with $\alpha=1,2,..,N^2-1$. 
$\mu^2$ is the mass of the boson 
which drives a quantum phase transition.
For $\mu^2 > 0$ the system is described by the Fermi liquid 
with the unbroken symmetry.
For $\mu^2 < 0$ a condensation of $\phi_\alpha$ 
breaks the flavor symmetry spontaneously.
We will later on focus on 
the critical point at which $\mu=0$.
However, it is helpful to keep a nonzero positive mass
temporarily in order to avoid spurious IR singularities 
that may show up in intermediate steps 
of the RG analysis.
It is noted that the boson is non-chiral
because a chiral boson can not be gapped 
even away from the critical point.
The non-chiral kinetic energy for the boson is generated
once the gapped non-chiral modes near the edge
are integrated out.
In the absence of time-reversal or parity invariance,
one may add local terms that are linear in $\nu$ or $q_x$
in the boson action.
However, $\phi_\alpha \partial_\mu \phi_\alpha$ 
are total derivative terms and do not affect dynamics.
Terms of the form $\phi_\alpha \partial_\mu \phi_\beta$ 
with $\alpha \neq \beta$
are prohibited by flavor symmetry.
While $( \partial_\tau \phi^\alpha) (\partial_x
\phi^\alpha)$
is allowed by symmetry, it is irrelevant
compared to $( \partial_y \phi^\alpha) (\partial_y
\phi^\alpha)$.

Although the minimal theory has two patches, 
one can reinterpret the two-patch theory as an
one-patch theory with a larger representation of the flavor 
group.
In order to see this,
we define a fermion field $\psi_j$ with doubled flavor 
indices $j=1,2,....,2N$ as
\bqa
\psi_j(\om, \vec{k}) & \equiv & \td \psi_{+,j}(\om, \vec k) 
~~\mbox{for $1 \leq j \leq N$}, \nn
\psi_j(\om, \vec k) & \equiv & \td \psi_{-,j-N}^*(-\om, - 
\vec k) ~~\mbox{for $N+1 \leq j \leq 2N$}.
\label{eq:trans}
\eqa
In terms of the $2N$-component fermion field, 
the action is written as
\eqn{
S &=   \int \frac{d\omega ~d^2 \vec{k}}{(2\pi)^3} ~ \lt(i
\eta \omega +  k_x +  \gamma k_y^2 \rt) 
\psi_{j}^*(\omega,\vec{k})  \psi_{j}(\omega,\vec{k})
\nn
& + \frac{1}{2} \int \frac{d\nu ~d^2 \vec{q}}{(2\pi)^3} ~
( \nu^2 + |\vec q|^2  + \mu^2 )^{\theta/2} 
~\phi_{\alpha}(-\nu,-\vec{q} ) \phi_{\alpha}(\nu,\vec{q}) 
\nn
& + g \int \frac{d\om ~d^2 \vec{k}}{(2\pi)^3} ~
\frac{d\nu ~d^2 \vec{q}}{(2\pi)^3} ~ \nn
& \quad \times \phi_{\alpha}(\nu,\vec{q}) ~
 \psi_{i}^*(\om + \nu, \vec{k}+ \vec{q}) ~
 T^{\alpha}_{ij} ~   \psi_{j}(\om,\vec{k}).
 \label{eq: b.f.action}
}
Here $T^{\alpha}$ is $2N \times 2N$ reducible 
representation
of the $SU(N)$ group given by 
\begin{align}
T^{\alpha} \equiv
\begin{pmatrix}
     \td T^{\alpha} & 0 \\
     0 & -\lt[\td T^{\alpha}\rt]^T
\end{pmatrix}.
\label{doubleT}
\end{align}
We also consider a generalized boson action of the form
$( \nu^2 + q_x^2 + q_y^2 + \mu^2 )^{\theta/2}$,
where $\theta$ is treated as a free parameter.
The theory in Eq. (\ref{eq: b.f.action}) 
is nothing but a chiral one-patch theory
where fermions transform in an enlarged
reducible representation
of the flavor group.
In non-chiral two-patch theories,
the transformation in \eq{eq:trans} 
leads to a Dirac fermion with two 
components\cite{dalidovich}.
In this respect, 
the nature of the chiral two-patch theory
is fundamentally different from 
the non-chiral two-patch theory.
We choose the normalization of $T^{\alpha}$ such that 
$
\mbox{Tr}\lt[ (T^{\alpha})^2 \rt] = 1
$
for all $\alpha$.

At the Gaussian fixed point, 
the scaling dimensions 
of momentum components, fields and the coupling
are given by
\begin{eqnarray}
{[} \om {]}  & = &  1, \nn
{[}k_x{]}  & = &  1, \nn
{[}k_y{]}  & = & \frac{1}{2}, \nn
{[}\psi_{i}{]} & = & -\frac{7}{4}, \nn
{[}\phi_\alpha{]} & = & -\frac{5 + {\theta}}{4}, \nn
{[}g{]} & = & \frac{{\theta} - 1}{4}, \nn
{[}\mu {]}  & = & \frac{1}{2}.
\label{eq:GS}
\end{eqnarray}
It is noted that $\nu^2, q_x^2$ in the boson action is 
irrelevant compared to $q_y^2$.
Therefore, we drop the dependencies on $\nu$ and $q_x$
in the boson dispersion to consider the following boson 
propagator,
\bqa
\chi_{\theta}(\vec{q}) =  \frac{1}{|q_y^2 + 
\mu^2|^{\theta/2}}.
\label{eq:IRchi}
\eqa
The fermion-boson coupling $g$ is relevant (irrelevant) 
for ${\theta} > 1$ (${\theta} < 1$) 
and marginal for $\theta = 1$. 
Consequently, the (2+1)-dimensional theory
is at the upper critical dimension when $\theta = 1$,
and below (above) the upper critical dimension when
${\theta} > 1$ (${\theta} < 1$).
The earlier works \cite{nayak, mross} developed a 
perturbative expansion for the critical point with $\mu=0$
near ${\theta} = 1$.
The most generic value for local theories,
including the theory for the system discussed in the 
previous section, 
is $\theta=2$.
For a generic ${\theta} > 1$ not close to $1$,
the critical theory flows to a strongly interacting 
non-Fermi liquid state at low energies.
However, we can obtain exact dynamical information of the 
theory thanks to the chiral nature of the theory. 
Note that we could have added a $\lambda \phi^4$ term to 
the action in \eq{eq: general_action_1}. 
At the Gaussian fixed point the scaling dimension of the
coupling $\lambda$ 
is $[\lambda] = \theta - 5/2$.
Since it is irrelevant for  $\theta < 5/2$, we drop it.

Upon integrating out the boson, we obtain a four fermion 
interaction,
\begin{align} 
& \int \frac{d\om_1 d^2\vec k_1 ~ d\om_2 d^2\vec k_2 ~ 
d\nu d^2\vec q}{(2\pi)^9} ~ 
V_{ij;ln} \chi_\theta(q)  
~ \psi^*_{i}(\om_2, \vec k_2)  \nn
& \times \psi_{j}(\om_2 + \nu, \vec k_2 + \vec q) 
\psi^*_{l}(\om_1 + \nu, \vec k_1 + \vec q) \psi_{n}(\om_1, 
\vec k_1),
\end{align}
where
$V_{ij;ln} = - \frac{g^2}{2} \sum_{\alpha} T^{\alpha}_{ij}
T^{\alpha}_{ln}$.
In this paper, we take the viewpoint that 
the Fermi surface is made of a collection of 
one-dimensional chiral fermions 
labeled by $y$-momentum $k_y$. 
In order to simplify notation, 
henceforth, we will use $k$ without subscript
to represent $y$ component of momentum.
We go over to the $(1+1)$-dimensional real space   
which is conjugate to $(\om,\epsilon_k)$, 
\bqa
\psi_{j, k}(\tau,x)  =  \int \frac{d\om}{2\pi} 
\frac{d\eps_k}{2\pi} ~ e^{i\om \tau + i\epsilon_k x} 
~ \psi_j(\om, \eps_k - \gamma k^2, k). 
\eqa
Note that $x$ is conjugate to 
$\epsilon_k = k_x + \gamma k^2$,
but not to $k_x$.
This implies that slowly varying modes in $x$ 
carry momenta close to the Fermi surface, 
that is, $k_x \sim - \gamma k^2$.
In this basis, the action takes the form of 
a {\it local} (1+1)-dimensional theory,
\begin{align}
 S  &=  
\int \msr{k} \int d^2r ~ \psi_{i, {k}}^*(r) \Bigl[ \eta 
~  \partial_{\tau}  - i\partial_x \Bigr] \psi_{i, {k}}(r)  
\nn 
& +  \int \msr{{k}_1} ~ \msr{{k}_2} ~ \msr{{q}} \int 
d^2r ~~ e^{i 2 \gamma ( k_1 -  k_2 )  q x } ~  
V_{ij;ln} \chi_\theta(q)  \nn
& \qquad \times \psi^*_{i, k_2}(r) ~ \psi_{j, 
k_2+q}(r) ~ \psi^*_{l,k_1+q}(r) ~ 
\psi_{n,k_1}(r). \label{eq: action}
\end{align}
Here 
$r=(\tau, x)$ and
$q$, $k$, $k_i$ represent $y$-momenta, 
which are now interpreted as continuous flavors
carried by the 1+1D chiral fermions
in addition to the discrete flavor. 
The long-range interaction mediated by the critical boson
becomes a four-fermion vertex.  
In this representation, 
the sliding symmetry\cite{metlitski} 
is realized by a simple shift in $y$-momentum,
$\psi_{i,k} \rightarrow \psi_{i,k+\Delta k}$.

It is interesting to note that the curvature of Fermi 
surface
completely drops out from the kinetic term.
This is due to the fact that 
the kinetic energy is ultra-local in momentum space. 
Instead the curvature shows up 
in the phase factor of the four-fermion vertex.
The phase factor plays a crucial role in 
describing the low energy physics correctly.
In particular, the phase factor is the only term
that `remembers' the $(2+1)$-dimensional nature of the 
theory.
Without the phase factor, 
the theory becomes an $(1+1)$-dimensional theory
which is fundamentally different from the 
$(2+1)$-dimensional theory.
In other words, a nonzero curvature is a relevant 
perturbation
to the $(1+1)$-dimensional theory which qualitatively
modifies the low energy behaviors.
This feature can be checked from the fact that 
various expressions that characterize low energy properties 
of the system
become singular in the zero curvature limit,
as will be shown later.

In \eq{eq: action}, the translational symmetry appears to 
be 
explicitly broken by the phase factor.
However, one can check that the symmetry is still intact
once the fermion field $\psi(\tau,x)$ transforms 
projectively as
\eqn{
  \psi_{i,k}(\tau,x) \rightarrow 
e^{-i\gamma k^2 x_0} ~ \psi_{i,k}(\tau,x+x_0)
\label{eq:trans1}
}
under a translation along the $x$-direction.
Before we delve into the renormalization group analysis, 
we make a small digression in the next section 
to discuss about
the origin and implications of the phase factor 
in the four-fermion interaction.

\section{Emergent non-commutativity} \label{sec: non-comm}

\begin{figure}[!ht]
	
\subfigure[]{\includegraphics[width=0.6\columnwidth]
{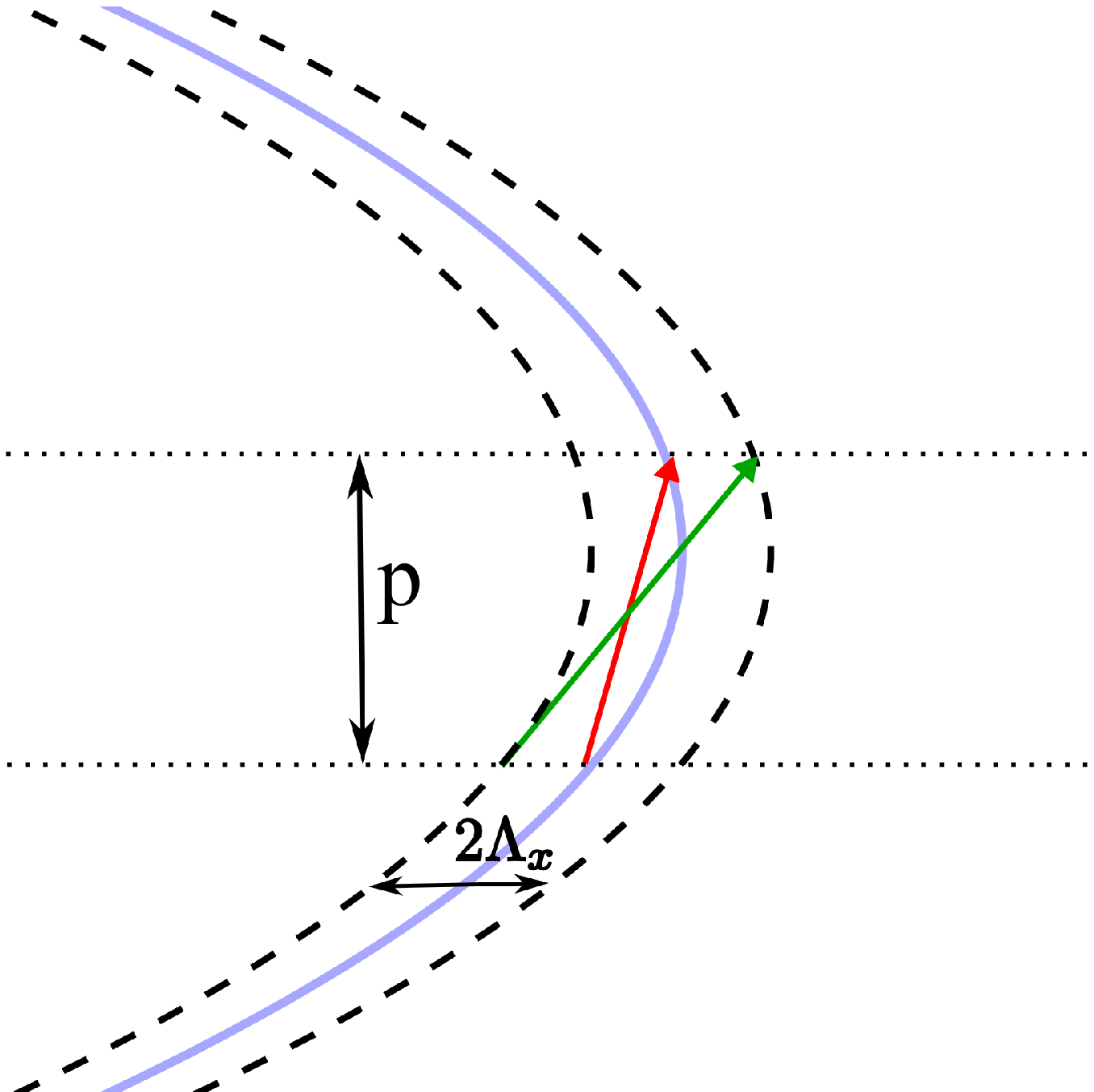}} 
\subfigure[]{\includegraphics[width=0.6\columnwidth]
{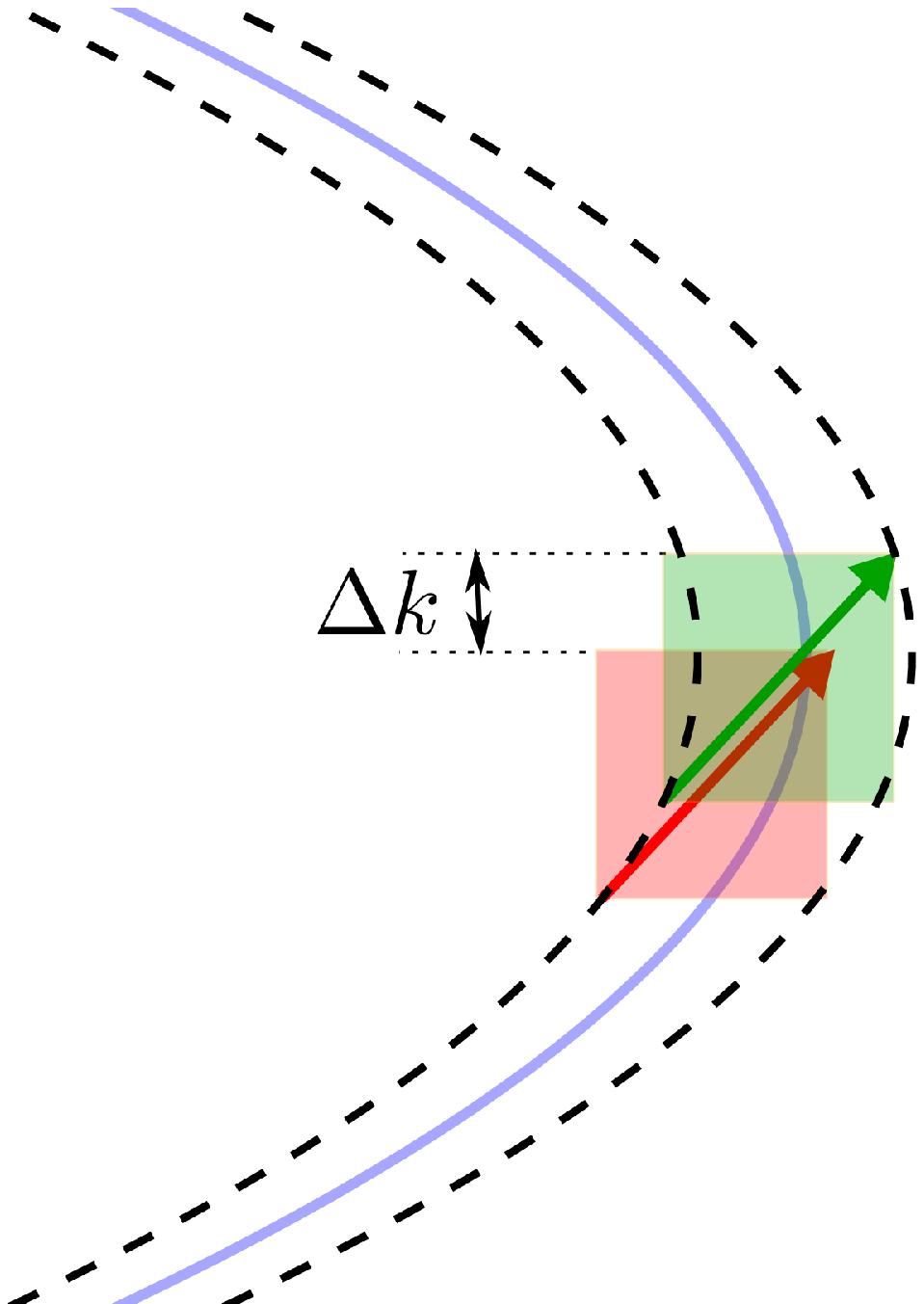}}
\subfigure[]{\includegraphics[width=0.6\columnwidth]
{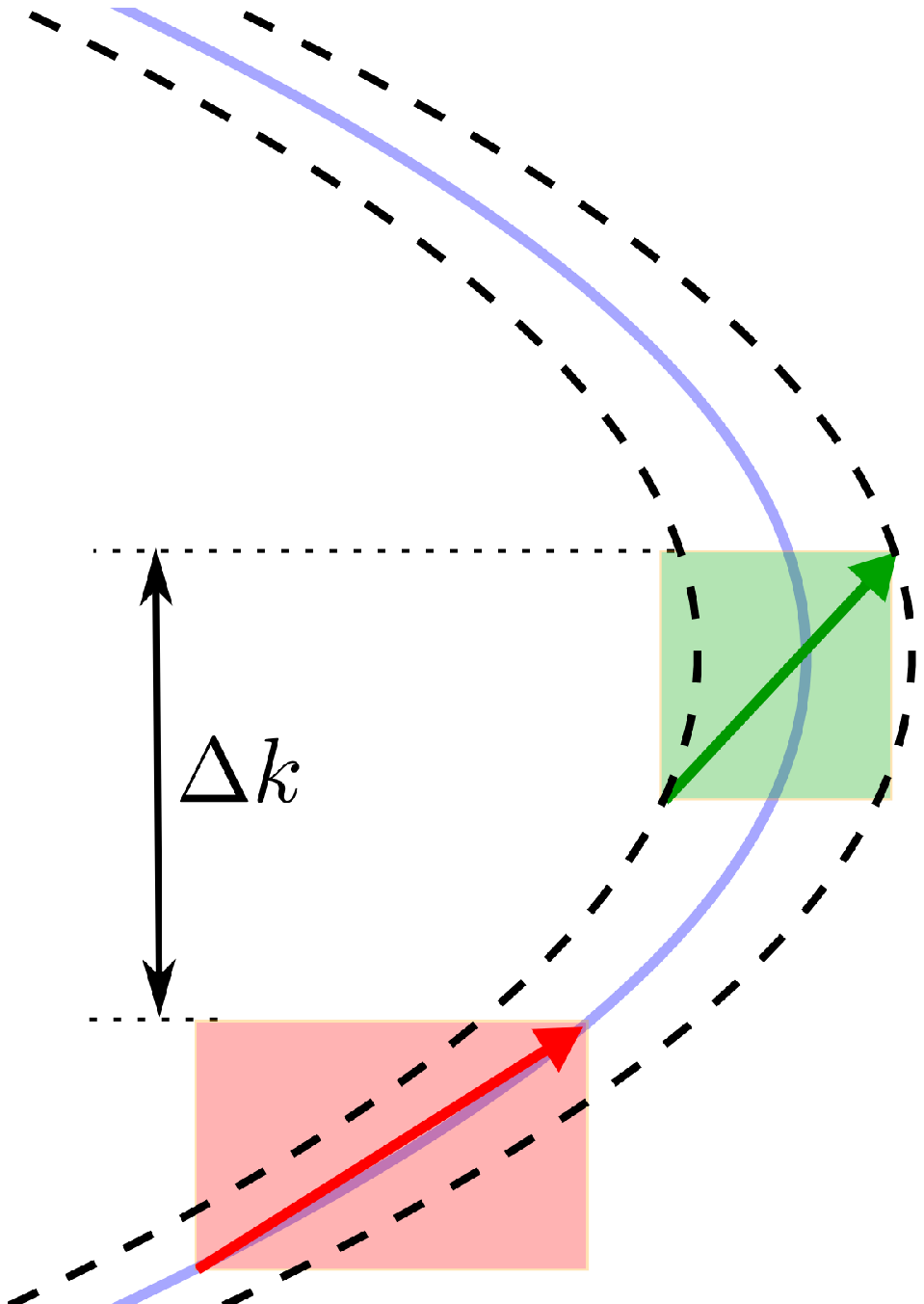}}
\caption{
The solid curve is the Fermi surface, 
and the dashed curves represent the Fermi surface shifted 
by $\pm \Lambda_x$ along the $k_x$ direction.
(a) The two arrows represent 
the minimum and maximum $x$-momenta
a particle-hole pair 
$\psi^*_{i,  k +  p} \psi_{j, k}$
with fixed $k$ and $p$
can have 
within the thin shell of width $2 \Lambda_x$. 
(b) The x-momenta of two pairs of particle-hole excitations 
created by $\psi^*_{i,  k + p} \psi_{j, k}$ and 
$\psi^*_{i,  k + \Delta k +  p} \psi_{j, k  + \Delta k}$
can not be resolved within the uncertainty $2 \Lambda_x$
if $\Delta k$ is too small.
This is because the allowed $x$-momenta for the first pair 
overlap with those for the second.
(c) If $\Delta k$ is large enough, the x-momenta of the two 
pairs can be resolved because the largest possible 
$x$-momentum of the first particle-hole pair is smaller 
than 
the smallest possible momentum of the second pair.
}
\label{fig:UC}
\end{figure}

The phase factor in the four-fermion vertex 
is reminiscent of the Moyal product in non-commutative 
field theories\cite{NC}
although it is not exactly of the same form.
Indeed, it suggests that 
there is an emergent non-commutative structure 
between $x$-coordinate and $y$-momentum at low energies.
In this section and Appendix \ref{app: 1}, 
we will elaborate on this point.
This section is rather orthogonal to the rest of the paper,
and it may be skipped by the readers
who want to reach
the main conclusion of the paper quickly.

Physically, the phase factor represents 
a mismatch of $x$-momenta
between low-energy fermions 
which carry different $y$-momenta. 
This is a consequence of the fact that $x$-momentum is tied 
with $y$-momentum at low energies due to the curvature of 
the Fermi surface.
Suppose $\Lambda_x$ is a UV cut-off in energy, 
which is equivalent to the largest $x$-momentum 
a particle or hole can have away from the Fermi surface.
Let us consider low-energy particle-hole pairs 
that can be created within the thin shell with width
$2 \Lambda_x$ near the Fermi surface.
When $\Lambda_x=0$, 
the $x$-momentum is completely determined 
from the $y$-momenta of the particle and hole :
a particle-hole pair created by
$\psi^*_{i,  k +  p} \psi_{j, k}$
carries $x$-momentum, 
$-\gamma[ (k+p)^2 - k^2 ]$.
When $\Lambda_x \neq 0$, 
there is an uncertainty $2 \Lambda_x$
in the $x$-momentum of the particle-hole pair 
with fixed $p$ and $k$
as is illustrated in Fig. \ref{fig:UC} (a).
Now we consider a second particle-hole pair
created by $\psi^*_{i,  k +\Delta k +  p} \psi_{j, k+\Delta 
k}$, which carries the same net $y$-momentum $p$
but is made of different constituents.
Due to the non-zero curvature of Fermi surface, 
the second pair has a different net $x$-momentum. 
However, if $\Delta k$ is too small, 
the $x$-momentum of the second pair 
can not be distinguished from that of
the first pair within the
uncertainty $2 \Lambda_x$,
as is shown in  Fig. \ref{fig:UC} (b).
Therefore, $\Delta k$ has to be large enough 
for the $x$-momenta of the two particle-hole pairs 
to be resolved (see  Fig. \ref{fig:UC} (c)).
In order for the difference 
in the $x$-momenta of the two pairs 
to be greater than $2 \Lambda_x$,
$\Delta k$ has to satisfy the inequality,
\begin{align}
& \abs{\gamma \left[ ( k + \Delta k +  p )^2 
- (  k + \Delta  k )^2 \right] 
- \gamma \left[ (  k +  p )^2 
-   k^2 \right]}  \nn
& = \abs{2 \gamma p \Delta k} > 2 \Lambda_x.
\end{align}
This implies 
\bqa
\Delta  k \Delta x > \frac{1}{\gamma p},
\label{UC}
\eqa
where we use the fact that the uncertainty
in the real space  
is given by $\Delta x \sim \frac{1}{\Lambda_x}$.
Without loss of generality, $\Dl k$, $p$ and $\Dl x$ 
are 
assumed to be positive.
Therefore, there is a non-trivial uncertainty relation
between the $x$-coordinate and the $y$-momentum,
which is inversely proportional to the curvature
and the net $y$ momentum of a particle-hole pair.
This indicates that the zero curvature limit is singular.
It is also interesting to note that the `Planck constant' 
in the uncertainty relation is dimensionful.

This uncertainty relation implies that the theory has an IR 
scale which is inversely proportional to a UV scale.
For example, we can construct an IR scale 
for the momentum of a particle-hole pair,
\bqa
p^* \sim \frac{1}{\gamma \Delta k^{max} \Delta x} \sim 
\frac{\Lambda_x}{\gamma \Lambda}
\label{eq:uvir}
\eqa
from a UV scale $\Lambda$, 
where we use the fact that the largest uncertainty in 
$k$ is given by the size of Fermi surface $\Lambda$.
It is noted that $\Lambda$ 
which corresponds to the Fermi momentum 
is in general much 
larger than $\Lambda_x$ 
which is set by the low energy scale, say temperature. 
Conversely, a small transverse momentum 
of a particle-hole pair $p$
sets a UV momentum scale given by
\bqa
\Lambda^* \sim \frac{\Lambda_x}{\gamma p}.
\label{eq:iruv}
\eqa
%

\begin{figure}[!ht]
   \centering
\includegraphics[width=0.5\columnwidth]{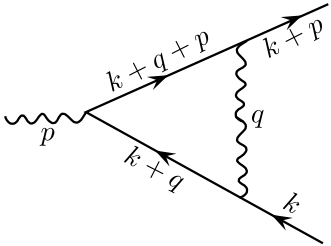}
 \caption{One-loop vertex function.}
 \label{f.v1}
\end{figure}

\begin{figure}[!ht]
\centering
\subfigure[]{\includegraphics[width=0.6\columnwidth]
{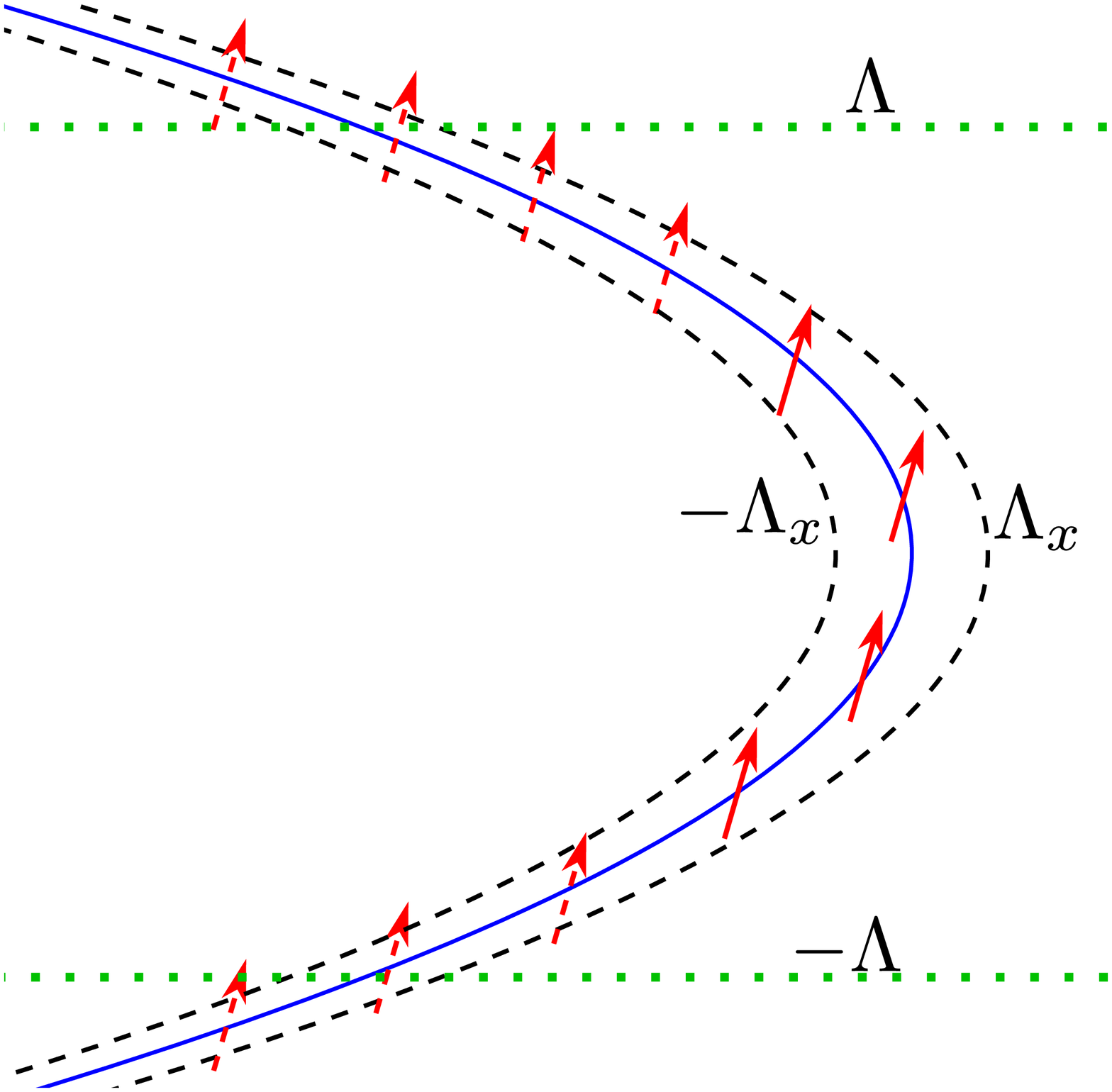} \label{fig: finite_FS_big}}
 ~ \\
\subfigure[]{\includegraphics[width=0.6\columnwidth]
{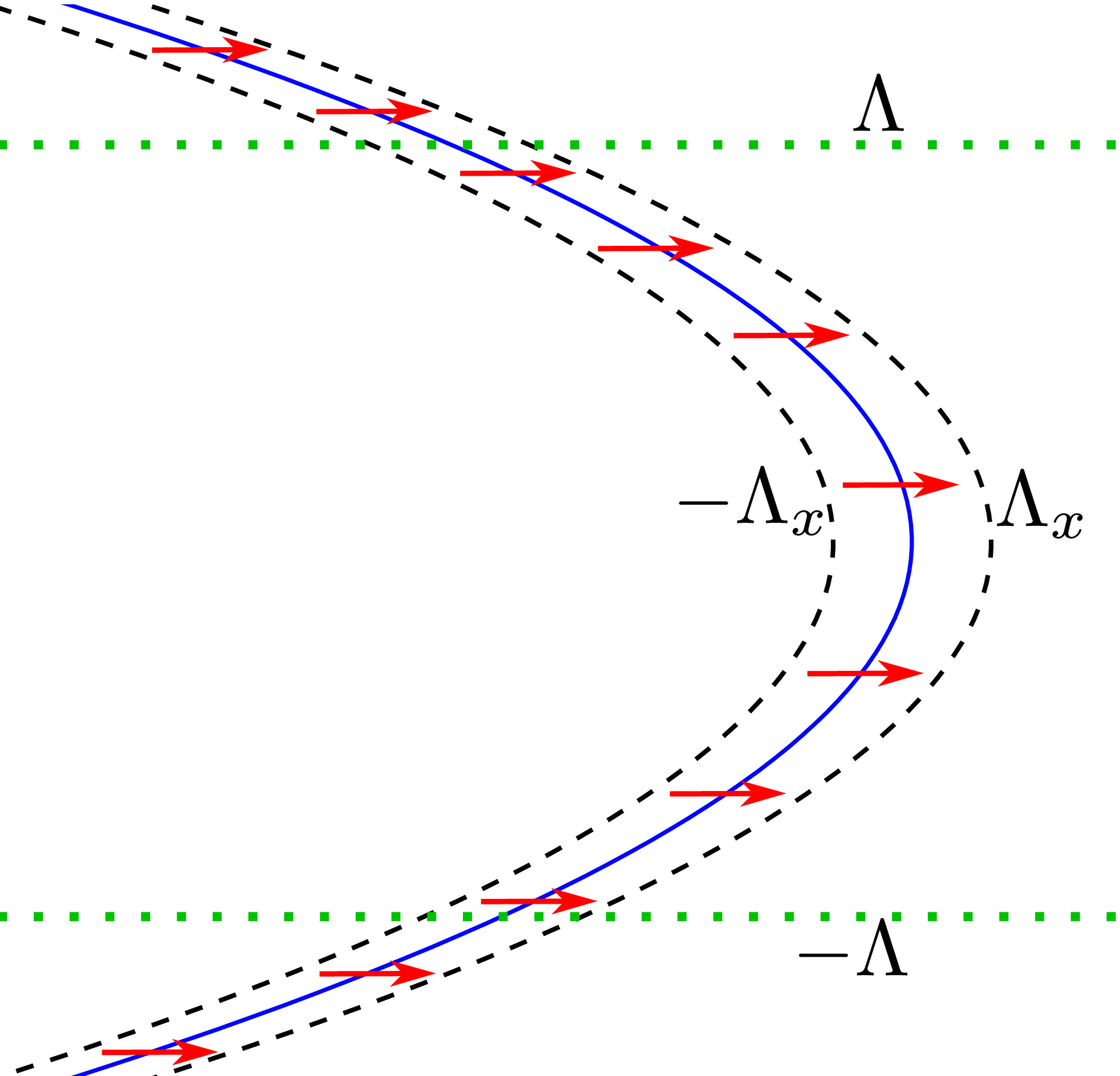}  \label{fig: finite_FS_small}}
     \caption{ The solid curve is the Fermi surface, 
and the dashed curves represents the Fermi surface shifted 
by $\pm \Lambda_x$ along the $k_x$ direction.
The (red) arrows denote particle-hole pairs 
with momentum $\vec p$ created near 
the Fermi surface.
The dotted horizontal lines represent the UV cut-off in the
$y$-momentum $(\pm \Lambda)$,
which is set by the size of the Fermi surface.
(a) If the $y$-momentum of a particle-hole pair, $p$, is 
large, then the phase space available to the pair within
energy $\Lambda_x$ is limited by $\frac{\Lambda_x}{\gamma
p}$.
Therefore the arrows go outside the thin shell 
before they sense the full extension of the Fermi surface.
(b) If $p$ is small, then the available phase space of the
particle-hole pairs is only limited by the size of Fermi
surface.
}
\label{fig: finite_FS}
\end{figure}

In order to appreciate the physical meaning
of Eqs. (\ref{eq:uvir}) and (\ref{eq:iruv}),
let us consider an one-loop vertex function
shown in Fig.  \ref{f.v1}.
Here a boson with three-momentum 
$(\omega_p, \vec p)$ creates
virtual particle-hole excitations 
with $(\om_k + \om_p + \nu, \vec k + \vec p + \vec q)$ and 
$(\om_k + \nu,\vec k + \vec q)$
before the intermediate state
settles at the final state with $(\om_k + \om_p, \vec k + 
\vec p)$ and $(\om_k, \vec k)$.
The virtual particle-hole excitations that 
contribute to this scattering amplitude
are the ones whose net momentum is $\vec p$
with an energy cut-off $\Lambda_x \sim \om_p$.
In Fig. \ref{fig: finite_FS}, 
the arrows represent the momentum $\vec p$
which is shifted along the Fermi surface to 
show possible particle-hole pairs 
that can be excited with the net momentum $\vec p$.
Those with energy less than the cut-off,
which fit inside the shell with width $\Lambda_x$, 
are denoted as solid arrows.
Those with energy greater than the cut-off
are drawn as dashed arrows.
If $p$ is large (relative to what we will explain below),
only a small region near the Fermi surface can
accommodate the arrows within the shell
as is shown in Fig. \ref{fig: finite_FS_big}.
In this case, 
the largest momentum that the constituent
particle/hole can have is cut off by \eq{eq:iruv}
which is independent of $\Lambda$.
Namely, the virtual particle-hole excitations
do not sense the full extent
of the Fermi surface. 
As $p$ becomes smaller, 
more arrows can fit inside the shell.
Eventually the largest momentum of the constituent 
particle/hole
is set by the UV cut-off $\Lambda$
as is shown in Fig. \ref{fig: finite_FS_small}.
The momentum scale of $p$ at which this 
crossover occurs is given by \eq{eq:uvir}.

As a result of the interplay between IR and UV scales, 
IR behaviors of the theory can depend on
UV scales in a non-trivial way\cite{Minwalla}.
This is particularly the case 
if there is UV divergence in the theory.
Since the theories at and above the upper critical dimension
are sensitive to UV,
we expect a non-trivial UV/IR mixing 
in Fermi liquids (marginal Fermi liquids) with ${\theta} < 
1$ (${\theta} = 1$).
Let us ask how the scattering amplitude at a fixed $\vec p$ 
behaves as the UV cut-off $\Lambda$ is increased.
When $p$ is small compared to $p^*$ defined in \eq{eq:uvir},
the phase space for the intermediate states
increases as $\Lambda$ increases. 
As a result, the scattering amplitude 
 grows with a positive power of $\Lambda$
in Fermi liquids with ${\theta} < 1$,
as is shown in Appendix \ref{app: 1}.
For sufficiently large $\Lambda$, 
this UV divergence is eventually cut off by 
the scale given in \eq{eq:iruv}, 
giving a singular dependence on $p$.
We can also view this from a different perspective.
For a fixed $\Lambda$, 
the scattering amplitude grows 
as $p$ decreases as long as $p$ is larger than $p^*$.
When $p$ becomes smaller than $p^*$,
the putative IR singularity in $p$ is eventually cut off by $p^*$.
Therefore, 
the $\Lambda \rightarrow \infty$ limit
and
the $p \rightarrow 0$ limit
do not commute.

In Fermi liquids, the UV and IR singularities 
of the amplitude are closely connected.
This is not surprising because  
the modes that carry large momenta can have 
arbitrarily small energy.
The UV/IR mixing is one way of understanding 
why `UV structures' of the theory
should be specified 
in the low energy effective theory for Fermi liquids.
The shape of the entire Fermi surface and
the Landau parameters which are non-local in momentum space 
are among the UV data without which 
even the properties that are local in momentum space
can not be determined in Fermi liquids.
On the contrary,  
the UV/IR mixing is suppressed
in non-Fermi liquid states with ${\theta} > 1$ 
because the theory is UV finite 
and insensitive to the UV cut-offs.
As a result, properties that are local in momentum space,
such as the vertex functions with small momentum transfers,
can be obtained only from the patch which is local in 
momentum space without invoking the knowledge of the entire 
Fermi surface.
The suppression of  UV/IR mixing is what makes the 
local-patch
description possible in non-Fermi liquid states.
We illustrate this difference between Fermi liquids
and non-Fermi liquids in Appendix \ref{app: 1}
through an explicit calculation of
the vertex function.

\section{Regularization and RG prescription}
\label{sec:regularization}

\begin{figure}[!ht]
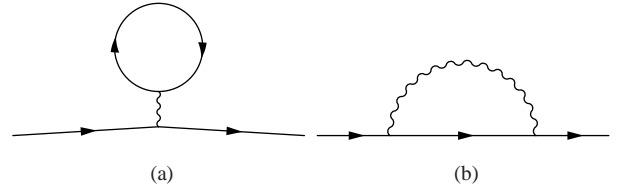

\centering     
\subfigure[]{\includegraphics[width=0.45\columnwidth]{HF.2} 
\label{fig:hatree}}         
\subfigure[]{\includegraphics[width=0.45\columnwidth]{HF.1} 
\label{fig:fock}}
\caption{Corrections to the quadratic vertex due to 
normal ordering of the quartic vertex.}
\label{fig: HF}
\end{figure}

In the following sections,
we will perform a renormalization group (RG) analysis of
the action in \eq{eq: action} 
to show that the theory flows 
to a stable interacting fixed point 
in the low energy limit.
As a first step, we regularize the theory.
For this, we will use
the `mixed' space representation 
which consists of 
$x$ coordinate and frequency.
This representation is convenient
because we will adopt a RG prescription
where the Wilsonian effective action is local in $x$ 
but not in $\tau$. 
In the mixed space,
the bare action 
in \eq{eq: action}
is written as
\begin{align}
 S  & =  
\int \msr{k} ~ \msr{\om} \int dx ~ \psi_{i, {k}}^*(\om,x) 
\Bigl[  i\eta \om  - i\partial_x \Bigr] \psi_{i, 
k}(\om,x)  \nn 
& +  \int \msr{k_1} ~ \msr{k_2} ~ \msr{q} ~ 
\msr{\om_1} ~ \msr{\om_2} ~ \msr{\nu} \int dx
~~ e^{2i \gamma ( k_1 -  k_2 )  q x }  \nn
& \qquad \times   V_{ij;ln}~ \chi_\theta(q) ~ 
\psi^*_{i, k_2}(\om_2,x) ~ \psi_{j, k_2+q}(\om_2+\nu,x) \nn
& \qquad \times ~ \psi^*_{l, k_1+q}(\om_1+\nu,x) ~
\psi_{n,k_1}(\om_1,x).
\label{eq: mixed_action}
\end{align}
Potential divergences present in the composite operators 
are removed
by writing them in terms of 
normal ordered operators\cite{polch_book} 
defined by
\begin{align}
 & \ccolon \mathcal{O} \ccolon ~    = \exp{\left[ - \sum_i 
\int dk ~ d\om \int dx_1 ~ dx_2 ~ \right. } \nn
&  \times (2\pi)^2 ~ G_0(\om,x_{12}) { \left. 
\frac{\dl}{\dl \psi^*_{i,k}(\om, x_2)} ~
\frac{\dl}{\dl \psi_{i,k}(\om,x_1)} \right]} \mathcal{O}, 
\label{eq: norm_ordr}
\end{align}
where the bare Green's function is 
\begin{align}
      G_0(\om,x_{12}) &= \int_{-\infty}^{\infty} 
\msr{\eps_k} ~ e^{i  x_{12} \eps_k}
~ \frac{1}{i \eta \om + \eps_k}  \nn
   &= - i ~ \sgn{\om} ~ \Theta(-x_{12} \om) 
~e^ {-\eta \abs{x_{12}} \abs{\om} } \label{eq: bare_G}
\end{align}
with  $x_{12} \equiv x_1-x_2$. 
The chiral nature is manifest in the fact that
the fermion propagator vanishes for $x_{12} \omega > 0$ :
particles (anti-particles) propagate only in one (the 
other) direction.
This feature will play a crucial role in 
proving the stability of the chiral non-Fermi liquid state 
later.
It is convenient to use the diagrammatic representation
to visualize various channels in which fields are 
contracted in the normal ordering and the operator product 
expansion (OPE). 
Two contracted fields are represented by an internal line.
External lines represent uncontracted fields.
Wiggly lines represent the four-fermion vertices 
or the propagator of the boson which has been integrated
out. 
Upon normal 
ordering, the quartic vertex produces
normal ordered 
quartic and quadratic vertices along with a constant.
The quadratic vertex is generated from the quartic vertex 
as a pair of fermion fields are contracted as is shown in 
\fig{fig: HF}. 
In order to contract two fermion fields within 
one composite operator without ambiguity,
we introduce a point splitting in the four-fermion composite
operator in \eq{eq: mixed_action} as
\begin{align}
& \psi^*_{i, k_2}(\om_2,x + \eps) ~ \psi_{j,
k_2+q}(\om_2 + \nu, x + \eps)  \nn
& \quad \times ~ \psi^*_{l, k_1 + q}(\om_1 + \nu, x) ~
\psi_{n, k_1}(\om_1, x).
\end{align}
The Hartree contribution from \fig{fig:hatree} vanishes 
due to the traceless condition of the interaction vertex, 
$\sum_{ij}
\dl_{ij} V_{ij;nl} = 0$.
The Fock term in \fig{fig:fock} is non-vanishing and given 
by
\begin{align}
S_2^{'} &=  \int \frac{dk ~d\om}{(2\pi)^2} dx ~ \ccolon
\psi^*_{i,k}(\om,x) \psi_{i,k}(\om,x) \ccolon \nn
& \quad \times i g^2 v ~ \sgn{\om}  \int
\frac{dq}{2\pi} \int_{0}^{|\om|} \frac{d\nu }{2\pi}
~\chi_{\theta}(q), \label{eq: fock}
\end{align}
where the constant $v$ is defined by
\eqn{
v = \frac{1}{N} \sum_{\alpha} 
\tr{  T^{\alpha} T^{\alpha} }.
\label{eq: v_defn}
}
Here the $\eps \rtarw 0$ limit is taken
before the UV cut-off for frequency
is taken to be infinite.

The partition function can be formally expanded as
\begin{eqnarray}
Z &=& \int \mathcal{D}\psi ~
e^{-S_2}
\Bigl(1 -  (S_2^{'} + S_4)  + \frac{1}{2}(S_2^{'} + S_4)^2  
- \ldots
\Bigr),  \nn
\label{z}
\end{eqnarray}
where 
\begin{align}
S_2 &= \int \frac{d k ~ d\om}{(2\pi)^2} \int dx ~ 
 \psi_{i, k}^*(\om,x) \Bigl[  i\eta
\om - i\partial_x \Bigr] \psi_{i, k}(\om,x), \nn
S_2^{'} &=  \int \frac{dk ~d\om}{(2\pi)^2} \int dx ~ \ccolon
\psi^*_{i,k}(\om,x) \psi_{i,k}(\om,x) \ccolon
\nn
& \quad \times i g^2 v ~ \sgn{\om}  \int
\frac{dq}{2\pi} \int_{0}^{|\om|} \frac{d\nu }{2\pi}
~\chi_{\theta}(q), \label{eq:S2} \\
S_4 &= \int \frac{dk_1 ~ dk_2 ~ dq ~ d  \om_1 ~ 
d\om_2 ~ d{\nu}}{(2\pi)^6} \int dx
~~ e^{i 2 \gamma ( k_1 -  k_2 )  q x }  \nn
& \qquad \times   V_{ij;ln} ~\chi_\theta(q) ~ 
\ccolon \psi^*_{i, k_2}(\om_2,x) ~ \psi_{j, 
k_2+q}(\om_2+\nu,x) \nn
& \qquad \times ~ \psi^*_{l, k_1+q}(\om_1+\nu,x) ~
\psi_{n,k_1}(\om_1,x) \ccolon. 
\label{eq:S4}
\end{align}
We view this as a grand canonical ensemble 
for a gas of vertices $S_2^{'} + S_4$ 
evaluated with respect to the quadratic action\cite{Cardy}.
It is noted that the action is local
in the one-dimensional space in $x$,
and operators in the ensemble 
can be arbitrarily close to each other in $x$-direction.
This can in principle give rise to UV divergences.
However, we will see that UV divergence is absent 
in the present theory due to chirality for  $\theta > 1$.
Therefore we proceed without imposing a short distance 
cut-off in the $x$-direction.

\begin{figure}[!ht]
 \centering
 \includegraphics[width=0.95\columnwidth]{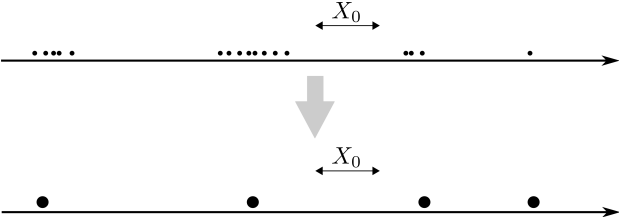}
 \caption{Any group of operators, represented by small 
dots in the upper panel, where separations 
between nearest neighbors are less than $X_0$ in $x$ 
direction are 
fused into a composite operator which are  denoted by big 
dots in the lower panel.}
 \label{fig: op_gas}
\end{figure}

Now, we consider a Wilsonian effective action
with a running cut-off length scale $X_0$.
The Wilsonian effective action 
is constructed by fusing all operators 
whose relative distances
in $x$-direction is smaller than $X_0$
in the ensemble of operators, 
$\int dx_1 dx_2 \ldots dx_n  
 O(x_1) ~ O(x_2) \ldots  O(x_n) $.
Here the frequencies and other indices are suppressed.
For example, let us consider the $n$ normal ordered 
operators $ O(x_1), O(x_2), \ldots, O(x_n)$ in $( S_4
)^n$ of \eq{z}
located at positions $x_1, x_2, ..., x_n$.
If there is a group of operators 
$ O_{i_1},  O_{i_2}, \ldots, O_{i_m}$ in $(S_4 )^n$ such
that for every operator in the group, say $O_{i_p}$, 
there exists another operator $ O_{i_q}$ in
the group
with $\abs{x_{i_p} - x_{i_q}} < X_0$, 
then the cluster of the $m$ operators are 
fused into a series of 
normal ordered operators according to the OPE,
\begin{align}
 \prod_{a=1}^{m} O_{i_a} ~ = 
~ e^{\widehat{\mc{D}}_m} ~ \ccolon \prod_{a=1}^{m}
 O_{i_a} \ccolon, 
\label{eq: fusing_D}
\end{align} 
where
\begin{align}
  \widehat{\mc{D}}_m \equiv 
\sum_{c = 2}^{m} 
\sum_{s = 1}^{c-1} 
\widehat{d}_{c,s} \label{eq: D}
\end{align}
with
\begin{align}
& \widehat{d}_{c,s} = \sum_j \int dk d\om \int dx_1 
dx_2 \nn
& \times  \lt[ G_0(\om,x_{12}) 
\frac{\dl}{\dl \psi^{(c)*}_{j,k}(\om, x_2)} ~
\frac{\dl}{\dl \psi^{(s)}_{j,k}(\om,x_1)} + ~ \mbox{c. c.} 
\rt]. \label{eq: d}
\end{align}
Here the role of $\widehat{d}_{c,s}$ is to contract a pair
of fermion fields, one from  $ O_{i_c} $ 
and the other from $ O_{i_s} $.
Fusion of operators is illustrated in Fig. \ref{fig: 
op_gas}. 
The Wilsonian effective action
should include 
the vertices generated from the OPE.
In particular, four-fermion vertices 
are generated from 
contracting $2(n-1)$ pairs of fermion fields 
in $\lt(S_{4} \rt)^n$, 
\begin{align}
\dl S_4 = - \sum_{n=2}^{\infty}
\frac{(-1)^n}{n!}  \frac{( \what{\mc{D}}_n 
)^{2(n-1)}}{(2n-2)!} ~
\ccolon 
\lt[ \lt( S_{4} \rt)^n \rt]_{X_0}
\ccolon, 
\label{eq: schem4}
\end{align}
where 
$\lt[ \lt( S_{4} \rt)^n \rt]_{X_0}$ denotes
the configurations for a group of $n$ quartic vertices 
where the separation between nearest neighbor vertices 
are less than $X_0$ in $x$-direction.
For example,
\begin{align}
& \lt[ \lt( \int dx ~ O(x)  \rt)^2 \rt]_{X_0} 
 = \int \limits_{|x_1-x_2|<X_0} d x_1 dx_2
 O(x_1)~ O(x_2).
\label{eq:OX}
\end{align}
Extension of \eq{eq:OX} to 
$\lt[ \lt( S_{4} \rt)^n \rt]_{X_0}$ with general $n$
is straightforward, if more complicated.
Similarly, quadratic vertices are generated 
by fusing $2n-1$ pairs of fermion fields,
\begin{align}
\dl S_2 & =  -  \sum_{n=2}^{\infty}
\frac{(-1)^n}{n!} 
\frac{( \what{\mc{D}}_n)^{2n-1}}{(2n-1)!} ~
\ccolon 
\lt[ \lt( S_{4}  \rt)^n  \rt]_{X_0}
\ccolon.
\label{eq: schem2}
\end{align}

\begin{figure}[!ht]
 \centering
 \includegraphics[width=0.9\columnwidth]{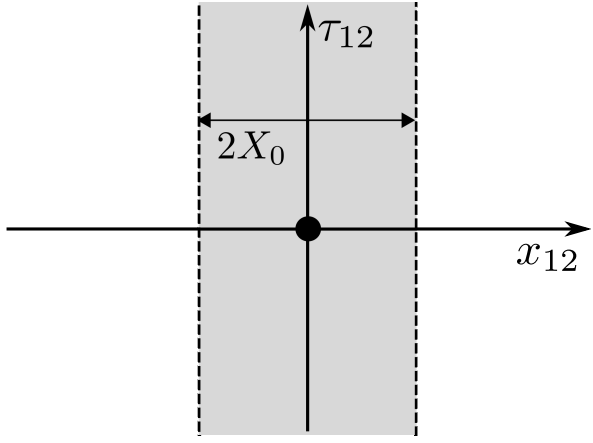}
 \caption{The origin represents the position of one 
operator 
in the $(1+1)$-dimensional real space. If the second 
operator 
is within the shaded region in the space 
of the relative coordinate $(x_{12}, \tau_{12})$,
two operators are fused into one normal ordered operator.}
 \label{fig: cutoff_scheme}
\end{figure}

It is noted that we have employed an unconventional cut-off 
scheme.
In the $(1+1)$-dimensional real space, 
two operators that are far from each other in the temporal
direction are fused as far as their spatial separation is 
less than $X_0$
as is shown in Fig. \ref{fig: cutoff_scheme}.
The reason why we choose this unusual cut-off scheme is 
that the dynamical
critical exponent $z$ is not known a priori.
Since $z$ should be determined dynamically,
we don't know yet how to re-scale the temporal direction 
relative to the spatial
direction under scale transformation.
The present cut-off scheme is convenient 
because it is guaranteed to be invariant under the scale 
transformation with
any $z$.
The price we have to pay is that the fusion processes 
generate operators that
are non-local in the temporal direction.
For this reason, 
we have to explicitly compute the non-local terms 
that are generated from the fusion processes, 
and add them to the effective action.
This is different from the usual 
procedure for relativistic field theories
where UV cut-off is imposed in all space-time directions,
and a fusion of operators
generates only local operators that are already in the bare 
action.

\section{The Wilsonian effective action} \label{sec:con}

\begin{figure}[!ht]
 \centering 
\subfigure[]{\includegraphics[width=0.8\columnwidth]
{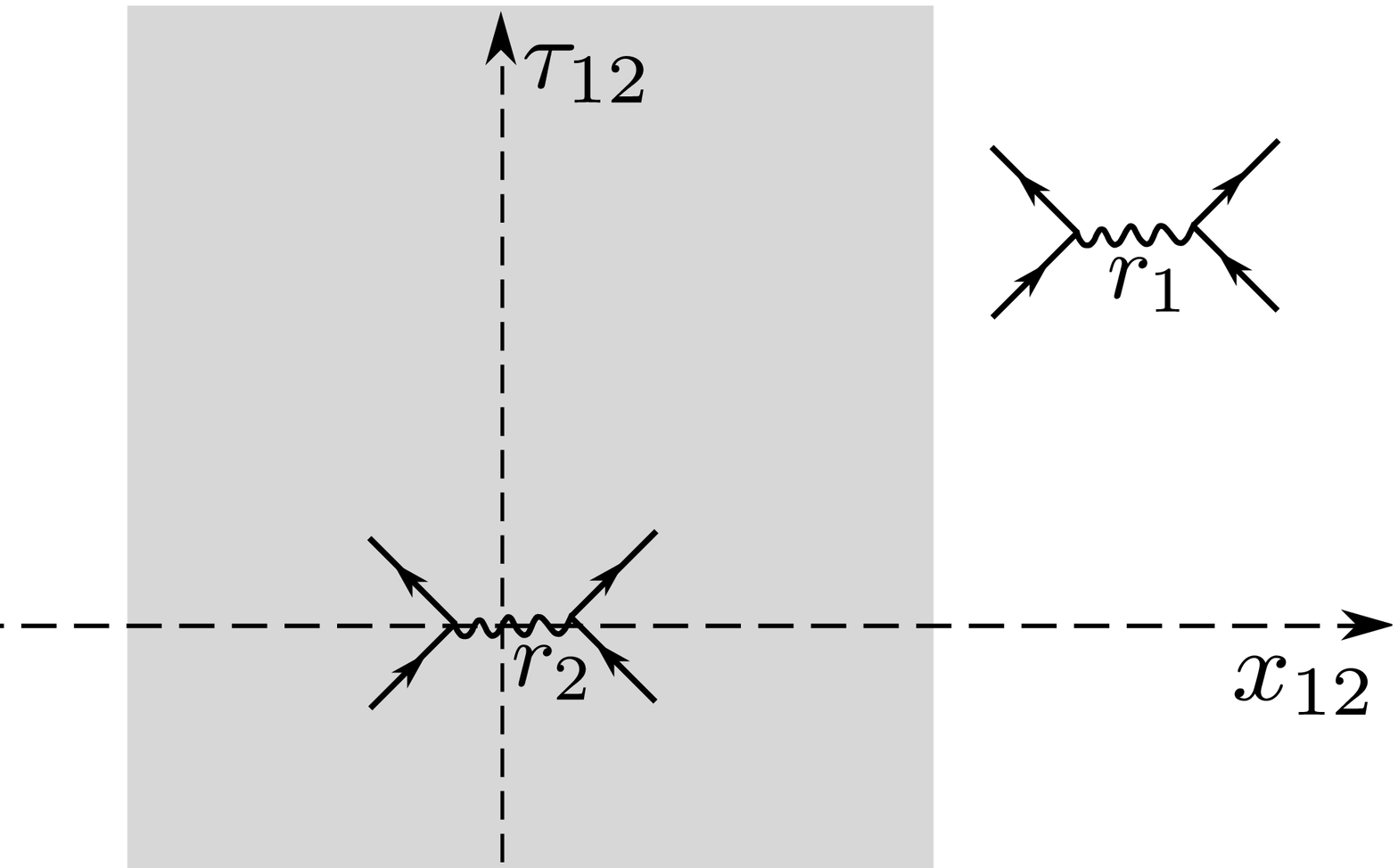} \label{fig: outside_strip}}
\subfigure[]{\includegraphics[width=0.65\columnwidth]
{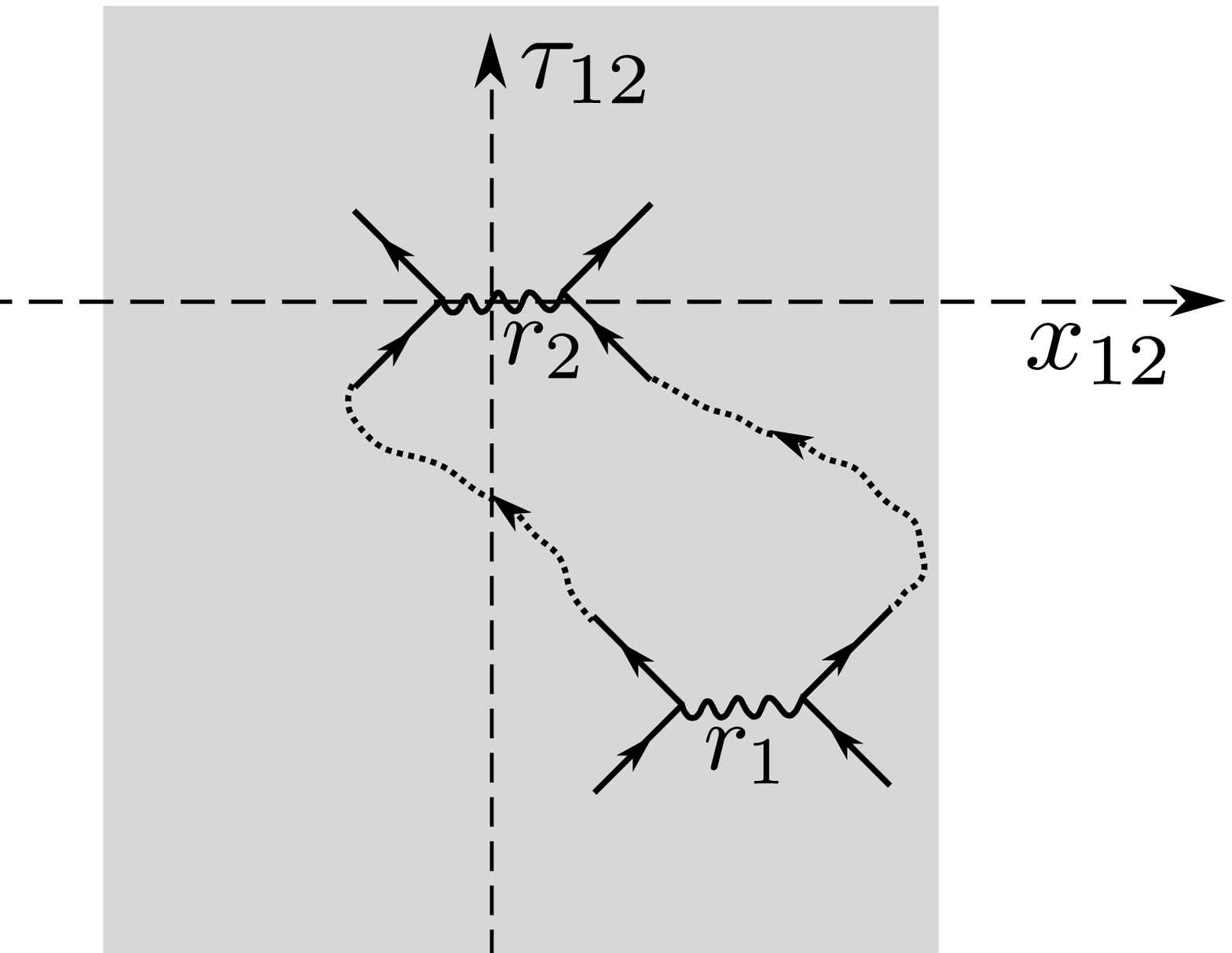} \label{fig: inside_strip}}
\caption{ 
The space of relative coordinate between two operators.
Each quartic vertex is drawn as an extended object 
although there is no spatial extension of the wiggly line
in the $x$-direction.
The shaded region 
represents $\abs{x_{12}} < X_0$ (see Fig. \ref{fig: 
cutoff_scheme}). 
(a) If the separation between two
operators along the $x$-axis is greater than $X_0$,
the two operators are 
considered as independent operators. 
(b) If the separation 
between the two operators along the $x$-axis is less than 
$X_0$, 
they are fused into one operator. 
Here we show a particular channel of fusion
where two quartic operators are fused into 
one quartic operator. 
Each internal (dotted) line represents a pair of contracted 
fields
whereas external (solid) lines represent uncontracted 
fields. 
}
\label{fig: diagram_analogy}
\end{figure}

In this section, 
we prove the stability of 
the chiral patch theory.
For this, we consider the Wilsonian effective action 
at scale $X_0$ constructed in the following way.
When the separation between two operators along the 
$x$-axis is larger than $X_0$ 
(see Fig. \ref{fig: outside_strip}), 
they are treated as independent operators. 
When their separation is less than $X_0$ 
(see Fig. \ref{fig: inside_strip}),
they are fused through OPE. 
Different ways of contracting fields are 
represented by Feynman diagrams.
Before we compute the Wilsonian effective action explicitly,
we first discuss about the general structure of 
the effective action inferred from scaling analysis.
In particular, we will show that 
the Wilsonian effective action is UV finite and 
has no scale except for the running cut-off scale $X_0$
in the low energy limit.
As a result, the exact scaling behavior of the theory
can be obtained.
After the general consideration, 
we will compute the Wilsonian effective action explicitly,
and confirm the general conclusion obtained from the 
scaling analysis.

\subsection{UV finiteness and scale invariance} 
\label{sec:UV_SC}

In general, 
the Wilsonian effective action at scale $X_0$
can be written as
\begin{widetext}
\bqa
S_{X_0} & = & 
\sum_{m=1}^\infty ~
\sum_{i_1,j_1,\ldots,i_m,j_m=1}^N ~
\sum_{s_1,t_1,\ldots,s_m,t_m=0}^\infty
\int dq_1 dk_1 \ldots dq_m dk_m
\int d\om_1 d\nu_1 \ldots d\om_m d\nu_m
\int dx \nn
&&
\times S^{s_1,t_1,\ldots,s_m,t_m}_{i_1,j_1,\ldots,i_m,j_m}
\left(x, q_1,k_1,\ldots,q_m,k_m,\om_1,\nu_1,\ldots,\om_m,
\nu_m; X_0,\Lambda,\eta,V_{ij;ln}, \mu \right)  
\delta \left( \sum_{a=1}^m ( q_a - k_a ) \right)
\delta \left( \sum_{a=1}^m ( \om_a - \nu_a ) \right)
\nn
&&
\times \left( \partial_x^{s_1} \psi^*_{i_1, q_1}(x,\om_1)
\right) 
\left( \partial_x^{t_1} \psi_{j_1, k_1}(x,\nu_1) \right) 
\ldots 
\left( \partial_x^{s_m} \psi^*_{i_m, q_m}(x,\om_m) \right) 
\left( \partial_x^{t_m} \psi_{j_m, k_m}(x,\nu_m) \right).
\eqa
Here $\Lambda$ is the UV cut-off for $y$-momentum.
One can reduce the number of independent arguments of
the effective action by one using scaling.
It is straightforward to check that there exists
no scaling under which all terms in \eq{eq: mixed_action}
are invariant for $\theta > 1$.
However, there are two natural choices of scaling.
The first is the Gaussian scaling in \eq{eq:GS}
under which the quadratic term is scale invariant
while the quartic vertex grows at low energies.
Under the Gaussian scaling,
$S^{s_1,t_1,\ldots,s_m,t_m}_{i_1,j_1,\ldots,i_m,j_m}$ has
the scaling dimension $-(3m-5)/2 - \sum_{i=1}^m (s_i + 
t_i)$,
and it can be written as
\bqa
&& S^{s_1,t_1,\ldots,s_m,t_m}_{i_1,j_1,\ldots,i_m,j_m}
\left(x,q,k,\om,\nu; X_0,\Lambda,V,\eta, \mu \right)   \nn
&& = X_0^{ (3m-5)/2 + \sum_{a=1}^m (s_a + t_a) }
\bar S^{s_1,t_1,\ldots,s_m,t_m}_{i_1,j_1,\ldots,i_m,j_m}
\left( x X_0^{-1}, q X_0^{1/2}, k X_0^{1/2}, \om X_0, \nu
X_0; \Lambda X_0^{1/2}, V X_0^{({\theta} - 1)/2}, \eta, \mu
X_0^{1/2} \right),
\label{eq:GSF}
\eqa 
\end{widetext}
where the subscripts in 
$q_a,k_a,\om_a,\nu_a,V_{ij;ln}$ are omitted to avoid 
clutter in notation.
In the long-distance limit with $X_0 \rightarrow \infty$,
$\Lambda X_0^{1/2}$ and $V X_0^{({\theta} - 1)/2}$ diverge 
for 
${\theta} > 1$.
This is expected because the interaction is relevant
at the Gaussian fixed point for ${\theta} > 1$.
If the effective action has singular dependence on 
the divergent parameters,
which is certainly the case for the effective action 
computed perturbatively in $V$,
the scaling dimensions in \eq{eq:GS} are modified
by quantum corrections.
However, the scaling form in \eq{eq:GSF} is not useful 
in extracting the low energy behavior
of strongly interacting theories 
unless the singular dependence of 
$\bar S^{s_1,t_1,..,s_m,t_m}_{i_1,j_1,..,i_m,j_m}$
on the divergent parameters
are exactly known.

\begin{figure}[!ht]
 \centering
 \includegraphics[width=0.25\columnwidth]{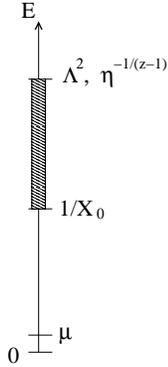}
 \caption{
In the Wilsonian effective action,
`short-distance modes' 
between UV scales set by $\Lambda^2$, $\eta^{-2/({\theta} - 
1)}$
and IR scale $X_0^{-1}$ are integrated out.
In the chiral theory, the resulting Wilsonian effective 
action
is regular in the $\Lambda \rightarrow \infty$
and $\eta \rightarrow 0$ limit.
As a result, only the running cut-off scale $X_0$
enters as a scale of the effective action
in the $\mu X_0^{1/2} \rightarrow 0$ limit.
}
 \label{fig: scale}
\end{figure}

There exists an alternative scaling 
from which we can extract 
exact scaling behaviors in the chiral theory.
This is the scaling where 
the interaction is kept  invariant
at the expense of making $\eta$ irrelevant.
The requirement that the quadratic term,
$\psi^* \partial_x \psi$ 
and the quartic interaction 
in \eq{eq: mixed_action}
remain invariant uniquely fixes the dimension of frequency 
to be $z=\frac{{\theta} + 1}{2}$.
Under this scaling, we assign the following scaling 
dimensions
to momenta and fields,
\begin{align}
{[}x{]}  & =   -1, \nn
{[}k{]}  & =  \frac{1}{2}, \nn
z \equiv {[}\om{]}  & =   \frac{{\theta} + 1}{2}, \nn
{[}\psi_{i}{]} & =  - \frac{{\theta} + 2}{4}, \nn
{[}V{]} & =  0, \nn
{[} \eta {]} & =  -\frac{{\theta} - 1}{2}, \nn
{[} \mu {]}  & =  \frac{1}{2}.
\label{eq:IS}
\end{align}
Here $\psi_{i,k}(\omega,x)$ is in the 
mixed space representation. 
It is noted that the dynamical critical exponent $z$ is defined to be the scaling dimension of frequency 
measured in the unit of the scaling dimension of $\eps_k$.
The scaling in \eq{eq:IS} allows one to write the
coefficients of the effective action as
\begin{widetext}
\begin{align}
&
S^{s_1,t_1,..,s_m,t_m}_{i_1,j_1,..,i_m,j_m}
\left(x,q,k,\om,\nu; X_0,\Lambda,V,\eta, \mu \right) \nn
& = X_0^{ m({\theta} + 2)/2 -({\theta} + 4)/2 + 
\sum_{a=1}^m (s_a + t_a) } \nn
& \qquad \times 
{\tilde S}^{s_1,t_1,..,s_m,t_m}_{i_1,j_1,..,i_m,j_m}
 \left( x X_0^{-1}, q X_0^{1/2}, k X_0^{1/2}, 
\om X_0^{({\theta} + 1)/2},
\nu X_0^{({\theta} + 1)/2};
\Lambda X_0^{1/2}, V, \eta X_0^{- (\theta - 1)/2}, \mu
X_0^{1/2} \right).
\label{eq:ISF}
\end{align}
\end{widetext}
In this scaling, 
$\eta$ enters as a scale
in addition to $\Lambda$,
while $V$ is deemed dimensionless.
In other words,
$\Lambda$ and $\eta^{-1}$
play the role of UV cut-off scales
whereas $\mu$ is an IR cut-off,
as is shown in Fig. \ref{fig: scale}.
If the effective action was singular in the limit of
$\Lambda X_0^{1/2} \rightarrow \infty$,
$\eta^{-1} X_0^{({\theta} - 1)/2} \rightarrow \infty$
and $\mu X_0^{1/2} \rightarrow 0$,
anomalous dimensions would arise
relative to the `bare' scaling dimensions shown in 
\eq{eq:IS}.
However, it turns out that the chiral nature of the theory
puts strict constraints on the way the effective action
depends on $\Lambda$ and $\eta$,
and the effective action is finite
in the $\Lambda X_0^{1/2}, \eta^{-1}  X_0^{({\theta} - 1)/2}
\rightarrow \infty$ limit.
This brings us to the two key results
of this paper.  
\begin{itemize}
 \item 
{\it 
The Wilsonian effective action 
is finite in the limit of 
$\Lambda X_0^{1/2} \rightarrow \infty$, 
$\eta^{-1} X_0^{({\theta} - 1)/2} \rightarrow \infty$ 
and $\mu X_0^{1/2} \rightarrow 0$
for ${\theta} > 1$.
Since the UV and IR cut-off scales can be dropped, 
the Wilsonian effective action
at the critical point with $\mu = 0$
is invariant under the coarse graining
associated with an increase of $X_0$ 
and the re-scaling dictated by \eq{eq:IS},
which gives the exact scaling dimensions.}
\item
{\it
The effective action
is dominated by the RPA diagrams
in the limit, 
$q X_0^{1/2},
k X_0^{1/2},
\om X_0^{({\theta} + 1)/2},
\nu X_0^{({\theta} + 1)/2} \rightarrow 0$ 
while other diagrams
become important as well
when
$q X_0^{1/2},
k X_0^{1/2},
\om X_0^{({\theta} + 1)/2},
\nu X_0^{({\theta} + 1)/2} \ge 1$ .
}
\end{itemize}
We will prove these statements
in generality and through explicit calculations 
in the following sections.

\subsection{General proof of UV finiteness} 
\label{sec: UV_finite}

In this section 
it is shown that all quantum corrections 
in the effective action 
are UV finite
in the $\Lambda \rightarrow \infty$
and $\eta \rightarrow 0$ limit.
In particular,  
integrations over internal frequencies and $y$-momenta
are separately UV finite for all diagrams.

\subsubsection{UV finiteness of internal frequency 
integrations} 
\label{app: int_freq}

\begin{figure}[!ht]
 \centering
\subfigure[]{\includegraphics[scale=.4]{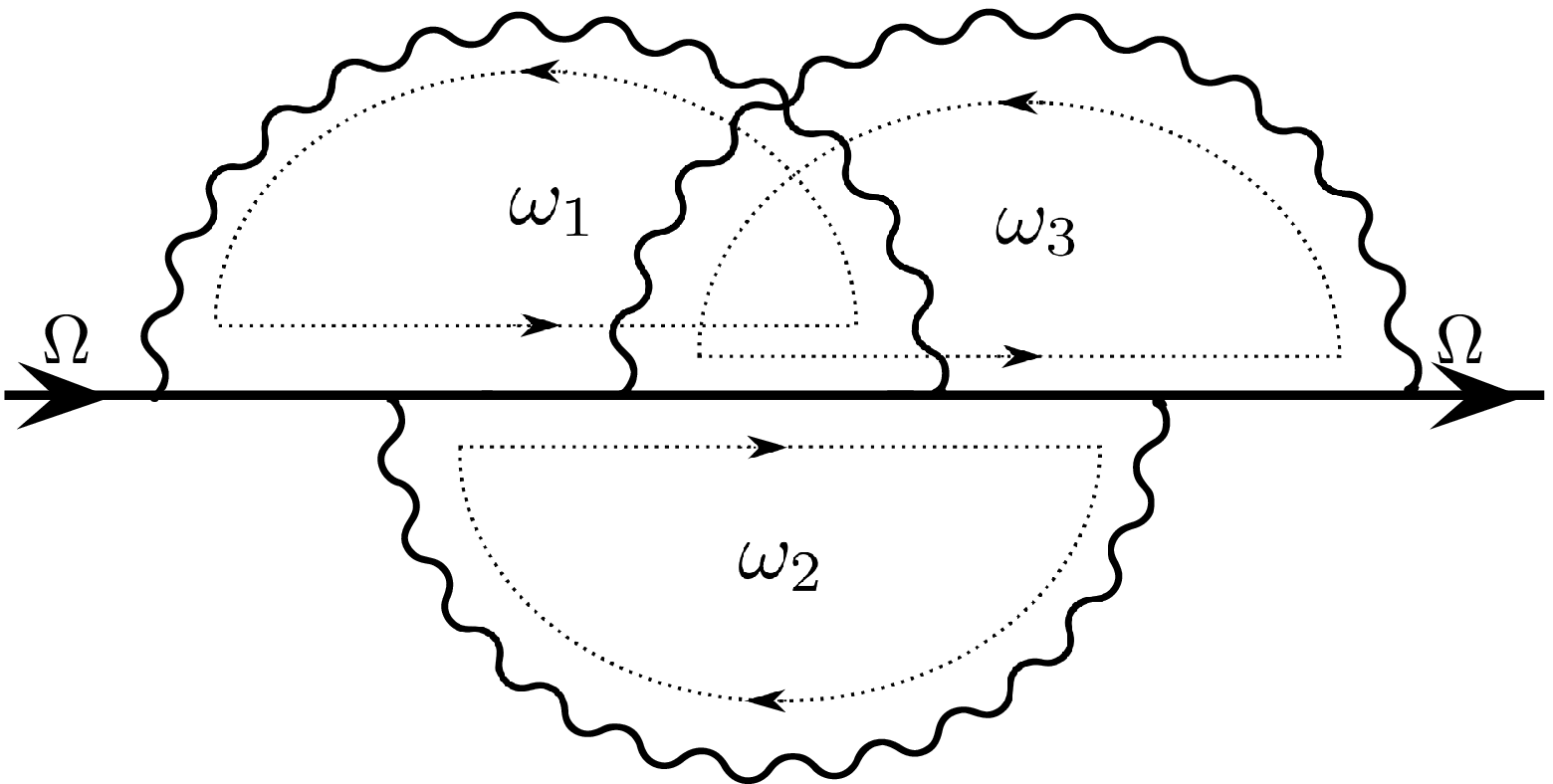} 
\label{fig: 
LC1}} \hspace{0.2in} 
\subfigure[]{\includegraphics[scale=.4]{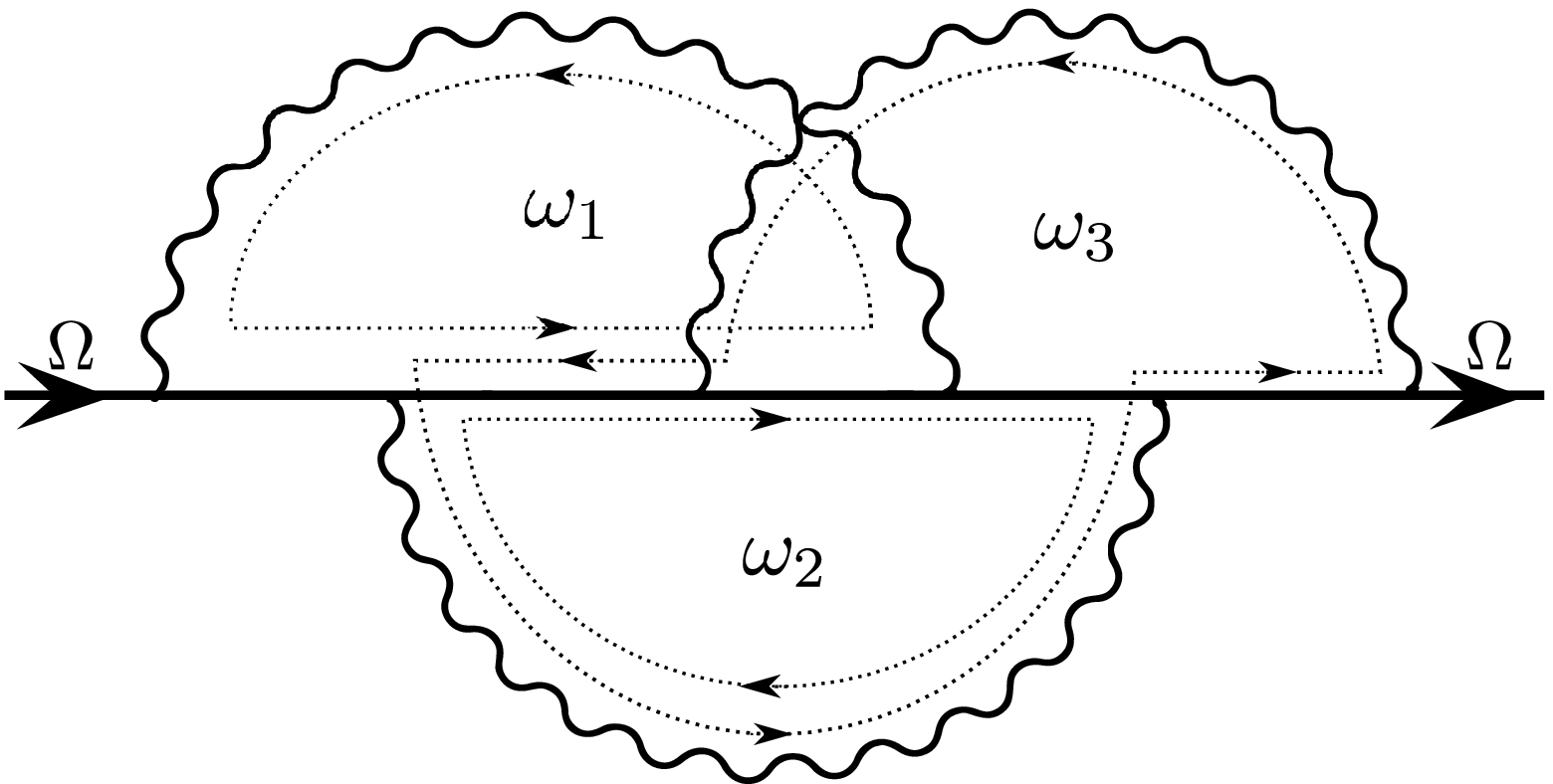} 
\label{fig: LC2}}
\\
\subfigure[]{\includegraphics[scale=.35]{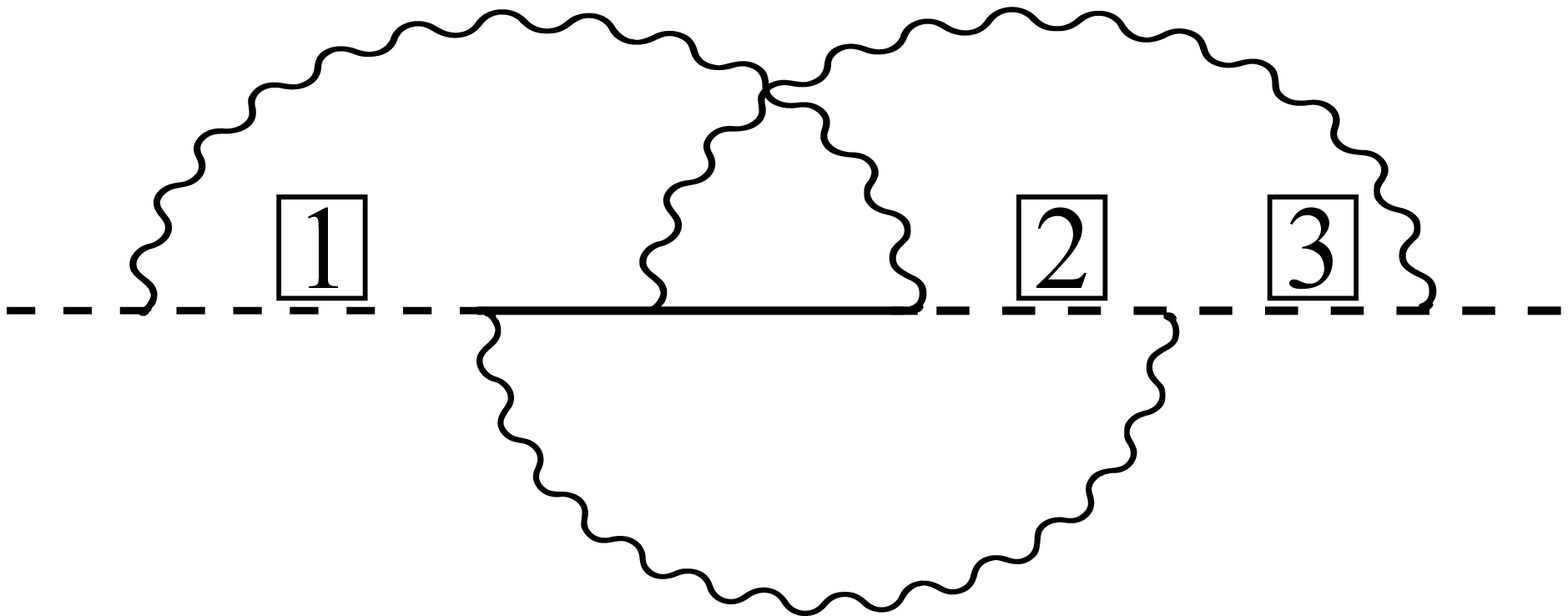}
 \label{fig: LC2-1}}
\hspace{0.2in}
\subfigure[]{\includegraphics[scale=.3]{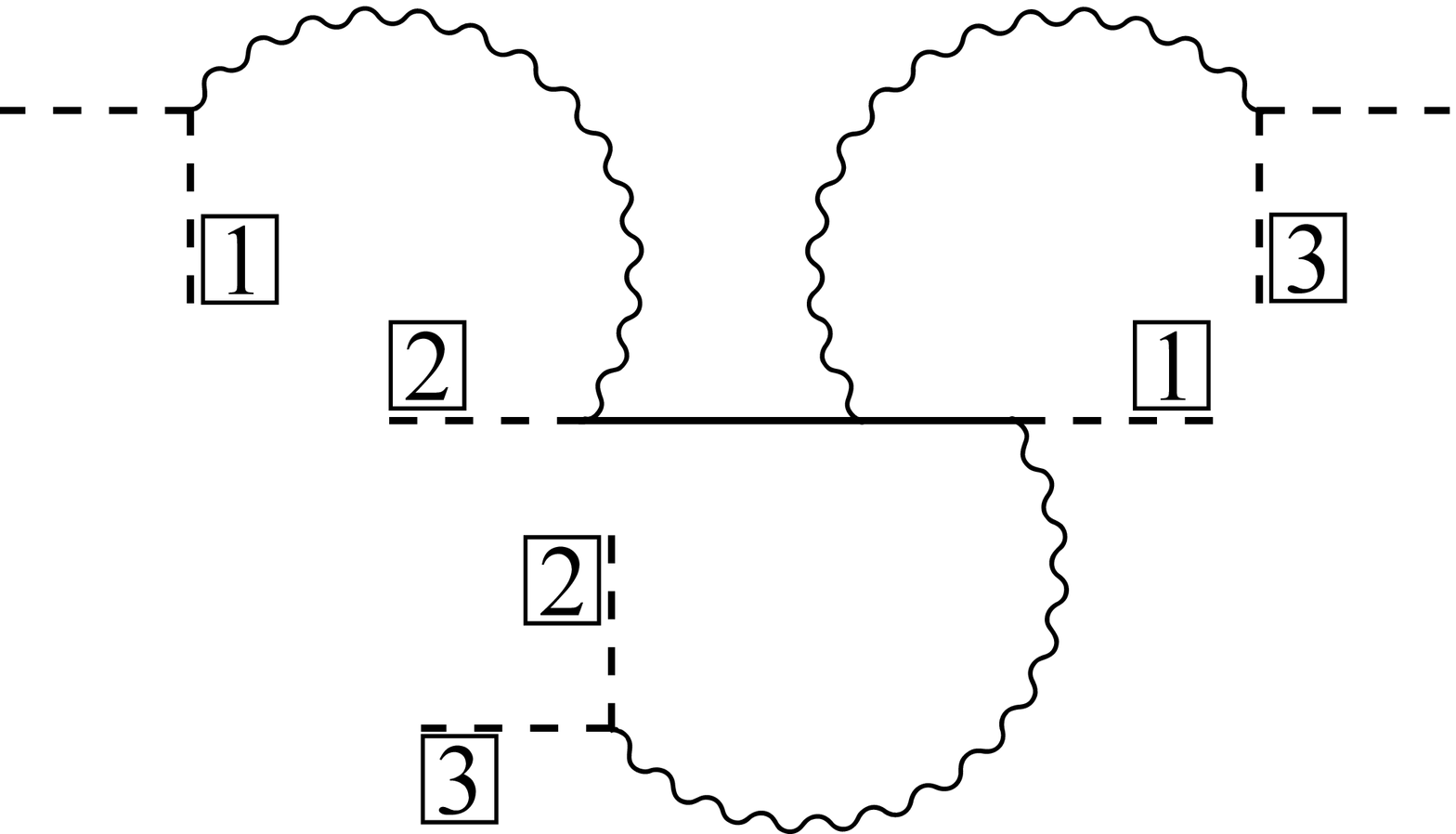} 
\label{fig: tree}}\\
\caption{
A three-loop correction to the quadratic action.
There are different ways of assigning internal 
frequencies, 
where internal frequencies are assigned to run 
along  the dotted lines in (a) and (b).
(a) All the fermion propagators in the loop 
of internal frequency $\omega_2$ 
are shared with other loops associated with 
the internal frequencies, $\omega_1$ and $\omega_3$. 
(b) Each loop denoted by the dotted lines 
has an exclusive fermion propagator 
that is not shared with other loops. 
(c) The propagators that are exclusive to each loop in Fig. 
\ref{fig: LC2} are
represented by dashed lines. 
The numbers in the boxes indicate the indices for the 
internal
frequencies that run through the exclusive propagators.
(d) A tree diagram whose legs are contracted to yield Fig. 
\ref{fig: LC2}. 
The legs with matching numbers are contracted to give the 
corresponding dashed
propagator in Fig. \ref{fig: LC2-1}.}
\label{fig: loop_cover}
\end{figure}

Here we prove that frequency integrations are UV finite 
in the $\eta \rightarrow 0$ limit.
In this limit, the fermion Green's function in \eq{eq: 
bare_G}
becomes $G(\omega, x) = -i~ \sgn \omega \Theta( -x \omega)$.
At first, it appears dangerous to take the $\eta 
\rightarrow 0$ limit
because the fermion propagator is not suppressed at large 
frequencies.
It turns out that this does not cause any UV divergence
because all internal frequencies in a given diagram
are bounded by the external frequencies.
To begin with the proof, we note that there are different 
ways 
to label internal frequencies for a given diagram.
As an example, we show two different ways 
of assigning  internal frequencies within 
a three-loop fermion self-energy diagram
in Figs. \ref{fig: LC1} and \ref{fig: LC2}.
It turns out that there is a special choice
which is more convenient for our proof.
This is the choice where each loop 
(denoted by the dotted lines 
in Fig. \ref{fig: LC2})
associated with an internal frequency 
contains at least one fermion propagator
which does not belong to other loops.
We call this choice an `exclusive loop covering'.
Fig. \ref{fig: LC1} is not an exclusive loop covering
because the loop for $\omega_2$ does not have any
`exclusive' fermion propagator which carries only 
$\omega_2$.
On the other hand, Fig. \ref{fig: LC2} is 
an exclusive loop covering because 
for every $\omega_i$ there is at least one exclusive 
fermion propagator
which carries only $\omega_i$.
The exclusive fermion propagator for each internal frequency
is denoted as dashed lines in Fig. \ref{fig: LC2-1}.

Does an exclusive loop covering exist for every diagram ?
To show that the answer is yes, 
we note that 
any  diagram 
can be constructed out of a connected tree diagram by
contracting some of its legs. 
This is always possible because 
one can keep cutting internal lines 
until all loops disappear
without cutting the diagram into two disconnected ones.
For example, the three-loop fermion self-energy diagram in
Fig. \ref{fig: LC2-1} is constructed by joining three pairs 
of legs
in the tree diagram shown in Fig. \ref{fig: tree}. 
In Fig. \ref{fig: LC2-1}, 
we represent the internal lines of the parent tree diagram 
by solid lines 
and the new internal lines created through the joining 
procedure by dashed lines. 
A loop is formed by contracting a pair of legs,
where an internal frequency is assigned to run 
through solid lines and the dashed line created
from a new contraction.
For each loop formed in this way, 
the solid propagators  
are in general shared by multiple loops 
while each of the dashed propagators, 
i.e. the ones formed by contracting external legs of the 
tree graph, 
are exclusive to one loop. 
Since this is true for all loops, 
we obtain an exclusive loop covering for the diagram. 

\begin{figure}[!ht]
 \centering
 \includegraphics[width=0.9\columnwidth]{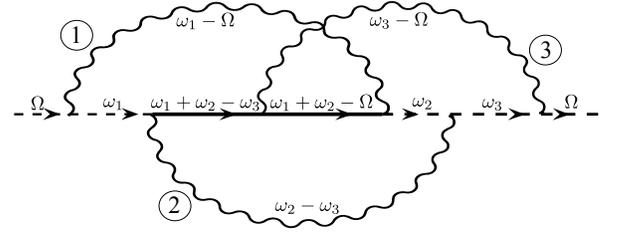}
 \caption{A three loop contribution to self-energy.}
 \label{fig: self_energy_freq-1}
\end{figure}

Now we show that every internal frequency is bounded by
the external frequency using the exclusive loop covering. 
As a simple example, let us examine 
the three-loop fermion self-energy diagram more closely.
Each internal line carries
a linear combination of the external frequency
and internal frequencies 
as is shown in Fig. \ref{fig: self_energy_freq-1}. 
The key reason for the existence of an upper bound for 
internal frequencies
is chirality.
Since the fermions are chiral,
the fermion propagator in \eq{eq: bare_G}
vanishes for positive (negative) frequency with $x>0$ 
($x<0$).
As a result, 
there is a set of constraints that internal frequencies 
have to satisfy
for a given set of relative coordinates $x_{ij}$ between 
vertices.
For the three-loop self-energy diagram,
the constraints read
\begin{align}
 \omega_1 ~ x_{12} & > 0, \label{eq: a1}\\
 \omega_2 ~ x_{12} & > 0, \label{eq: a2}\\
 \omega_3 ~ x_{23} & > 0, \label{eq: a3} \\
 (\omega_1 + \omega_2 - \omega_3) ~ x_{23} & > 0, 
\label{eq: a4}\\
 (\omega_1 + \omega_2 - \Omega) ~ x_{31} & > 0. \label{eq: 
a5} 
\end{align}
Here $x_{ij} \equiv x_i - x_j$ is the separation between 
the vertices $i$ and $j$
which are labeled in Fig. \ref{fig: self_energy_freq-1}. 
The constraints in Eqs. (\ref{eq: a1}) - (\ref{eq: a3}) 
are the constraints from the exclusive propagators
each of which carries only one internal frequency.
Let $S(\{\omega_i\})$ be the set of internal frequencies
that satisfy Eqs. (\ref{eq: a1}) - (\ref{eq: a5}).
Once the signs of $x_{ij}$'s are fixed, 
the set does not depend on the magnitudes
of $x_{ij}$'s.
Our goal is to show that the set is bounded by the external 
frequency
in all directions in the space of internal frequencies.
To see this, we first add all non-exclusive constraints
that contain $\omega_1$.
In this case, they are Eqs. \eqref{eq: a4} and \eqref{eq: 
a5}.
This leads to an inequality for $\omega_1$,
\begin{align}
 (\omega_1 + \omega_2) x_{21} - \omega_3 x_{23} - \Omega ~ 
x_{31} > 0.
\label{eq: a6}
\end{align}
Together with \eq{eq: a1}, \eq{eq: a6} 
limits the range of $\omega_1$ as
\begin{align}
 & 0 < \omega_1 ~ x_{12} < \omega_3 x_{32} + \Omega ~ 
x_{13} - \omega_2 ~ x_{12}. \label{eq: a7} 
\end{align}
Eqs. (\ref{eq: a2}) and (\ref{eq: a7})
constrain the range of $\omega_2$ as
\begin{align}
 &  0 < \omega_2 ~ x_{12} < \omega_3 x_{32} + \Omega ~ 
x_{13}. \label{eq: a8}
\end{align}
Finally, \eq{eq: a8} together with \eq{eq: a3} leads to 
\begin{align}
 0 < \omega_3 ~ x_{23} < \Omega~ x_{13}. \label{eq: a9}
\end{align}
This implies that for a fixed set of $\{x_{ij}\}$, 
$\omega_3$ is bounded by $\Omega$. 
Applying this to 
Eqs. \eqref{eq: a7} and \eqref{eq: a8},
we find that $\omega_1$ and $\omega_2$ 
are also bounded by $\Omega$. 

Let $\wtil{S}(\{\omega_i\};\{x_{ij}\})$ be
the set of internal frequencies 
that satisfy Eqs. (\ref{eq: a7}) - (\ref{eq: a9}).
It is of note that $\wtil{S}(\{\omega_i\};\{x_{ij}\})$ not 
only depends
on the signs of $x_{ij}$'s 
but also on their magnitudes unlike $S(\{\omega_i\})$.
An important property 
of $\wtil{S}(\{\omega_i\};\{x_{ij}\})$ is that
$\wtil{S}(\{\omega_i\};\{x_{ij}\}) \supseteq 
S(\{\omega_i\})$
for any $\{x_{ij}\}$.
This is due to the fact that
Eqs. (\ref{eq: a7}) - (\ref{eq: a9}) 
are necessary (not sufficient in general) conditions 
of Eqs. (\ref{eq: a1}) - (\ref{eq: a5}).
Since $\wtil{S}(\{\omega_i\};\{x_{ij}\})$ is bounded
for any fixed set of $\{x_{ij}\}$,
$S(\{\omega_i\})$ is also bounded
by the external frequency in all directions
in the space of internal frequencies.
This proves that the integration over the internal 
frequencies
is UV finite.

This argument can be easily generalized to all other 
diagrams.
The rule is as follows.
Consider a diagram with $L$ loops
with $I$ internal fermion propagators
and $V$ vertices.
For a fixed set of relative coordinates $\{ x_{ij} \}$ of 
the vertices,
there exist $I$ constraints for $L$ internal frequencies.
Since vertices are normal ordered, 
fermion propagators can connect only distinct vertices.
The existence of an exclusive loop covering implies that
for every internal frequency $\omega_q$ 
there exists an exclusive constraint of the form,
\bqa
F_q( \omega_q ) = x_{i_q j_q} \omega_q > 0,
\label{Fq}
\eqa
where $q=1,2,..,L$ and 
$x_{i_q j_q}$ is the separation between the two 
vertices
connected by the fermion propagator which carries only 
$\omega_q$.
Therefore the $I$ constraints can be divided into 
the $L$ `exclusive' constraints and
$I-L$ `non-exclusive' constraints.
Non-exclusive constraints in general 
contain external frequencies and more than one internal 
frequency,
\bqa
f_p( \{ \omega_q \}, \{ \Omega_r \} ) > 0,
\label{fq}
\eqa
where $\{ \Omega_r \}$ is a set of external frequencies
and $p=1,2,...,I-L$.
Without loss of generality, 
we can assume that only the first $k_1$ non-exclusive 
constraints, 
$f_1, f_2, ..., f_{k_1}$ contain $\omega_1$.
We add them up to create a new constraint, 
\bqa
C_1 = f_1 + f_2 + .. + f_{k_1}  > 0.
\eqa
Since the first exclusive constraint is written as
\bqa
F_1 = x_{i_1 j_1} \omega_1 > 0,
\label{F1}
\eqa 
it is guaranteed that $C_1$ is of the form
\bqa
C_1 = - x_{i_1 j_1} \omega_1 + C_1^{'}(\omega_2, \omega_3, 
.., \omega_L; \{ \Omega_r \} ) > 0
\label{G1}
\eqa
This is due to the fact 
the relative coordinates of the vertices around a loop
form a cyclic sum in $C_1 + F_1$, and it is independent of 
$\omega_1$.
With the aid of Eqs. (\ref{F1}) and (\ref{G1}), 
we see that $\omega_1$ is bounded by 
other internal frequencies and the external frequencies,
\bqa
0 <  x_{i_1 j_1} \omega_1 <  C_1^{'}(\omega_2, \omega_3, 
.., \omega_L ; \{ \Omega_r \} ).
\label{H1}
\eqa
Next we construct a set of non-exclusive constraints for 
$\omega_2, \omega_3, ..., \omega_L$ made of 
\bqa
&& F_1 + C_1 = C_1^{'}(\omega_2, \omega_3, .., \omega_L; \{ 
\Omega_r \} ) > 0, \nn
&& \mbox{and } f_{j} > 0 \mbox{ with $ j= k_1+1, k_1+2, 
..., L-I.$ }
\eqa
Using this set of non-exclusive constraints, 
we apply the same procedure for $\omega_2$.
Namely, 
we add all non-exclusive constraints that contain $\omega_2$
to construct a new constraint $C_2 > 0$.
Combined with $F_2 > 0$,
one can obtain a bound for $\omega_2$ of the form,
\bqa
0 <  x_{i_2 j_2} \omega_2 <  C_2^{'}(\omega_3, \omega_4, 
.., \omega_L ; \{ \Omega_r \} ).
\label{H2}
\eqa
In this way, one can show that the range of $\omega_l$ is 
bounded by a linear combination of 
the external frequencies and the internal 
frequencies $\omega_m$ with $m>l$,
\bqa
0 <  x_{i_l j_l} \omega_l <  C_l^{'}(\omega_{l+1}, 
\omega_{l+2}, .., \omega_L ; \{ \Omega_r \} ).
\label{Hl}
\eqa
The last internal frequency $\omega_L$ is bounded only by 
the external frequencies,
\bqa
0 <  x_{i_L j_L} \omega_L <  C_L^{'}( \{ \Omega_r \} ).
\label{HL}
\eqa
The set of $L$ inequalities given by \eq{Hl}
implies that all the internal frequencies are bounded by 
the external frequencies.
From the argument given below Eq. (\ref{eq: a9}),
the set of internal frequencies that satisfy 
the original constraints given by Eqs. (\ref{Fq}) and 
(\ref{fq}) 
are also bounded by the external frequencies.
Because all internal frequencies are bounded,
all frequency integrations are UV finite
even in the $\eta \rightarrow 0$ limit.
This completes the proof that integrations over internal 
frequencies 
are UV finite for general diagrams.
We emphasize that the UV finiteness is due to the chiral 
nature of the theory.
For non-chiral theories, internal frequencies
are not bounded by external ones.

\subsubsection{UV finiteness of 
\texorpdfstring{$y$}{y}-momentum integrations}
\label{app: int_flavor}

It is easier to see that $y$-momentum integrations are UV
finite.
Let us first consider fermion loops 
which refer to loops that are solely made of fermion 
propagators.
For example, a bubble in Fig. \ref{fig: v4B}
leads to an integration of the form,
\bqa
\int_{-X_0}^{X_0}  dx_{12} ~ \int_{-\Lambda}^{\Lambda} dp  
~~ e^{ 2 i \gamma x_{12}  q p },
\label{ph55}
\eqa
where 
$x_{12}$ is the relative coordinate between the two quartic 
vertices,
$p$ is the $y$-momentum that runs inside the loop,
and $q$ is the $y$-momentum transfer.
It is noted that fermion propagators and the
four-fermion vertices do not depend on 
the $y$-momentum $p$ that runs within the fermion loop
except for the phase factor that results in \eq{ph55}. 
This leads to large fluctuations 
in the $y$-momentum of the internal fermion. 
Due to the emergent uncertainty relation 
between the $x$-coordinate and $y$-momentum, 
as is manifest from the phase factor , 
wild fluctuations in $p$ leads to 
a `confinement' of the relative coordinate of the two
vertices $x_{21}$. 
In the $\Lambda \rightarrow \infty$ limit,
the integration over $p$ simply
generates a delta function that 
puts a constraint on the positions
of the vertices attached to the loop,
which leads to $\frac{1}{\gamma|q|}$
after the integration over $x_{12}$.
For a fermion loop with more than two external legs,
the integration over the $y$-momentum along the fermion loop
fixes one of the coordinates of the vertices
without UV divergence.

Now we consider mixed loops
which refer to loops that contain at least one boson 
propagator.
One can always assign internal momenta such that
every mixed loop has at least one
boson propagator which carries no other internal 
momenta except for the one associated with the loop.
This can be easily understood 
from an argument that is similar to the one 
used in the previous section to show the existence
of an exclusive loop covering for fermion propagator.
This time, we cut boson propagators
to remove all mixed loops without creating
disjoint diagrams. 
For each mixed loop removed from this procedure,
the boson propagator that is cut 
is identified as the exclusive propagator. 
Therefore, each internal $y$-momentum integration 
of the original diagram goes as
\bqa
\int_{-\Lambda}^\Lambda dq ~\frac{1}{|q^2 + 
\mu^2|^{\theta/2}}
\label{eq:da}
\eqa
at most in the large $q$ limit.
For ${\theta} > 1$ this is UV convergent
in the $\Lambda \rightarrow \infty$ limit.

\subsection{IR finiteness}
\label{sec: IR_finite}

There are two reasons for the UV finiteness of the Wilsonian
action.
First, $(2+1)$-dimension is 
below the upper critical dimension 
for ${\theta} > 1$.
As a result, only a finite number of
diagrams are potentially UV divergent.
Second, even the potentially divergent diagrams
are finite due to chirality,
which strictly limits the ranges of internal 
frequencies by the external ones.

On the other hand, the IR finiteness
in the $\mu \rightarrow 0$ limit is less obvious.
For example, the integration over the $y$-momentum
in \eq{eq:da} is IR divergent for $\mu=0$ with ${\theta} > 
1$.
While the UV finiteness is controlled
by kinematic constraints,
the way IR finiteness is restored 
in the $\mu \rightarrow 0$ limit
is through dynamical mechanism.
It turns out that the theory 
cures the IR singularity dynamically
such that the theory flows to an IR fixed point
governed by the scaling in \eq{eq:IS}.
In the following section,
we will see that this is indeed the case 
through an explicit computation of the 
Wilsonian effective action.

\subsection{An explicit computation of the Wilsonian 
effective action}
\label{sec: EC}

Because the theory flows to a strongly interacting fixed
point,
it is not easy to compute the full Wilsonian effective
action exactly.
However, one can compute the effective action 
in the limit the running cut-off length scale is small 
compared to the length scales associated with
external momenta.
The main outcome of the explicit calculation is that 
the operators that are generated from the RPA channels 
shown in Fig. \ref{fig: RPA} are the only ones 
that are $\ordr{1}$ in this limit.
All other channels generate operators 
with extra factors of $X_0$ 
accompanied by additional factors of frequency or momentum.
Therefore, those contributions are suppressed 
in the small momentum limit with fixed $X_0$
(equivalently small $X_0$ limit with fixed momenta).
Once the $\ordr{1}$ corrections 
are consistently taken into account in the Wilsonian 
effective action,
the IR singularity encountered in 
the $\mu \rightarrow 0$ limit is cured. 
We first illustrate the dominance of 
the RPA diagrams in the small $X_0$ limit
by computing two representative diagrams,
followed by a generalization to all diagrams.

\subsubsection{\texorpdfstring{$(X_0)^0$}{X_0^0} order}

\begin{figure}[!ht]
     \centering
     \includegraphics[scale=.8]{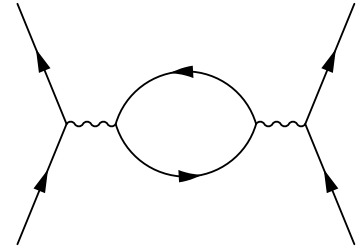} 
     \caption{
A fusion of two quartic vertices
which results in an operator which is independent of $X_0$.
}
     \label{fig: v4B}
\end{figure}

As a first example, let us consider the diagram shown
in Fig. \ref{fig: v4B}, where two quartic vertices are fused
into one quartic vertex as two pairs of fermion fields are
contracted. 
The one-loop contribution is given by
\begin{widetext}
\begin{align}
& \dl  S^{(RPA,1)}_{4} 
= 2\int  \frac{ dk_1 dk_2  dq d\om_1 d\om_2 
d{\nu}}{(2\pi)^{6}} ~  dx_1 ~ e^{i 2 \gamma ( 
k_1 -  k_2 )  q x_1 } 
V_{ij;ln} \chi_\theta(q) ~ \nn
& \qquad \times \frac{(-g^2)}{2} \chi_\theta(q) \int 
\msr{k'} 
~ \msr{\om'}\int_{-X_0}^{X_0} dx_{21} ~  e^{-2i \gamma
k' q x_{21} }  G_0(\om'+\nu,x_{21}) ~
G_0(\om',-x_{21}) \nn
& \qquad  \times \ccolon
\psi^*_{i,k_2}(\om_2,x_1) 
~ \psi_{j,k_2+q}(\om_2+\nu,x_1) ~ \psi^*_{l, k_1 +
q}(\om_1+ \nu,x_1+x_{21}) ~
\psi_{n,k_1}(\om_1,x_1+x_{21}) \ccolon \label{eq:
calc_1B-1} \\
	& =  \int  \frac{ dk_1 dk_2  dq d\om_1 d\om_2 
d{\nu}}{(2\pi)^{6}} ~  dx_1 ~ e^{i 2 \gamma ( 
k_1 -  k_2 )  q x_1 } V_{ij;ln}  \chi_\theta(q) ~ 
\lt[- c_B  \chi_\theta(q) \frac{\abs{\nu}}{|q|}\rt] \nn 
	& \qquad \times   \ccolon
\psi^*_{i,k_2}(\om_2,x_1) 
~ \psi_{j,k_2+q}(\om_2+\nu,x_1) ~ \psi^*_{l, k_1 +
q}(\om_1+ \nu,x_1) ~ \psi_{n,k_1}(\om_1,x_1) 
\ccolon, \label{eq: calc_1B-2}
\end{align} 
\end{widetext}
where $ c_B = (g^2/(8 \pi \gamma))$ 
and $x_{21} = x_2 - x_1$ 
is the relative coordinate between the two vertices.
The singular dependence of $c_B$ on $\gamma$ again 
signifies the importance of the local curvature of Fermi 
surface.
The second line in \eq{eq: calc_1B-1} represents the
contribution from the fermion loop. 
The integration over the $y$-momentum 
$k^{'}$ that runs within the fermion loop
results in $\frac{1}{|q|}
\dl(x_{21})$. Due to the delta function, 
the non-zero contribution is concentrated at $x_{21}=0$.
As a result, \eq{eq: calc_1B-2}
is independent of $X_0$.

\begin{figure}[!ht]
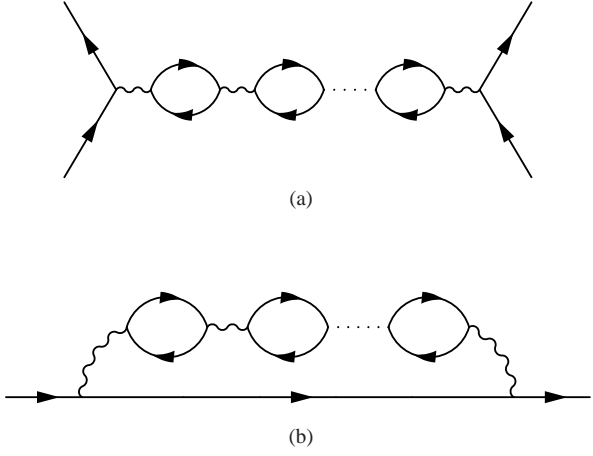

\centering
\subfigure[]{\includegraphics[width=0.9\columnwidth]
{chain.2} \label{fig: RPA_4}} \\
\subfigure[]{\includegraphics[width=0.9\columnwidth]
{chain.1} \label{fig: RPA_2}} \\
\caption{Fusion channels that give rise to $O(1)$ 
corrections to the Wilsonian effective action in the small
$X_0$ limit.}
\label{fig: RPA}
\end{figure} 

Similarly, the fusion of $L+1$ quartic vertices
in the RPA channel, as is shown in Fig. \ref{fig: RPA_4}, 
generates a $L$-loops diagram which is given by
\begin{align}
      & \dl  S^{(RPA,L)}_{4} = \int  \frac{dk_1 dk_2 dq
d\om_1 d\om_2 d\nu }{(2\pi)^6}  dx ~ e^{2i 
\gamma ( k_1 -  k_2 )  q x } V_{ij;ln} \nn 
      & \times    \chi_\theta(q) 
\lt[-c_B \chi_\theta(q) \frac{ \abs{\nu}}{|q|} \rt]^L 
~ \ccolon \psi^*_{i, k_2}(\om_2,x)  
\psi_{j,k_2+q}(\om_2+\nu,x) \nn
      &  \times \psi^*_{l, k_1 + q}(\om_1+\nu,x) ~
\psi_{n,k_1}(\om_1,x) \ccolon . 
\label{eq: 21}
\end{align}
For the derivation of \eq{eq: 21}, 
see Appendix \ref{app: u2_V}.
As is the case for the fusion of two vertices 
shown in Fig. \ref{fig: v4B}, 
$L$ relative coordinates between
$L+1$ vertices are completely fixed by 
the $L$ delta functions that result
from the wildly fluctuating flavors within the $L$ fermion 
loops.
Therefore, all operators generated from fusions  
in the RPA channel are independent of $X_0$. 
The infinite series of operators that are 
$O(1)$ can be summed over 
to renormalize the boson propagator to 
\begin{align}
\chi_\theta^{(RPA)}( \nu, q ) &=
\frac{1}{|q^2 + \mu^2|^{\theta/2} + c_B ~
\dsty{\frac{\abs{\nu}}{|q|}}}. 
\label{eq: RPA_dress_V}
\end{align}
Note that the $q \rightarrow 0$ limit of the quartic vertex
is now well defined even in the $\mu \rightarrow 0$ limit.
As a result, the IR divergence we encountered in 
integrations over $y$-momentum (for example in \eq{eq:da})
in the $\mu \rightarrow 0$ limit is cured by the quantum 
corrections.
Henceforth, we set $\mu =0$ to focus on the critical point.

There are $O(1)$ contributions to the quadratic vertex as 
well.
These are generated by contracting an extra pair of fermion 
fields 
in the diagrams that generate $O(1)$ correction to the 
quartic vertex.
Diagrammatically, these are nothing but the RPA diagrams for
the fermion self-energy as is shown in Fig.
\ref{fig: RPA_2}, where the number of fermion loops matches
with the number of relative coordinate between vertices. 
The $O(1)$ correction to the quadratic action is given by
\begin{align}
 \dl S_2^{(RPA)} &= \int dx  \frac{d \om}{2\pi} 
\frac{dk}{2\pi} ~ \psi^*_{i,k}(x,\om) ~ 
\Sigma^{(RPA)}_{ij}(\om) ~ 
\psi_{j,k}(x,\om),
\end{align}
where the self-energy is 
\begin{align}
     & \Sigma^{(RPA)}_{ij}(\om)  
= i ~ \dl_{ij} ~ c_F ~ \sgn{\om} \abs{ \om}^{2 / ({\theta} +
1)} \label{eq: RPA_self-energy}
\end{align}
with 
\eqn{
c_F = \frac{g^2 v}{2 \pi^2} ~
c_B^{\frac{1-\theta}{1+\theta}}  \int_0^{\infty} dy ~ y ~  
\ln \lt(1 + \frac{1}{y^{\theta + 1}} \rt). \label{eq: c_F}
}
It is noted that the integration 
over the $y$-momentum (represented by $y$)
in \eq{eq: c_F}
is finite with $\mu=0$
because the IR divergence is cured by
the series of RPA diagrams. 
We emphasize that 
{\it the RPA diagrams are dynamically selected, not by
hand.}
As will be shown in the following section, 
all other diagrams vanishes at least linearly in $X_0$ in 
the small $X_0$ limit
due to the chiral nature of the theory. 
Those contributions necessarily contain 
larger powers of momentum or frequency,
which are suppressed at low momentum/frequency.

\subsubsection{Higher order in \texorpdfstring{$X_0$}{X_0}}

\begin{figure}[!ht]
\centering
\subfigure[]{\includegraphics[width=0.6\columnwidth]
{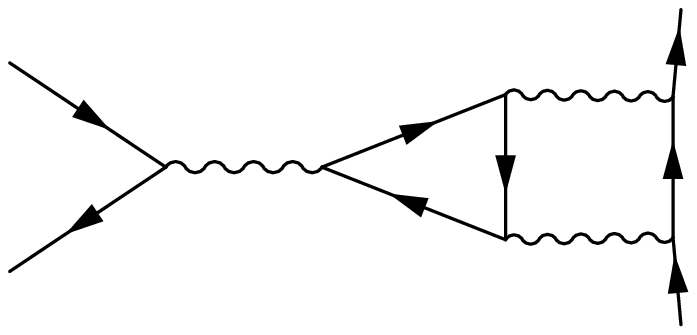} \label{fig: 4f2L}}
\subfigure[]{\includegraphics[width=0.9\columnwidth]
{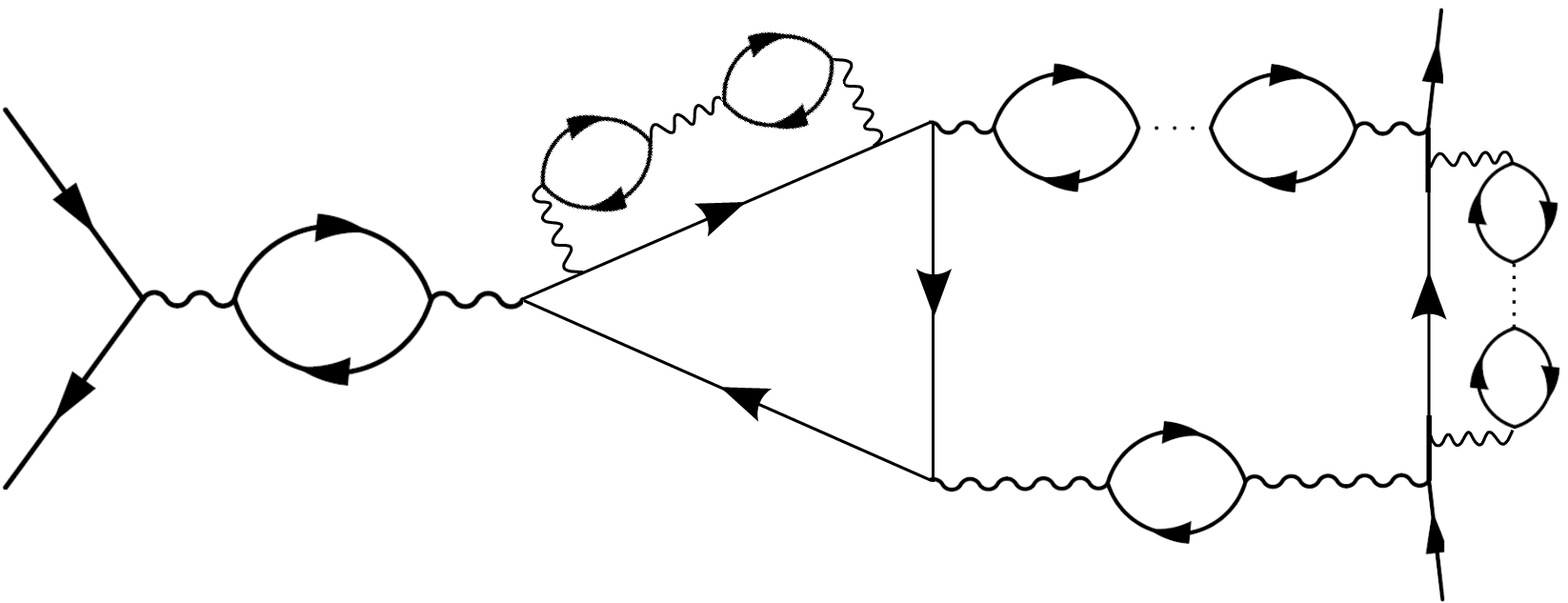} \label{fig: 4f2L_RPA}} 
\caption{A sub-leading correction to the quartic vertex. 
(a) With bare vertices and propagators, the resulting 
operator is IR divergent in the $\mu \rtarw 0$ limit.
(b) Once the vertices and propagators are dressed by all 
other $O(1)$ corrections, the resulting operator is finite
as $\mu \rtarw 0$.
}
\end{figure}

To illustrate the generic feature of operators 
generated from fusion in non-RPA channels,
we compute the two-loop vertex
correction shown in Fig. \ref{fig: 4f2L}. 
Here three quartic operators are fused into 
one quartic vertex, which results in 
\begin{widetext}
\begin{align}
& \dl S_4^{(nRPA,3)} 
 = -  8 \int \frac{dk_1  dk_2  dq  
d\om_1  d\om_2  d{\nu}}{(2\pi)^6} ~ dx_1
~ e^{2 i \gamma ( k_1  -  k_2 )  q x_1 } 
V_{i' j';l n} ~ V_{i l';n' i'} ~ V_{l' j;j' n'} 
\chi_{\theta}(q) \int
\frac{dq' d\nu'  dk'  d\om'}{(2\pi)^4} ~
\chi_\theta(q')  \chi_\theta(q'-q) \nn
& ~ \times \int_{\mc{C}} dx_{21} ~ dx_{32}  ~ e^{2i\gamma
(k' - k_2)q x_{21}} ~
e^{2i\gamma (k' - k_2)(q - q')x_{32}}  G_0(\om',x_{21}) ~ 
G_0(\om' + \nu',x_{32}) ~ G_0(\om'+\nu, -x_{32} - x_{21}) 
~ G_0(\om_2+ \nu',-x_{32}) \nn 
&  ~ ~ \times \ccolon \psi^*_{i,k_2}(\om_2,x_1 + x_{21}) 
~ \psi_{j, k_2+q}(\om_2+\nu,x_1 + x_{32} + x_{21}) 
\psi^*_{l, k_1 + q}(\om_1+ \nu ,x_1) ~ 
\psi_{n,k_1}(\om_1,x_1) 
\ccolon, \label{eq: nRPA-1} 
\end{align}
\end{widetext}
where $x_{21}$ and $x_{32}$ are the two relative 
coordinates 
of the three vertices
integrated over the region of $x$-space,
$\mc{C} \equiv \cup_{\alpha=1}^6 
\mc{C}_\alpha$
with $\mc{C}_\alpha$ defined in \eq{eq:sets}.
The $y$-momentum $k'$ 
running in the fermion loop generates
$\dl(q x_{21}  - (q'- q)x_{32})/2\gamma$, 
which fixes one of the relative coordinates ($x_{32}$). 
The integration over the other relative coordinate
($x_{21}$) gives rise to a factor of $X_0$. 
Since the remaining internal momentum and frequency 
integrals are UV finite,
the resulting operator vanishes linearly
in the $X_0 \rightarrow 0 $ limit, 
\begin{widetext}
\begin{align}
\dl S_4^{(nRPA,3)} & = - \int \frac{dk_1  dk_2  dq  
d\om_1  d\om_2  d{\nu}}{(2\pi)^6} ~ dx_1
~ e^{2i \gamma ( k_1  -  k_2 )  q x_1 } 
V_{i' j';l n} ~ V_{i l';n' i'} ~ V_{l' j;j'n'}
\chi_\theta(q) ~ \Gamma^{(nRPA,3)}(\om_2,\nu,q,X_0) 
\nn
& \quad \times  \ccolon \psi^*_{i,k_2}(\om_2,x_1) 
~ \psi_{j, k_2+q}(\om_2+\nu,x_1) \psi^*_{l,
k_1 + q}(\om_1+ \nu ,x_1) ~ \psi_{n,k_1}(\om_1,x_1) 
\ccolon,
\label{eq:nRPA-2} 
\end{align} 
where 
$\Gamma^{(nRPA,3)}(\om_2,\nu,q,X_0)$ is proportional to 
$X_0$. 
The expression becomes simpler when one of the frequencies 
vanishes, 
\begin{align}
 \Gamma^{(nRPA,3)}(\om_2=0,\nu,q,X_0) 
 & = \frac{1}{2 \pi^3 \gamma} \int dq' ~
\frac{ \chi_\theta(q')  \chi_\theta(q'+q) }{|q'|}
\int_{0}^{X_0} 
 dx_{R} ~ \step{X_0 - \frac{q}{q'} x_R } 
\step{\frac{q}{q'}}\int^{\abs{\nu}}_0 d\om' 
\int_0^{\om'} d\nu'  \nn
& \qquad \qquad  \times \exp \lt[ -\eta \abs{x_{R}} \lt( 
\abs{\om'}+ \abs{1 + \frac{q}{q'}} \abs{|\nu| - \om'} +  
\abs{\frac{q}{q'}} \lt\{|\om' - \nu'| + |\nu'| \rt\} \rt) 
\rt].
\label{eq:nRPA-3} 
\end{align} 
\end{widetext}

Although the integration over $q'$ in the above expression 
is IR divergent in the $\mu \rightarrow 0$ limit, 
the divergence disappears 
once other operators which are also $O(X_0)$ are 
consistently included.
The other diagrams which are $O(X_0)$ are the ones where 
the vertices and the fermion propagators
in Fig. \ref{fig: 4f2L}
are dressed by the RPA diagrams as is shown 
in Fig. \ref{fig: 4f2L_RPA}.
Including all the $O(X_0)$ contributions 
amounts to replacing the bare vertices and the bare 
propagators
in \eq{eq: nRPA-1}  with the dressed ones 
shown in Eqs. (\ref{eq: RPA_dress_V}) and (\ref{eq: 
RPA_self-energy}).
Taking this into account, we obtain a finite expression,
\begin{align}
  & \Gamma^{(nRPA,3)}(\om_2=0,\nu,q,X_0) = \frac{1}{2
\pi^3} ~ ~ \frac{ X_0 \abs{\nu}^{2/({\theta} + 1)}}{\gamma  
~ c_B^{{2\theta}/({\theta} + 1)}
} \nn
& \times f_1\lt(c_F X_0 \abs{\nu}^{2/({\theta} + 1)},
\frac{q}{(c_B\abs{\nu})^{1/({\theta} + 1)}} \rt),
\end{align}
where $f_1(x,y)$ is a finite universal function 
which has the following asymptotic behavior,
\begin{align}
& \lim_{x\rtarw 0} f_1(x,1) \sim 1,\\
 & \lim_{x\rtarw \infty} f_1(x,1) \sim x^{-1}.
\end{align}
It is noted that the non-RPA correction 
is suppressed by an extra factor of 
$X_0 |\nu|^{2/({\theta} + 1)}$ in 
the small $\nu$ limit 
with fixed $X_0$.

\subsubsection{General arguments}

In this section we provide a general argument for the 
statement 
that all non-RPA diagrams contain positive powers of $X_0$ 
in the small $X_0$ limit. 
Consider a cluster of $(V+1)$ vertices 
which are to be fused into one vertex
through a fusion channel with $L_f$ fermion loops. 
They have $V$ relative coordinates to be integrated 
over the range which is order of $X_0$.
Integrating out the $y$-momenta running in these loops 
yields 
$L_f$ $\dl$-functions for the relative coordinates, 
$x_{ij}$.
After $L_f$ relative coordinates are fixed, 
the remaining $V - L_f$ relative coordinates 
give rise to a factor of $X_0^{V-L_f}$.
This implies that 
the only $O(1)$ contributions are
the diagrams with $V = L_f$. 
These are exactly the RPA diagrams.
All other diagrams, 
including higher order vertices 
generated from the quartic vertex,
necessarily include positive powers of $X_0$.

If the Wilsonian effective action were UV divergent 
in the $\Lambda X_0^{1/2}, \eta^{-1} X_0^{(\theta -1)/2}
\rightarrow \infty$ limit, then
$\Lambda X_0^{1/2}, \eta^{-1} X_0^{(\theta -1)/2}$ should be
kept finite.
In such a case, 
the extra power of $X_0$ in the non-RPA diagrams
could be saturated by $\Lambda$ or $\eta^{-1}$.
In the present theory, there is no UV divergence due to 
chirality and $\theta > 1$.
As a result, one can take 
the $\Lambda X_0^{1/2}, \eta^{-1} X_0^{(\theta
-1)/2} \rightarrow \infty$ limit.
Since there is no scale in the Wilsonian effective action,
the extra powers of $X_0$ in the non-RPA diagrams
should be accompanied by extra powers of
momenta or frequency.
This is why all non-RPA diagrams are suppressed
in the low momentum/frequency limit
with fixed $X_0$.

\subsubsection{The  Wilsonian effective action}
\label{sec: EWEA}

Including the corrections to the zeroth order in $X_0$, 
we obtain the Wilsonian effective action 
which is exact modulo irrelevant terms,
\bqa
&&   S_{X_0} = \int \msr{k} ~ \msr{\om} ~ dx ~ \ccolon 
\psi_{i, 
k}^*(\om,x) \nn
&&  \quad \times \Bigl[  i c_F \sgn{\om}
\abs{\om}^{2/({\theta} + 1)} - i\partial_x
\Bigr] \psi_{i, k}(\om,x) \ccolon \nn
&& + \int \frac{d  k_1 ~ dk_2 ~ dq  ~ d\om_1 ~
d\om_2~ d\nu}{(2\pi)^6} ~ dx ~~ e^{ 2 i \gamma ( k_1 -  k_2 
) 
q x } \nn 
&& \quad \times \frac{V_{ij;ln}}{|q|^{\theta} + c_B
\frac{\abs{\nu}}{|q|} } ~ \ccolon
\psi^*_{i,k_2}(\om_2,x) ~ \psi_{j,
k_2+q}(\om_2+\nu,x) \nn  
&& \qquad \qquad \times \psi^*_{l,k_1+q}(\om_1+\nu,x)
~ \psi_{n, k_1}(\om_1,x)\ccolon, \label{eq: eff_act}
\eqa
where $c_B$ and $c_F$ are positive constants. 
The effective action is local
in the x-direction but not in the $\tau$-direction 
as can be seen from the terms that are 
non-analytic in frequency. 
Once non-analytic terms are allowed in the effective action,
usually the standard RG procedure becomes less useful 
because, in principle, infinitely many marginal or 
relevant non-local operators are generated.
In the present case, however, the
chiral nature of the theory puts a strong constraint 
on the form of non-local terms 
that can be generated. 
The contributions from non-RPA diagrams 
are systematically 
suppressed by positive powers
of $ (\abs{\om}^{2/({\theta} + 1)} X_0)$, $(q^2 X_0)$ or 
$X_0 
\partial_x$ compared to the RPA contributions that are 
included in Eq. (\ref{eq: eff_act}),
where $\om$ and $q$ are external frequency and 
$y$-momentum of the operator.
Higher order vertices such as $\psi^6$ are also generated. 
These vertices can be obtained by cutting open some
internal lines from quartic vertices.
As a result, they necessarily have less constraints 
on the relative coordinates of the vertices
compared to the RPA diagrams. 
Since they are accompanied by positive powers of $X_0$,
they are negligible in the low energy limit. 
Therefore, the effective action in \eq{eq: eff_act} includes
all terms apart from the terms that are irrelevant by power 
counting. 
It is emphasized that  
RPA diagrams are dynamically selected to 
generate the leading order terms
in the Wilsonian effective action.

With the renormalized action, 
it is more convenient to use a new normal ordering scheme
based on the dressed Green's function,
\begin{align}
G(\om,x_{12}) &= - i~ \sgn{\om} ~ \Theta(-x_{12} 
\om) \nn
& \quad \times \exp \lt[-c_F \abs{x_{12}} 
\abs{\om}^{2/({\theta} + 1)} \rt]. \label{eq: eff_G}
\end{align}
The new normal ordering is related to the old one through
\begin{align}
     : \mathcal{O} : &=  \exp{\lt[- \sum_i \int dk ~ d\om 
dx_1  dx_2  ~ (2\pi)^2 \rt.}\nn
 & \qquad \quad \times \lt\{ G(\om,x_{12}) - 
G_0(\om,x_{12}) \rt\}  \nn 
	 & \qquad \quad \times ~ \lt. \frac{\dl}{\dl 
\psi^*_{i,k}(\om,x_2)}
\frac{\dl}{\dl
\psi_{i,k}(\om,x_1)} \rt]  \ccolon \mathcal{O} \ccolon.
 \label{eq: norm_ordr2}
\end{align}
This transformation modifies only the irrelevant operators 
in the action
because the difference in the propagator vanishes in the 
low energy limit,
\begin{align}
	& G(\om,x_{12}) - G_0(\om,x_{12}) \nn
	&= - i ~ \Theta(-\om x_{12}) \sgn{\om} \nn 
	& \quad \times \lt[ 
\exp(-c_F\abs{x_{12}}\abs{\om}^{2/({\theta} + 1)}) -
\exp(-\eta\abs{x_{12} \om})\rt] \nn
	&=  - i ~ \Theta(-\om x_{12}) \sgn{\om} \nn
	& \quad \times \lt[ 
-c_F\abs{x_{12}}\abs{\om}^{2/({\theta} + 1)} 
+ \ldots \rt].
\end{align}
From now on, all composite operators are understood to be
normal ordered according to Eq. (\ref{eq: norm_ordr}) with
$G(\om,x_{12})$ replacing $G_0(\om,x_{12})$.

\subsection{Wilsonian effective action vs. quantum 
effective action}
\label{sec: WQ}

\begin{figure}[!ht]
     \centering
     \includegraphics[scale=0.5]{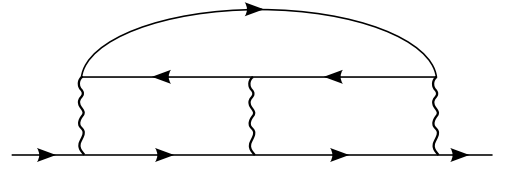}
     \caption{A three-loop correction to the quadratic
vertex.
In the full quantum effective action, 
this is one of the infinite series of diagrams 
which is of the same order as the RPA corrections.
In the Wilsonian effective action, 
this is sub-leading (see the text). 
}
\label{fig: 3L_SE}
\end{figure}


How was it possible for us to construct 
the Wilsonian effective action in \eq{eq: eff_act}
in the strongly coupled field theory ?
The answer to this question lies in the difference
between the quantum effective action
and the Wilsonian effective action.
In the quantum effective action
computed in Ref. \cite{slee}, 
quantum fluctuations are fully incorporated, 
including the contributions from the gapless modes
right on the Fermi surface.
On the contrary, in the Wilsonian effective action
constructed in \eq{eq: eff_act},
only the short-distance quantum fluctuations
up to the length scale $X_0$ are included.
Therefore, the diagrams that 
are of the same order as the RPA diagrams
in the quantum effective action
do not necessarily
come in the same order in the Wilsonian effective action.
In this section, we show that 
non-RPA diagrams are 
indeed sub-leading in the Wilsonian effective action 
although they are not suppressed in the full
quantum effective action.

As a concrete example,
we consider a three-loop diagram shown in Fig. \ref{fig: 
3L_SE}, where three quartic vertices are fused into a 
quadratic vertex
in the Wilsonian effective action.
The resulting vertex is given by
\begin{align}
 \dl S_2^{(3)} &= \int dx ~ \frac{d\om ~ 
dk}{(2\pi)^2} ~~ 
\psi^*_{i,k}(\om,x)
 ~ \Sigma^{(3)}_{ij}(\om,X_0) ~ \psi_{j,k}(\om,x),
\end{align}
where the self-energy is
\begin{widetext}
\begin{align}
 \Sigma^{(3)}_{ij}(\om,X_0) &= \frac{i}{4\pi^5} 
\frac{c_B^{(1-3{\theta})/({\theta} + 1)}}{c_F \gamma} 
\sgn{\om} 
\abs{\om}^{2/({\theta} + 1)} V_{ij';l'n'} V_{m'l';j'i'} 
V_{n'm';i'j} 
f_2\lt(c_F X_0 \abs{\om}^{2/({\theta} + 1)} \rt).
\end{align}
The dimensionless function $f_2(s)$ is 
\begin{align}
 f_2(s) =& s \int_0^1 dx \int_0^1 dt_1 \int_0^{1-t_1} dt_2 ~
dt_3 \int_{-\infty}^{\infty} dy_1 ~ dy_2 ~ 
\step{-\frac{y_1}{y_2}}  \step{1 + \frac{y_1}{y_2} x }
\nn
& \times \frac{|y_1|}{|y_1|^{{\theta} + 1} + (t_1 + t_2)} ~
\frac{1}{|y_2|^{{\theta} + 1} + (t_1 + t_3)} ~ 
\frac{\abs{y_1 - y_2}}{\abs{y_1 - y_2}^{{\theta} + 1} + 
|t_2 - 
t_3|} 
\nn
& \times \exp{\lt[- s x  \lt\{ 
\abs{1 - \frac{y_1}{y_2}} t_1^{\frac{2}{({\theta} + 1)}}  + 
\lt(t_2^{\frac{2}{({\theta} + 1)}} + 
(1-t_1-t_2)^{\frac{2}{({\theta} + 1)}}
\rt) +  \abs{\frac{y_1}{y_2}} \lt( t_3^{\frac{2}{({\theta} 
+ 1)}} +
(1-t_1-t_3)^{\frac{2}{({\theta} + 1)}} \rt) \rt\} \rt]},
\label{eq: f_2}
\end{align}
\end{widetext}
where $f_2(s)$ is finite for all $s$.
It has the following asymptotic behaviors,
\bqa
\lim_{s \rtarw 0} f_2(s) & \sim & s, \label{38} \\
\lim_{s \rtarw \infty} f_2(s) & \sim & 1. \label{limits}
\eqa
We note that  
the exponential factor in the last line of \eq{eq: f_2}
is less than $1$.
In order to obtain 
an upper bound in the small $s$ limit in \eq{38},
we can simply replace the exponential factor by $1$.
Since the rest of the integrals are finite,
$f_2(s)$ should be proportional to $s$
in the small $s$ limit.
The limit in \eq{limits} follows from the 
observation that the function 
multiplying $sx$ in the exponent in \eq{eq: f_2} 
is strictly greater than $0$ and is $\ordr{1}$. 
Therefore, as $s \rtarw \infty$
the leading order contribution to the $x$ integral comes 
from
the region $0 \leq x \lesssim 1/s$,
resulting in the asymptotic 
behavior in \eq{limits}.
\begin{figure}[!ht]
\centering
\includegraphics[width=0.9\columnwidth]{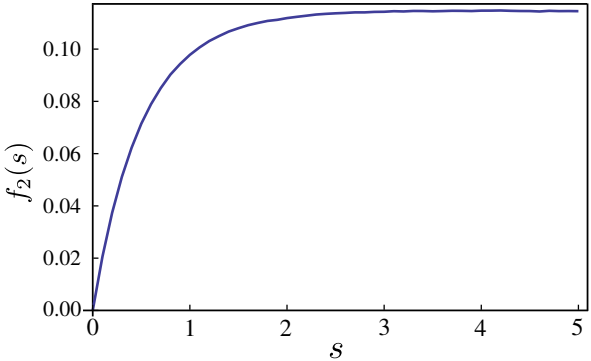}
\caption{A plot of the function $f_2(s)$ in \eq{eq: f_2} 
for ${\theta} = 2$.}
\label{fig: 3L_self-energy}
\end{figure}
We numerically compute $f_2(s)$,
and confirm the asymptotic behaviors
as is shown in Fig. \ref{fig: 3L_self-energy}.

In the Wilsonian effective action,
we consider the low energy limit with fixed $X_0$.
In this case, the limit in Eq. (\ref{38}) applies, 
and the three-loop self-energy has an extra 
factor of $c_F X_0 \abs{\om}^{2/({\theta} + 1)}$ 
compared to the leading order terms in \eq{eq: 
RPA_self-energy}. 
Therefore, the diagram do not contribute to the 
Wilsonian effective action to the leading order.
In the full quantum effective action, on the other hand, 
we consider the limit $X_0 \rightarrow \infty$ with fixed 
external frequency.
In this case, the limit in Eq. (\ref{limits}) applies,
and the diagram has the same scaling behavior as the 
RPA contributions.
In the renormalization group, 
only short distance quantum fluctuations are included
in every step of coarse graining.
In combination with the chiral nature of the theory
which constrains the degree of UV/IR singularity of the 
theory,
this allows one to compute the exact Wilsonian
effective action to the leading order.

Having said that the Wilsonian effective action can be
computed perturbatively in the 
low momentum/frequency limit with fixed $X_0$,
we note that physical observables are given
by the full quantum effective action.
If one wants to compute physical observables with 
external $y$-momenta $k$ using the effective action 
defined at a scale $X_0$ with $k X_0^{1/2} << 1$,
one still has to include the quantum fluctuations
between scales $X_0^{-1/2}$ and $k$, 
which are not included in the Wilsonian effective action.
Therefore, the exact form of the $n$-point function,
which is not dictated by the scaling, in general 
can not be computed perturbatively.

\section{Renormalization group}
\label{sec:RG}

\begin{figure}[!ht]
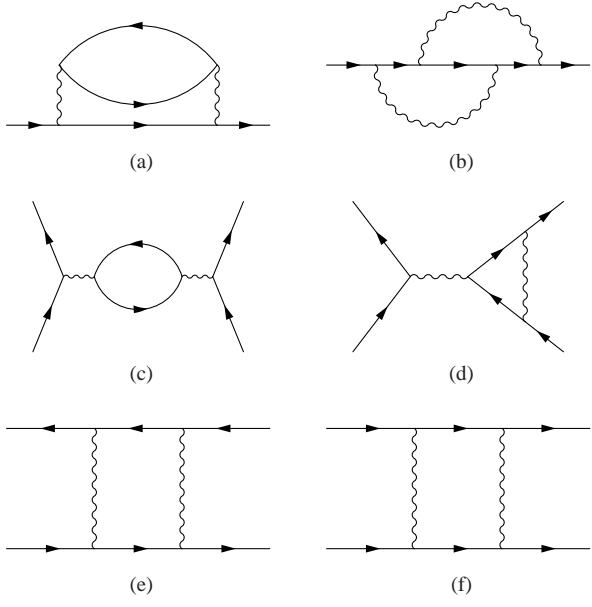

\centering
\subfigure[]{\includegraphics[scale=.5]{dia.1} 
\label{fig: v2B}} \hspace{0.2in}
\subfigure[]{\includegraphics[scale=.5]{dia.2} 
\label{fig: v2nB}} \\
\subfigure[]{\includegraphics[scale=.5]{dia.3}
\label{fig: v4B2}} \hspace{0.2in} 
\subfigure[]{\includegraphics[scale=.5]{dia.4} 
\label{fig: v4P}} \\
\subfigure[]{\includegraphics[scale=.5]{dia.5} 
\label{fig: v4PH}} \hspace{0.2in}
\subfigure[]{\includegraphics[scale=.5]{dia.6} 
\label{fig: v4PP}}
\caption{
The list of quantum corrections
generated from the leading order terms 
in the Wilsonian effective action 
at each step of coarse graining.
}
\label{fig: diagrams}
\end{figure}

It is now straightforward to perform the renormalization
group
analysis by increasing $X_0$ by a factor of $e^{dl}$
in the Wilsonian effective action. 
The quantum correction obtained by increasing $X_0$
can be easily read from
\bqa
\delta S = dl \frac{\partial  S_{X_0}}{\partial \ln X_0},
\eqa
where $S_{X_0}$ is the effective action 
in Eq. (\ref{eq: eff_act}).
The key observation is that $S_{X_0}$ is
independent of $X_0$ apart from irrelevant terms.
This implies that there is no quantum correction 
to the leading order terms in the Wilsonian effective 
action.

Here we show that there is indeed no quantum corrections
to the leading order through an explicit calculation.
We divide the configuration space of operators 
into two parts. 
The first part represents the
configurations where no two composite operators are closer 
than $X_0 e^{dl}$
along the $x$-direction for an infinitesimally small $dl$. 
The second part
represents the configurations where there is at least one 
pair of operators 
whose separation along the $x$-direction is in 
\eqn{
\Omega \equiv \Bigl\{ (x_1,x_2) ~ | ~ X_0 < |x_1 - x_2| 
\leq 
X_0e^{dl} \Bigr\}.
}
Two operators whose relative separation 
is in $\Omega$ are fused into
one composite operator. The volume of the phase space where 
more than two
operators fuse simultaneously is at most order of $dl^2$, 
which can be ignored.

Now we compute the quantum corrections explicitly. 
Fig. \ref{fig: diagrams} shows the channels 
where two quartic operators fuse into quadratic or 
quartic operators. 
The fusions generate the following vertices,
\begin{widetext}
\begin{align}
\Gamma^{(a)}_2 =& 0, \label{GA} \\
& ~ \nn
 \Gamma^{(b)}_2 =& 
 -  \frac{i ~ dl}{4\pi^4}  ~ c_B^{-2\frac{{\theta} - 
1}{{\theta} + 1}}  \int dx ~ \frac{dk d\om}{(2\pi)^2} ~  
V_{n' m';i l'} V_{m' j;l' n'}   
~~  \sgn{\om} ~ \abs{\om}^{2/({\theta} + 1)} 
:\psi^*_{i,k}(\om) ~ \psi_{j,k}(\om):\nn
& \qquad \qquad \times   X_0 \abs{\om}^{\frac{2}{{\theta} 
+ 1}} ~ f^{(b)} \lt(X_0 \abs{\om}^{\frac{2}{{\theta} + 1}} 
\rt), \label{GB} \\
& ~ \nn
 \Gamma^{(c)}_4 =& 0, 
\label{GC} \\
& ~ \nn
 \Gamma^{(d)}_4 =& - \frac{dl}{\pi^2}  ~
c_B^{-\frac{{\theta} - 1}{{\theta} + 1}} 
\int dx ~ \frac{dk_1  dk_2 dq d\om_1 d\om_2  
d\nu}{(2\pi)^6} 
   X_0 \abs{\nu}^{\frac{2}{{\theta} + 1}} ~   
f^{(d)}\lt(X_0 \abs{\nu}^{\frac{2}{{\theta} + 
1}},\frac{q}{\abs{c_B \nu}^{\frac{1}{{\theta} + 1 } }}
,\frac{\om_1}{\nu} \rt)    
\nn
& \times \frac{V_{ij;l'n'} V_{ll';n'n}}{|q|^{\theta} + c_B 
\abs{\nu}/|q|} ~ e^{2i \gamma (k_1 - 
k_2)q x} ~ : \psi^*_{i,k_2}(\om_2) ~  \psi_{j,k_2+q}(\om_2+ 
\nu) \psi^*_{l,k_1+q}(\om_1+\nu) \psi_{n,k_1}(\om_1):, 
\label{GD} \\
& ~ \nn
\Gamma^{(e)}_4 =&  -  \frac{dl}{2\pi^2} ~
c_B^{\frac{1-2{\theta}}{{\theta} + 1}}  
\int dx ~ \frac{dk_1  dk_2 dq ~ d\om_1 d\om_2  
d\nu}{(2\pi)^6 } ~  e^{2i \gamma (k_1 - k_2)q x} ~
\frac{X_0\abs{\Dl \om^{(e)}}^{\frac{2}{{\theta} + 
1}}}{\abs{\Dl
\om^{(e)}}^{\frac{{\theta}}{{\theta} + 1}}}\nn
& \times f^{(e)} \lt(X_0\abs{\Dl
\om^{(e)}}^{\frac{2}{{\theta} + 1}}, 
\frac{q~ \sgn{\Dl \om^{(e)}}}{\abs{c_B \Dl
\om^{(e)}}^{\frac{1}{{\theta} + 1}}}, \frac { 
k_2-k_1}{\abs{c_B
\Dl \om^{(e)}}^{\frac{1}{{\theta} + 1}}}, 
\frac{\om_2}{\Dl \om^{(e)}}, 
\frac{\nu}{\Dl \om^{(e)}} \rt)    \nn
& \times V_{ij';l' n} V_{j'j;l l'}  
~   : \psi^*_{i,k_2}(\om_2) ~ 
\psi_{j,k_2+q}(\om_2 + \nu)
\psi^*_{l,k_1+q}(\om_1+\nu) \psi_{n,k_1}(\om_1)
:,\label{GE} \\
& ~ \nn
 \Gamma^{(f)}_4 =& \frac{dl}{4 \pi^2} ~
c_B^{\frac{1 - 2{\theta}}{{\theta} + 1}} \int dx ~ 
\frac{dk_1  dk_2  dq 
~ d\om_1 d\om_2  d\nu }{(2\pi)^6 } ~  e^{2i \gamma (k_1 - 
k_2)q x} ~ \frac{X_0 \abs{\Dl \om^{(f)}}^{2/({\theta} + 
1)}}{\abs{\Dl
 \om^{(f)}}^{{\theta} / ({\theta} + 1)}} \nn
& \times  f^{(f)} \lt(X_0\abs{\Dl
\om^{(f)}}^{\frac{2}{{\theta} + 1}}, \frac{q}{\abs{c_B \Dl  
\om^{(f)}}^{\frac{1}{{\theta} + 1}}}, 
\frac{k_2-k_1}{\abs{c_B \Dl \om^{(f)}}^{\frac{1 
}{{\theta} + 1}}}, \frac{\om_2}{\Dl
\om^{(f)}}, \frac{\nu}{\Dl \om^{(f)}}, \sgn{\Dl \om^{(f)}}
\rt ) \nn  & \times V_{j'j;l'n} V_{ij';ll'} ~
:\psi^*_{i,k_2}(\om_2) \psi_{j,k_2+q}(\om_2 + \nu)
\psi^*_{l,k_1+q}(\om_1+\nu) \psi_{n,k_1}(\om_1)
:,\label{GF} 
\end{align}
\end{widetext}
where $\Dl \om^{(e)} =  \om_2 - \om_1$ and $\Dl
\om^{(f)} = \om_1 + \om_2 + \nu$.
Here we suppress the reference to the $x$ coordinate in
$\psi_{i,k}(\om,x)$
to simplify the notation. 
The universal crossover functions, $f^{(b,d,e,f)}$'s, are
given by  
\begin{widetext}
\begin{align} 
& f^{(b)}(s)  = \int_{-\infty}^{\infty} dy_1 ~ dy_2 
\int_0^1 dt_1 ~ \int_0^{1-t_1} dt_2 ~ e^{2i\gamma 
c_B^{\frac{2}{{\theta} + 1}} s y_1  y_2} ~ e^{-c_F s \lt[
t_1^{\frac{2}{{\theta} + 1}} + t_1^{\frac{2}{{\theta} + 1}}
+ (1 - t_1 - t_2)^{\frac{2}{{\theta} + 1}} \rt]} 
\nn
& \qquad \qquad \qquad \qquad \qquad \times
\frac{\abs{y_1}}{\abs{y_1}^{{\theta} + 1} + 
\abs{1-t_1}} ~ \frac{\abs{y_2}}{\abs{y_2}^{{\theta} + 1} + 
\abs{1-t_2}}, \\
& ~ \nn
& f^{(d)}(s,u,v) = \int_{-\infty}^{\infty} dy \int_0^1 dt
 ~  e^{-c_F s \lt[ t^{\frac{2}{{\theta} + 1}} +
(1-t)^{\frac{2}{{\theta} + 1}} \rt]} ~~ e^{2i\gamma
c_B^{\frac{2}{{\theta} + 1}} s u  y } ~
\frac{\abs{y}}{\abs{y}^{{\theta} + 1} + \abs{t + v}}, \\
& ~ \nn
& f^{(e)} \lt( s, u_1, u_2, v_1, v_2 \rt)  =
\int_{-\infty}^{\infty} dy \int_{0}^{1} 
dt ~  e^{-c_F s \lt[ t^{\frac{2}{{\theta} + 1}} +
(1-t)^{\frac{2}{{\theta} + 1}} \rt]} ~~  e^{2i\gamma
c_B^{\frac{2}{{\theta} + 1}} s u_2  y } \nn
& \qquad  \qquad \qquad \qquad \qquad \times
 \frac{\abs{y+u_1}}{\abs{y+u_1}^{{\theta} + 1} + 
\abs{v_1 - t} } ~ \frac{\abs{y}}{\abs{y}^{{\theta} + 1} + 
\abs{v_1 + v_2 - t} }, \\
& ~ \nn
& f^{(f)} \lt( s, u_1, u_2, v_1, v_2, r \rt) =
\int_{-\infty}^{\infty} dy \int_{0}^{1}
dt ~  e^{-c_F s \lt[ t^{\frac{2}{{\theta} + 1}} +
(1-t)^{\frac{2}{{\theta} + 1}} \rt]} ~~ e^{2i\gamma
c_B^{\frac{2}{{\theta} + 1}} s r (u_1+u_2 + y)  y}\nn
& \qquad  \qquad \qquad \qquad \qquad \times \frac{\abs{y +
u_1}}{\abs{y+u_1}^{{\theta} + 1} + 
\abs{v_1-t} } ~ \frac{\abs{y}}{\abs{y}^{{\theta} + 1} + 
\abs{v_1 + v_2 - t} }.
\end{align}
\end{widetext}
The absence of quantum corrections 
in Eqs. (\ref{GA}) and (\ref{GC}) 
is a consequence of the fact that the RPA diagrams
are independent of $X_0$.
All non-vanishing quantum corrections are proportional to 
$X_0$.
Moreover, crossover functions are finite for ${\theta} > 
1$. 
From dimensional ground, this implies that all quantum 
corrections
come with an additional factor of momentum or frequency.
Therefore all quantum corrections
in Fig. \ref{fig: diagrams} are irrelevant relative to 
the terms that are already present in the effective action.

The scaling is easily determined from the scaling
that leaves the Wilsonian effective action in Eq. (\ref{eq: 
eff_act}) invariant.
In order to put the cut-off structure to the original form 
after the coarse graining,
we re-scale 
frequency, 
$x$-coordinate, 
$y$-momentum and 
the field as
\bqa
\om' &=& e^{{z ~ dl}} ~ \om, \label{scale1} \\
x' &=& e^{{-dl}} ~ x, \\ 
k' &=& e^{\alpha dl} ~  k, \\
\psi'_{i, k'}(\om',x') &=& e^{{\Delta_\psi ~dl}} ~
\psi_{i, k}(\om,x).
 \label{scale}
\eqa
The dynamical critical exponent $z$,
the scaling dimension of the fermion field $\Delta_\psi$, 
and the dimension of $y$-momentum $\alpha$
should be determined from the condition
that the action is invariant. 
The condition that the marginal term $\psi^*\partial_x 
\psi$ 
should be scale invariant 
fixes the scaling dimension of the field to be
\bqa
\Delta_\psi &= - \frac{1}{2}\Bigl(z+ \alpha \Bigr).
\label{A-D} 
\eqa
Then the beta functions for $c_F$, $c_B$, $\gamma$ and 
$V_{ij;ln}$ 
are given by
\bqa
&& \frac{d c_F}{dl}  = \lt(1 - \frac{2z}{{\theta} + 1} 
\rt)c_F, \\
&&   \frac{d c_B}{dl}  = \Bigl(\alpha({\theta} + 1) - z 
\Bigr)c_B, 
\\
&&  \frac{d \gamma}{dl} = \Bigl( 1- 2\alpha \Bigr)\gamma, \\
&& \frac{d V_{ij;ln}}{dl} =  \Bigl[ 1
- z + \alpha({\theta} - 1)\Bigr] V_{ij;ln}.
\label{V-D}
\eqa
One can find a fixed point for the beta functions 
if and only if we choose
\bqa 
\alpha &=& 1/2, \\
 z &=& ({\theta} + 1)/2.
\label{az}
\eqa
This uniquely fixes the dynamical critical exponent and
the scaling dimension of the field.

\begin{figure}[!ht]
     \centering
     \includegraphics[scale=0.4]{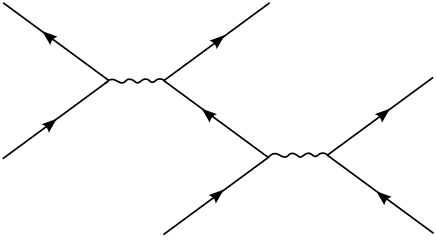}
     \caption{A six-fermion vertex generated from 
two quartic vertices.}
     \label{fig: 6_F}
\end{figure}

One can check that higher order vertices 
that are generated from the quartic vertices
are all irrelevant under the scaling 
in Eqs. (\ref{scale1}) - (\ref{scale}).
As an example,
we compute the quantum correction where
two quartic vertices fuse into a sixth order vertex
as is shown in Fig. \ref{fig: 6_F}. 
In the small $X_0$ limit, it becomes 
\begin{widetext}
\begin{align}
\Gamma_6 &= 4 i ~ X_0 dl \int \frac{dk_1 ~ dk_2 ~ dk_3~
dq_1 ~ dq_2 ~ d\om_1 ~ d\om_2 ~ d\om_3 ~ d\nu_1 ~ d\nu_2
}{(2\pi)^{10}} ~ dx
 ~~ e^{ 2 i \gamma \lt[ ( k_1 - k_2 )q_1 + (k_3 - k_1 )q_2
\rt] x } \nn
&  \times \frac{V_{ij;ln'} ~ V_{n' n; m s}}{|q_1|^{\theta} +
c_B ~ \abs{\nu_1}/|q_1| } ~
\frac{\sgn{\om_1}}{|q_2|^{\theta} + c_B ~ 
 \abs{\nu_2} / |q_2| } ~  e^{-X_0 \lt[
c_F \abs{\om_1}^{2/({\theta} + 1)} + 2i\gamma ~ {sgn}(\om_1)
(k_1 - k_3) q_2 \rt]}\nn 
&  \times : \psi^*_{i,k_2}(\om_2,x) 
\psi_{j, k_2 + q_1}(\om_2 + \nu_1, x)
\psi^*_{l, k_1 + q_1}(\om_1 + \nu_1, x)
\psi_{n, k_1+ q_2}(\om_1 + \nu_2, x)
\psi^*_{m, k_3 + q_2}(\om_3 + \nu_2, x) 
\psi_{s, k_3}(\om_3, x):.
 \label{eq: gamma_6}
\end{align}
\end{widetext}
According to the scaling in \eq{az} and the 
expression of $\Dl_{\psi}$ in \eq{A-D}, 
this is irrelevant.
This can be readily seen from the fact 
that the prefactor is proportional to $X_0$ 
which has scaling dimension $-1$.
This is true for any higher 
order vertices generated during the RG flow.
Contributions from these higher order vertices 
to the quadratic and quartic vertices
are also irrelevant.

As expected, the scaling form 
in the chiral non-Fermi liquid state
is fixed by the scaling in \eq{eq:IS}
where the interaction is kept invariant 
while the frequency dependent term
in the bare quadratic action is deemed strongly irrelevant.
This implies that the theory flows to a strongly 
interacting 
non-Fermi liquid fixed point in the low energy limit.
It is remarkable that it is possible to obtain the 
exact scaling relation for the strongly interacting 
non-Fermi liquid fixed points. 
The scaling relation in Eq. (\ref{az}) suggests that 
the exact fermion Green's function in the momentum space
has the form,
\begin{align}
G^{-1}(k) = \delta_k ~ 
g\lt( \frac{|\om|^{2/({\theta} + 1)}}{\dl_k} \rt),
\label{Gsc}
\end{align}
where $\delta_k = k_x + \gamma k^2$ and 
$g(x)$ is a universal function.
Note that the one-loop Green\rq{}s function
obeys the scaling form in \eq{Gsc}.
In other words, chirality allows us to extract 
the scaling form of the exact Green\rq{}s function, 
$G(k)$ 
from the one-loop Green's function. 
However, the dimensionless function $g(x)$ 
is not fixed by scaling, 
and the exact form of $g(x)$ can be, in principle, very 
different from
what is inferred from the one-loop Green's function.

\section{General patch theories}
\label{sec:general}

\begin{figure}[!ht]
 \centering
 \includegraphics[scale=0.5]{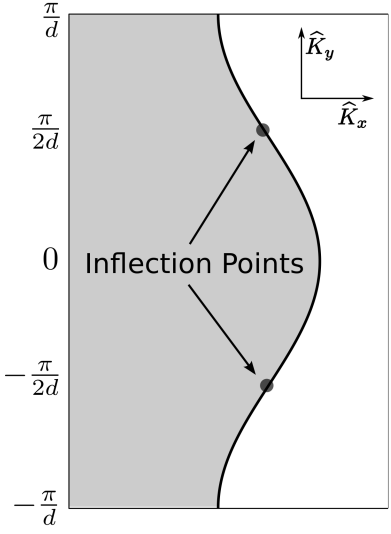}
 \caption{Two inflections points 
 on the chiral Fermi surface 
at which the quadratic curvature vanishes
for the dispersion in \eq{ek}.
}
\label{fig:inflection}
\end{figure}

The theory with the dynamical critical exponent
$z=\frac{{\theta} + 1}{2}$ 
captures the low energy physics near 
the Fermi surface with nonzero quadratic curvatures.
There exist special points 
where the quadratic curvature
vanishes.
In particular, 
the periodicity of the first Brillouin zone
in the $K_y$ direction guarantees 
that there exist inflection points
as is shown in Fig. \ref{fig:inflection}. 
For example, the dispersion in \eq{ek} 
has two inflection points at $K_y = \pm \pi/2d$. 
In the neighborhood of one of the inflection points
(say $K_y = -\pi/2d$),
the dispersion can be written as 
$\epsilon_{k} = k_x - 2td k_y + (td^3/3) ~ k_y^3 + 
\ordr{k_y^5}$,
where $\vec k$ is a deviation from the 
inflection point.
Defining
$ k'_x = k_x - 2td k_y $, $k_y' = k_y$,
the local dispersion is written as
\bqa
\epsilon_{k'} = k'_x + \gamma_3 {k_y'}^3
\eqa
with the cubic curvature given by
$\gamma_3 = \frac{t d^3}{3}$.
Henceforth, we will drop the prime in $k_x'$, $k_y'$. 
With some extra fine tuning, one can even have a higher 
inflection 
points with a local dispersion,
\bqa
\epsilon_k = k_x + \gamma_u  k_y^u
\label{localu}
\eqa
with $u>3$.
Therefore, a general patch theory for chiral non-Fermi 
liquids
can be parameterized by $(u,{\theta})$,
\eqn{
S^{(u,{\theta})} &=   \int 
\frac{ d\omega d^2\vec k}{(2\pi)^3} ~ 
\lt(i \eta \om + k_x + \gamma_u  k_y^u \rt)
\psi_{j}^*(\omega, \vec k) \psi_{j}(\omega, \vec k) \nn
& + \frac{1}{2} \int
\frac{ d\nu d^2 \vec q}{(2\pi)^3} ~ 
\chi_\theta^{-1}(q)
~\phi_{\alpha}(-\nu, -\vec q) 
\phi_{\alpha}(\nu, 
\vec q) \nn
& + g \int 
\frac{ d\omega d^2\vec k}{(2\pi)^3} 
\frac{ d\nu d^2 \vec q}{(2\pi)^3} ~  \nn
& \times ~~
\phi_{\alpha}(\nu, \vec q) 
\psi_{i}^*(\omega+\nu, \vec k+\vec q) ~ T^{\alpha}_{ij} ~ 
\psi_{j}(\omega, \vec k).
 \label{eq: b.f.action_n}
}
Here $u$ is a positive integer greater than $1$,
and ${\theta}$ is an even integer for local theories.

\subsection{UV finiteness and exact scaling}

In this section,
we show that each of the theory 
parameterized by $(\theta,u)$
in \eq{eq: b.f.action_n}
describes a distinct and stable 
non-Fermi liquid fixed point for
$1 < \theta < u + \half$ and $u\geq 2$. 
For $\theta > u + \half$, 
the theory should be modified by 
$\lambda \phi^4$ term 
which becomes relevant
at the Gaussian fixed point.
The RG analysis for general $u$ 
is similar to that for the theory with $u=2$
discussed in Secs. \ref{sec:regularization} -  
\ref{sec:RG}.
In this section, 
we will quickly recapitulate 
the key results from the previous sections
that apply to general $u \ge 2$,
and emphasize new features
that are absent in the case with $u=2$.
In the mixed-space of $x$-coordinate and frequency,
the action is written as
\begin{align}
 & S^{(u,{\theta})}  \nn
& =  \int \frac{dk ~ d\om}{(2 \pi)^2} \int dx ~ \psi_{i,
k}^*(\om,x) \Bigl[  i\eta \om  - i\partial_x \Bigr]
\psi_{i, k}(\om,x)  \nn 
& +  \int \frac{dk_1 ~dk_2 ~dq ~d\om_1 ~d\om_2 ~d\nu}{(2
\pi)^6} \nn 
& \times \int dx ~ \exp{\lt(i \gamma_u x 
\sum_{m=1}^{u-1}
\binom{u}{m} (k_1^m - k_2^m ) q^{u-m} \rt)} \nn
& \qquad \times   
V_{ij;ln} \chi_\theta(q) ~ 
\psi^*_{i, 
k_2}(\om_2,x) ~
\psi_{j, k_2+q}(\om_2+\nu,x) \nn
& \qquad \times ~ \psi^*_{l, k_1+q}(\om_1+\nu,x) ~
\psi_{n,k_1}(\om_1,x). 
\label{eq: mixed_action_n}
\end{align}
Again, $k_i, q$ refer to $y$-momenta.
Two comments are in order for 
the general theory in \eq{eq: mixed_action_n}. 
First, the phase factor in the quartic interaction
includes $u$ powers of $y$-momenta
which is inherited from the dispersion in \eq{localu}.
As a result, the sliding symmetry associated with
$\psi_{i,k} \rightarrow \psi_{i,k+\Delta k}$ is
absent for $u>2$.
Second, the number of points with a common tangent vector
determines the representation of the fermion field 
under the  flavor group.
For $u=3$, the two inflection points have different tangent 
vectors 
as is shown in Fig. \ref{fig:inflection}.
Since the two are decoupled at low energies,
only one point needs to be kept 
in the low energy effective theory.
As a result, one does not need to double the representation
as in \eq{doubleT}.

Now we consider the Wilsonian effective action 
with a running cut-off scale $X_0$,
which is obtained by fusing operators whose separation in 
$x$-direction is less than $X_0$. 
As is the case for $u=2$
discussed in Sec. \ref{sec:con}, 
the Wilsonian effective action is
finite in the $\Lambda X_0^{1/2} \rightarrow \infty, 
\eta^{-1} X_0^{(\theta - 1)/2} \rightarrow \infty$ 
and $\mu X_0^{1/2} \rightarrow 0$ limit.
Since the integrations over internal frequencies 
are insensitive to the value of $u$,
one can use the exactly same argument presented 
in Sec. \ref{sec: UV_finite}
to show that they are finite :
all internal frequencies are bounded
by the external frequencies due to chirality.
The integrations over $y$-momenta
are also UV convergent for ${\theta} > 1$.
As is discussed in Sec. \ref{app: int_flavor},
an integration over $y$-momentum
which goes through a boson propagator
is UV convergent
because it is suppressed as $q^{-\theta}$
at large momentum.
The only exceptions are the fermion loops 
which are solely made of fermion propagators.
This is because the fermion propagator does not depend
on $y$-momentum, and the integration over it 
diverges in the absence of the phase factor
which cuts off the divergence.
For example, 
the integration over $y$-momentum in 
the RPA bubble reads
\begin{align}
& \int_{-\Lambda}^\Lambda dp ~ \exp{\lt(i \gamma_u x_R 
\sum_{m=1}^{u-1} \binom{u}{m} p^m q^{u-m} 
\rt)} \nn 
& \sim \frac{1}{\abs{q ~ x_R}^{1/(u-1)}},
\label{eq:108}
\end{align}
in the $\Lambda \rightarrow \infty$ limit,
where 
$p$ is the $y$-momentum that runs in the loop,
$q$ is the  momentum transferred across the loop,  and
$x_R$ is the relative coordinate between two vertices.
The fact that the UV divergence from large $y$-momenta
is cut-off by a length scale in $x$-direction
is a consequence of the non-commutativity between
$x$-coordinate and $y$-momentum as discussed
in Sec. \ref{sec: non-comm}.
Because $x_R \sim X_0$, this gives rise to 
a UV enhancement factor of $X_0^{-1/(u-1)}$
in the $X_0 \rightarrow 0$ limit
for every fermion loop.
Although \eq{eq:108} is singular in the $x_R \rightarrow 0$ 
limit,
it is integrable for $u \geq 2$.
(For $u=2$, it gives rise to a delta function for the 
relative coordinate.)
Once the relative coordinate between vertices
is integrated over, the RPA bubble is order of 
\bqa
\int_{-X_0}^{X_0} dx_R ~ x_R^{-1/(u-1)} \sim 
X_0^{(u-2)/(u-1)}
\label{eq: 109}
\eqa
for $u>2$ and a constant independent of $X_0$ for $u=2$.

Because the Wilsonian effective action is UV finite, 
one can take the 
$\Lambda X_0^{1/2} \rightarrow \infty, 
\eta^{-1} X_0^{(\theta - 1)/2} \rightarrow \infty$ limit.
Since there is no scale in the full quantum theory,
the low energy physics should be invariant 
under the scale transformation which 
leaves the rest of the terms
in \eq{eq: mixed_action_n} invariant. 
In other words, the Wilsonian effective action should be
invariant under the coarse graining $X_0 \rightarrow X_0
e^{dl}$
followed by a re-scaling of the field and momenta dictated
by the following scaling dimensions,
\begin{eqnarray}
{[}x{]}  & = &  -1,  \label{eq:GIS1} \\
{[}k{]}  & = & \frac{1}{u}, \\
z_{\theta,u} \equiv {[}\om{]}  & = &  \frac{{\theta} + 
u-1}{u}, \\
{[}\psi_{i}{]} & = & -\frac{{\theta} + u}{2u}, \\
{[}V{]} & = & 0, \\
{[} \eta {]} & = & -\frac{{\theta} - 1}{u}.
\label{eq:GIS}
\end{eqnarray}
Eqs. (\ref{eq:GIS1}) - (\ref{eq:GIS})
give the exact dynamical critical exponent
and scaling dimensions.

\subsection{The Wilsonian effective action}

\subsubsection{The RPA correction}

In this section, 
we compute the Wilsonian effective action  
in the small $X_0$ limit. 
Let us estimate the magnitude of an operator 
generated by fusing $V$ quartic vertices
in the small $X_0$ limit.
The fusion of $V$ vertices involves $(V-1)$
relative coordinates that are integrated 
over the range of $X_0$.
This leads to a factor of $X_0^{V-1}$. 
As is discussed in the previous section, 
each fermion loop gives rise to 
a UV enhancement factor of $X_0^{-1/(u-1)}$.
This gives a net factor of 
\bqa
X_0^{V-1-L_f/(u-1)}
\label{eq:naive_power_counting}
\eqa
for a diagram with $V$ vertices
and $L_f$ fermion loops.
Indeed, \eq{eq:naive_power_counting} is the full answer
for $u=2$, as we have seen in 
in Sec. \ref{sec:con}.
For $u>2$ we will see that there are corrections
to \eq{eq:naive_power_counting}.
To see this, 
we first consider the RPA diagrams
which are largest 
in the small $X_0$ limit
according to \eq{eq:naive_power_counting}.
The RPA bubble dresses the quartic vertex to
(see Appendix \ref{app: general_n_1} for derivation) 
 \begin{align}
&  \chi^{(RPA)}_{u,{\theta}}(q,\nu,X_0) \nn
&= \frac{1}{ |q|^{\theta} + c_B^{(u)}(X_0)~ 
\xi_u(\sgn{\gamma_u \nu 
q})~\dsty{\frac{|\nu|}{|q|^{1/(u-1)}}} }
\label{eq: chi_n_main}
\end{align}
with
\eqn{
c_B^{(u)}(X_0) = \lt(\frac{g^2}{2\pi^2} \rt) 
\frac{X_0^{(u-2)/(u-1)}}{|u\gamma_u|^{1/(u-1)}} .
}
The key difference from the $u=2$ case is that 
the RPA correction in the dressed quartic vertex
vanishes in the $X_0 \rightarrow 0$ limit 
for $u>2$ as is shown in \eq{eq: 109}.
It is crucial to include the leading 
quantum correction to the boson self-energy
to dress the quartic vertex to cure the IR singularity
as we have already seen in the $u=2$ case.
Because the IR singularity is cut off by the RPA correction,
which vanishes in the small $X_0$ limit,
each integration over $y$-momentum
that goes through a boson propagator leads to
an IR enhancement factor,
\bqa
&& \int dq ~
\frac{1}{|q|^{\theta} + c_B^{(u)}(X_0)~ \xi_u(\sgn{\gamma_u
\nu q}) \dsty{\frac{\abs{\nu}}{|q|^{1/(u-1)}}}}  \nn
&&  \sim X_0^{-({\theta}-1)(u-2)/({\theta}(u-1)+1)}.
\label{eq:ml}
\eqa
Taking the IR enhancement factor into account,
the naive power counting in \eq{eq:naive_power_counting}
is modified as
\begin{align}
& X_0^{ (V-1) + \lt[-\frac{1}{u-1}\rt]  L_f + 
\lt[- \frac{({\theta}-1)(u-2)}{{\theta}(u-1)+1} \rt]  L_m }
\nn
& = 
X_0^{ 
\half ( E - 4 ) + \frac{u-2}{u-1} ~ L_f + \frac{{\theta} +
u -1}{{\theta}(u-1) + 1} ~ L_m },
\label{eq:mct}
\end{align}
where $L_m$ is the number of `mixed loops' 
that contain at least one boson propagator 
in contrast to fermion loops.
We use the identity $V = \frac{E}{2} + L_m + L_f - 1$,
and assume that only one boson propagator becomes singular
at a time within each mixed loop.
Since the coefficients of $L_f$ and $L_m$ are positive for
$u>2$, higher loop diagrams are suppressed 
in the small $X_0$ limit.
For the fermion self-energy,
the leading correction arises for $L_f=0, L_m=1, V=1$ 
which is order of 
$X_0^{-\frac{(u-2)({\theta} -1)}{{\theta}(u-1)+1}}$.
See Appendix \ref{app: general_n_2} for 
an explicit computation of the leading correction
to the quadratic action.
Therefore the Wilsonian effective action 
with the leading order quantum corrections
is written as
\begin{align}
& S^{(u,{\theta})}_{X_0} = \int \msr{k} ~ \msr{\om} ~ dx
~ \ccolon \psi_{i, k}^*(\om,x) \nn
&  \quad \times \Bigl[  i c_F^{(u)}(X_0)~ \sgn{\om}
|\om|^{\frac{u}{{\theta}(u-1)+1}} - i\partial_x
\Bigr] \psi_{i, k}(\om,x) \ccolon \nn
& + \int \frac{d  k_1 ~ dk_2 ~ dq ~ d\nu ~ d\om_1 ~
d\om_2}{(2\pi)^6} \int dx \nn 
& ~~~ \times  \frac{V_{ij;ln} ~ \exp{\lt(i \gamma_u x
\sum_{m=1}^{u-1} \binom{u}{m} (k_1^m - k_2^m ) q^{u-m}
\rt)}}{|q|^{\theta} + c_B^{(u)}(X_0)~ \xi_u(\sgn{\gamma_u
\nu q}) \dsty{\frac{\abs{\nu}}{|q|^{1/(u-1)}}}} \nn
& ~~~ \times   \ccolon \psi^*_{i,k_2}(\om_2,x) ~
\psi_{j,k_2+q}(\om_2+\nu,x)\nn  
& ~~~ \times ~ \psi^*_{l,k_1+q}(\om_1+\nu, x) 
\psi_{n,k_1}(\om_1,x)\ccolon, 
\label{eq: eff_act_general}
\end{align}
where
\begin{align}
c_F^{(u)}(X_0) &\propto 
X_0^{-\frac{(u-2)({\theta} -1)}{{\theta}(u-1)+1}}, 
\label{eq: cFn} \\
c_B^{(u)}(X_0) &\propto  X_0^{\frac{(u-2)}{(u-1)}} 
\label{eq: cBn} 
\end{align}
with
\begin{align}
 & \xi_u(\sgn{\gamma_u \nu q}) \nn 
 &= 
 \begin{cases}
  \dsty{ \frac{u-1}{u-2} \int_0^{\infty} dy ~
\cos\lt(y^{u-1} \rt)} & \mbox{for even } ~u\\
 ~ & \\
  \dsty{ \frac{u-1}{u-2} \int_0^{\infty} dy ~ e^{i 
~ sgn(\gamma_u \nu q) ~ y^{u-1}}} & \mbox{for odd } ~u.
 \end{cases}
 \label{eq: xi}
\end{align}
The expression in \eq{eq: xi} is well defined for any 
$u \geq 2$.
Note that $\xi_u$ is complex for odd 
$u$ and $\xi_u(\sgn{\gamma_u \nu q}) = 
\xi_u^*(-\sgn{\gamma_u \nu q})$.
This appears to break the inversion along the $y$ direction
under which $q$ is mapped to $- q$.
However, the full theory which includes the other patch
with the opposite $y$-momentum respects the inversion 
as is explained in Appendix \ref{app: general_n_3}.

It is noted that $c_F^{(u)}(X_0)$ 
in the fermion self energy 
is singular in the $X_0 \rightarrow 0$ limit for $u>2$, 
which reflects the fact that the IR divergence 
is cut off by the quantum correction 
which vanishes in the $X_0 \rightarrow 0$ limit
in \eq{eq: chi_n_main}.
Although $c_F^{(u)}(X_0)$ diverges 
in the $X_0 \rightarrow 0$ limit for $u>2$,
its feedback to the Wilsonian effective action
is sub-leading in the small $X_0$ limit.
This can be seen from the dressed fermion propagator,
\begin{align}
G(\om,x_{12}) &= 
- i~ \sgn{\om} ~ \Theta(-x_{12} \om) \nn
& \quad \times \exp \lt[-
c_F^{(u)}(X_0) \abs{x_{12}} 
|\om|^{\frac{u}{{\theta}(u-1)+1}} 
\rt]. 
\label{eq: eff_Gn}
\end{align}
Because $\abs{x_{12}} < X_0$ for propagators
that enter in OPE's for the Wilsonian effective action,
we have $c_F^{(u)}(X_0) \abs{x_{12}} 
< X_0^{\frac{\theta+u-1}{\theta(u-1)+1}} \rightarrow 0$ 
in the $X_0 \rightarrow 0$ limit.
As a result, one can ignore the exponential factor 
in \eq{eq: eff_Gn} to the leading order in $X_0$.

\subsubsection{Beyond the RPA correction}

\begin{figure}[!ht]
\includegraphics[width=0.6\columnwidth]{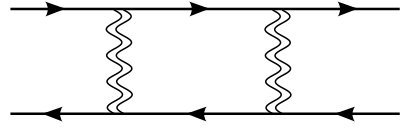}
\caption{A non-RPA contribution to the quartic vertex.
The double wiggly line represents the RPA dressed quartic
vertex. }
\label{fig: PH_ladder}
\end{figure}

Non-RPA corrections 
are suppressed 
in the small $X_0$ limit.
For example, a non-RPA diagram with $L_f=0$ and $L_m=1$ 
shown in Fig. \ref{fig: PH_ladder} contributes
\begin{align}
& \dl S_4^{(ph)} = - \int \frac{dk_1 dk_2 dq ~ d\om_1 d\om_2
d\nu}{(2\pi)^6} dx ~ V_{ij';l'n} V_{j'j;l l'} \nn
&\times  e^{i \gamma_u x \sum_{m=1}^{u-1} \binom{u}{m}
(k_1^m - k_2^m) q^{u-m} } ~~
\Gamma^{(ph)}(k_1,k_2,q,\om_1,\om_2,\nu) \nn
&\times \ccolon \psi^*_{i,k_2}(\om_2 ,x)
\psi_{j,k_2 + q}(\om_2 + \nu,x)  \nn
& \qquad \times \psi^*_{l,k_1 + q}(\om_1
+ \nu,x) \psi_{n,k_1}(\om_1,x) \ccolon,
\end{align}
where
\begin{widetext}
\begin{align}
& \Gamma^{(ph)}(k_1,k_2,q,\om_1,\om_2,\nu) =
\frac{1}{2\pi^2}\lt[\frac{g^2}{2 \pi^2 \abs{u
\gamma_u}^{1/(u-1)}}\rt]^{\frac{(u-1)(1 - 
2{\theta})}{{\theta}(u-1)+1}} ~
X_0^{\frac{({\theta} + 1) - ({\theta} - 
1)(u-2)}{{\theta}(u-1) + 1}} ~ \abs{\Dl
\om}^{\frac{1 - ({\theta} - 1)(u-1)}{{\theta}(u-1) + 1}} 
\nn 
& \times f^{(ph)} \lt( c_F X_0 \abs{\Dl 
\om}^{\frac{u}{{\theta}(u-1) +
1}},  \gamma_u \sgn{\Dl \om} X_0
\lt(c_B^{(u)} \abs{\Dl \om} 
\rt)^{\frac{u(u-1)}{{\theta}(u-1) +
1}}, \frac{q}{\lt(c_B^{(u)} \abs{\Dl \om}
\rt)^{\frac{u-1}{{\theta}(u-1) + 1}}}, 
\frac{k_1}{\lt(c_B^{(u)}
\abs{\Dl \om} \rt)^{\frac{u-1}{{\theta}(u-1) + 1}}}, \rt. 
\nn
& \qquad \qquad \qquad \lt. \frac{k_2}{\lt(c_B^{(u)}
\abs{\Dl \om} \rt)^{\frac{u-1}{{\theta}(u-1) + 1}}}, 
\frac{\nu}{\Dl
\om}, \frac{\om_2}{\Dl \om} \rt), \\
& \quad \nn
& \mbox{with } \Dl \om = \om_1 - \om_2 \mbox{ and} \nn
& \quad \nn
& f^{(ph)}(s_1, s_2, u_1, u_2, u_3, v_1, v_2)= \int_0^1 dx 
~dt \int_{-\infty}^{\infty} dy ~ e^{i s_2 x \sum_{m=1}^{u-1}
\binom{u}{m} (u_2^m - u_3^m) y^{u-m}} ~ e^{- s_1 x \lt(
t^{\frac{u}{{\theta}(u-1)+1}} + 
(1-t)^{\frac{u}{{\theta}(u-1)+1}} \rt)}
\nn
& \quad \times \frac{\abs{y- u_1}^{\frac{1}{u-1}}}{\abs{y -
u_1}^{{\theta} +  \frac{1}{u-1}} + \xi_u(\sgn{\gamma_u
(t+v_1+v_2)(y-u_1)}) \abs{t + v_1 + v_2}} ~
\frac{\abs{y}^{\frac{1}{u-1}}}{\abs{y}^{{\theta} +  
\frac{1}{u-1}}
+ \xi_u(\sgn{\gamma_u (t+v_2) y}) \abs{t + v_2} }.
\end{align}
\end{widetext}
Since the cross-over function has complicated dependence
on the dimensionless parameters, let us consider a 
particular limit to compare with the RPA correction. We
consider the small frequency limit with fixed
$y$-momenta,
where $s_i \ll 1$, $u_i \gg 1$ and $v_i \sim 1$.
In this limit, the cross-over function becomes
\begin{align}
f^{(ph)}(0,0,u_1,u_2,u_3,v_1,v_2) \sim 
\abs{u_1}^{-{\theta}},
\end{align}
resulting in the quantum correction, 
\begin{align}
& \Gamma^{(ph)}(k_1,k_2,q,\om_1,\om_2,\nu) \sim 
\lt[\frac{g^2}{\abs{\gamma_u}^{1/(u-1)}}\rt]^{\frac{
(u-1)(1 - 2{\theta})}{ {\theta}(u-1)+1}}  \nn
& \times   \lt( X_0 \abs{\Dl \om}^{\frac{u}{{\theta} +  u - 
1}}
\rt)^{\frac{{\theta} + u-1}{{\theta}(u-1)+1}} ~ 
|q|^{-{\theta}}.
\end{align}
This is smaller than the RPA dressed vertex 
in \eq{eq: eff_act_general} in the limit under
consideration.

\begin{figure}[!ht]
\includegraphics[width=0.9\columnwidth]{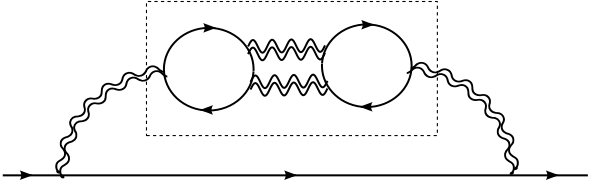}
\caption{
A diagram where two quartic vertices (boson propagator)
outside the box carry a same momentum.
The double wiggly line represents the RPA dressed quartic
vertex.
}
\label{fig: nRPA_1L}
\end{figure}
%
%
\begin{figure}[!ht]
\subfigure[]{\includegraphics[width=0.6\columnwidth]
{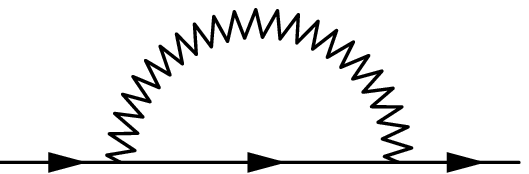} \label{fig: full_chi_a}} \\
\subfigure[]{\includegraphics[width=\columnwidth]{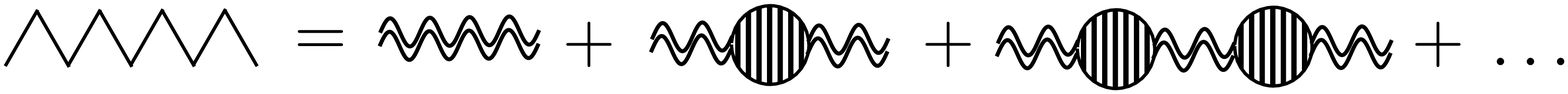}}
\caption{
(a) The diagram in Fig. \ref{fig: nRPA_1L} is 
a part of the `one-loop\rq{} fermion self-energy.
(b) The zig-zag line in (a) represents the quartic vertex
dressed with an infinite series of the three-loop boson
self-energy drawn inside the box in \fig{fig: nRPA_1L}.
}
\label{fig: full_chi}
\end{figure}

There are diagrams 
which do not obey the counting in \eq{eq:mct}.
For example, the diagram shown in \fig{fig: nRPA_1L}
has a smaller power of $X_0$ than predicted 
in \eq{eq:mct}.
In this diagram, there are two boson propagators
(the ones outside the box in \fig{fig: nRPA_1L})
which carry exactly same momentum.
In the absence of the RPA correction to the quartic vertex,
the two boson propagators become singular simultaneously.
As a result,
the integration over the $y$-momentum 
that goes through the two boson propagators
has a larger IR enhancement factor than predicted in
\eq{eq:ml}.
However, we can not consider this diagram by itself
because it is part of an infinite series of diagrams as is
shown in \fig{fig: full_chi}.
The quartic vertex dressed by
higher-loop boson self-energies
can be written as
\begin{align}
&  \int \frac{dk_1 dk_2 dq ~ d\om_1 d\om_2
d\nu}{(2\pi)^6} dx ~ \frac{V_{ij;ln}}{|q|^{{\theta}} + 
\Pi(k_1,k_2,q,\nu,X_0)} \nn
&\times e^{i \gamma_u 
x \sum_{m=1}^{u-1} \binom{u}{m} (k_1^m - k_2^m) q^{u-m} 
} \nn
&\times  \ccolon \psi^*_{i,k_2}(\om_2 ,x) \psi_{j,k_2 + 
q}(\om_2 + \nu,x) \nn
& \quad \times \psi^*_{l,k_1 + q}(\om_1 + \nu,x) 
\psi_{n,k_1}(\om_1,x) \ccolon,
\end{align}
where 
\eqn{
  \Pi(k_1,k_2,q,\nu,X_0) =& \Pi_{u}(k_1,q,\nu,X_0) \nn
& \quad + \Pi_{nRPA}(k_1,k_2,q,\nu,X_0). \label{eq: full_pi}
}
Here $\Pi_u$ is the RPA correction defined in \eq{eq: 
pi_gen_defn} and $\Pi_{nRPA}$ represents the corrections 
beyond the RPA level.
$\Pi_{nRPA}$ includes the sub-diagram inside the box in \fig{fig: nRPA_1L}. 
The one-particle irreducible non-RPA
correction is suppressed compared to $\Pi_u$ 
according to \eq{eq:mct}. 
Therefore, the RPA diagram dominates 
in the small $X_0$ limit. 

Although, the diagram in \fig{fig: full_chi_a} is consistent
with \eq{eq:mct}, we do not have a systematic
way of computing the exact dependence on $X_0$ for general diagrams.
However, we emphasize that 
the scaling dimensions in \eq{eq:GIS}
hold exactly 
irrespective of the magnitudes
of individual diagrams 
in the small $X_0$ limit.

\section{Thermodynamic response}
\label{sec:TR}

One full Fermi surface generally includes 
multiple patch theories with different values of $u$
that belong to different universality classes.
What is then the thermodynamic response of the whole system?
Here we consider the case where
the quadratic curvature is nonzero
except for an isolated inflection point of the $u$-th order.

Suppose there is an inflection,
$\mc{P}^* \equiv  (K_x^*, K_y^*)$ 
at which the fermion dispersion
goes as $\epsilon_k = k_x + k_y^u$.
The $u$-th curvature $\gamma_u$ is scaled to be one.
Let us consider a point $\mc{P}$ 
on the Fermi surface near $\mc{P}^*$. 
Let $q$ be the difference in the 
$y$-momentum between $\mc{P}$ and $\mc{P}^*$. 
The local  energy dispersion around $\mc{P}$ includes
the lower order terms as
\bqa
\epsilon_k = k_x + \sum_{n=2}^{u-1} \gamma_n(q) k_y^n + 
k_y^u,
\eqa
where 
the term linear in $k_y$ is absorbed into a redefinition of 
$k_x$, 
and the lower order curvatures
go to zero near $\mc{P}^*$ as
$\gamma_n(q) = \wtil{\gamma}_n q^{u-n}$
with $\wtil{\gamma}_n \sim 1$ .
Consider the free energy density
per unit $y$-momentum 
at  temperature $T$ :
$\tilde f(T, q )$
is the contribution to the free energy density
from a unit segment of the Fermi surface at point $\mc{P}$.
The total free energy density is given by 
an integration over the momentum along the Fermi 
surface\cite{Senthil2008,dalidovich},
\bqa
f(T) = \int_{-\Lambda}^\Lambda dq 
~ \tilde f(T,q),
\label{eq:fT}
\eqa
where $\Lambda$ is a UV cut-off
set by the size of the Fermi surface.
\eq{eq:fT} is a consequence of the fact that 
the $y$-momentum $q$ has a positive scaling dimension
and the theory is local in the momentum space\cite{LEE2008}.
The scaling at the inflection point, $\mc{P}^*$, fixes the 
form of the free energy density to be
\bqa
\tilde f(T, q) &=& 
T^{1+1/z_{\theta,u}} 
~h \left( 
\frac{ q^u }{T^{1/z_{\theta,u}}}
,\td \gamma_n
\right).
\label{62}
\eqa
Here we use the fact that the lower order curvatures
$\gamma_n(q)$'s are relevant 
perturbations with the scaling dimension $(u-n)/u$
to the inflection point which is described by the theory
with the dynamical critical exponent $z_{\theta,u}$.
$h(x)$ is a universal function that describes the 
crossover 
from the high temperature scaling controlled
by the inflection point
to the low temperature scaling 
controlled by the points with 
nonzero quadratic curvatures.
Its asymptotic behaviors are given by
\bqa
h(x, \wtil{\gamma}_n) & \sim & 1 ~~\mbox{for $x \rightarrow 
0$}, \label{63} 
\\
h(x, \wtil{\gamma}_n) & \sim &  
x^{\frac{ (u-2)(\theta-1)}{u(\theta+1)} } ~~
\mbox{for $x \rightarrow \infty$}. 
\label{64}
\eqa
\eq{63} is determined from the fact that
the scaling dimension of 
$\tilde f(T,q=0)$ is $z_{\theta,u}+1$
at the inflection point,
whereas \eq{64} follows from the fact that 
$\tilde f(T,q) \rightarrow T^{1+1/z_{2,\theta}}$
in the $T \rightarrow 0$ limit with $\wtil{\gamma}_2 \neq
0$.
If there was a hierarchy in the magnitudes of
$\wtil{\gamma}_n$,
there could be multiple crossovers.
But, in this case, 
there is only one crossover from the 
multi-critical point dictated by dispersion $k_y^u$
to the critical point with dispersion $k_y^2$
because $\wtil{\gamma}_n \sim 1$ for all $n$,
and the quadratic term is most relevant.
Upon integrating along the Fermi surface, 
we obtain two universal 
terms 
for the free energy density,
\begin{align}
f(T) & \sim  T^{1+\frac{1}{z_{\theta,u}}} 
\int_0^{T^{\frac{1}{u z_{\theta,u}}}}  dq \nn
& \qquad + T^{1+\frac{1}{z_{\theta,2}}} 
\int_{  T^{\frac{1}{u z_{\theta,u}}} }^\Lambda
dq ~q^{ \frac{ (u-2) (\theta-1)}{ (\theta+1)}  } \nn
& \sim   A ~ T^{\frac{\theta+3}{\theta+1}}
+ B ~ T^{ \frac{\theta+2u}{\theta+u-1}  }, 
\end{align}
where $A$, $B$ are constants.
The $A$ term is the contribution 
from the extended region
with non-zero quadratic curvatures whereas
the $B$ term is from the region near the inflection point.
Therefore the specific heat scales as
\bqa
c \sim   A ~ T^{\frac{2}{\theta+1}}
+ B ~ T^{ \frac{u+1}{\theta+u-1}  },
\eqa
where the first term dominates in the low temperature limit.
In the presence of the most generic inflection point with 
$u=3$
for the quadratic dispersion of boson with $\theta=2$,
the specific heat of the whole system goes as
\bqa
c \sim   A ~ T^{\frac{2}{3}} + B ~ T^{1}.
\eqa
This analysis can be extended to  other physical responses.

\section{Summary and discussions}
\label{sec:summary}

In this paper we considered a class of 
non-Fermi liquid states 
without time-reversal and parity symmetries
in $(2+1)$ dimensions. 
The chiral non-Fermi liquid states can be potentially
realized at quantum critical points where 
two dimensional chiral Fermi surface 
is coupled with a critical boson
associated with a spontaneous symmetry breaking.
Chiral Fermi surface naturally arises
on a stack of quantum Hall layers\cite{balents1}
or on the surface of three dimensional
Weyl metals\cite{Volovik,Wan,Yang,Burkov}.
The former example, however, is simpler
because there are no gapless bulk degrees of freedom.
In principle, chiral metals with multiple flavors can be 
realized by one stack of quantum Hall layers with $\nu > 1$.
Alternatively, multiple chiral modes can arise
at a junction of semiconductors with oppositely charged
carriers in a uniform magnetic field.


In two-dimensional non-Fermi liquid states,
the local patch description is valid
due to the emergent locality in the momentum 
space\cite{LEE2008}.
General patch theories for the chiral non-Fermi liquid 
states
can be classified by 
the local shape of the Fermi surface,
the dispersion of the critical boson
and the symmetry group.
Although the non-Fermi liquid fixed points are 
described by strongly interacting quantum field theories,
the stability of the fixed points can be established 
non-perturbatively, 
and the exact critical exponents can be computed.
The main ingredient that makes
an exact analysis possible
is the chiral nature of the theory.
Because of chirality, 
internal frequencies in scattering processes are 
strictly bounded by the external frequencies.
Exploiting this property, it is possible to prove
that the theory is UV finite 
below the upper critical dimension,
which is the case for non-Fermi liquid states.
The absence of UV divergence 
guarantees that the theory flows
to a fixed point governed by
the scaling which leaves the interaction
invariant in the bare action.
We also confirm the general conclusion
by computing the Wilsonian effective action
explicitly in the low momentum/frequency limit
with a fixed running cut-off.

For the RG analysis, 
we formulate the low energy excitations 
near the Fermi surface 
as a collection of one dimensional fermions 
with a  continuous flavor labelling 
the momentum along the Fermi surface. 
In this formalism, the curvature of the Fermi surface
manifests itself through 
a non-commutative structure 
between a coordinate and momentum 
in different directions.
The emergent non-commutativity 
leads to a UV/IR mixing in Fermi liquid states,
where IR (UV) behavior of the system
is sensitively controlled by 
UV (IR) structures.
On the other hand, 
there is no prominent UV/IR mixing
in non-Fermi liquid states
due to the UV finiteness of the theory.
The absence of non-trivial UV/IR mixing is 
what makes the patch description valid
in the non-Fermi liquid states.

The chiral non-Fermi liquid states are 
two-dimensional cousins
of the chiral Luttinger liquids in one dimension\cite{CLL} 
whose stability is guaranteed by the absence of back 
scatterings.
In the chiral Luttinger liquids, 
the scaling dimension of the fermionic operator is solely 
determined from the topological property of the system,
independent of the microscopic details.
Similarly, in the chiral non-Fermi liquids, 
the critical exponents only depend on the
geometrical properties of the local Fermi surface
and the dispersion of the critical boson.
This is the reason why exact dimensions can be obtained.

Despite the similarity, 
the two-dimensional state can not be
obtained from a finite number of coupled one-dimensional 
chains.
This is because the low energy limit 
and the limit of infinite chains do not commute.
The two-dimensional non-Fermi liquid state is obtained 
when one takes the limit of infinite chains 
before taking a low energy limit. 
This is manifest from the fact that the momentum 
along the Fermi surface is continuous, 
and  it has a non-trivial scaling dimension.

\section{Acknowledgment}

We thank 
Djordje Minic,
Subir Sachdev, 
Luiz Santos,
T . Senthil,
and Xiao-Gang Wen
for helpful comments and discussions.
The research was supported in part by 
the Natural Sciences and Engineering Research Council of 
Canada,
the Early Research Award from the Ontario Ministry of 
Research and Innovation,
and the Templeton Foundation.
Research at the Perimeter Institute is supported 
in part by the Government of Canada 
through Industry Canada, 
and by the Province of Ontario through the
Ministry of Research and Information.


\section{Appendix}
\appendix

\section{UV/IR mixing} 
\label{app: 1}

In this appendix 
we compute the one-loop vertex function
shown in Fig. \ref{f.v1}.
In this section, we focus on the critical point with 
$\mu=0$.
The one-loop vertex correction describes a process
where a boson with $(\omega_p, \vec p)$
creates a virtual particle-hole pair at 
$(\omega_k+\omega_p+\nu, \vec k+ \vec p + \vec q)$ and 
$(\omega_k+\nu,\vec k+ \vec q)$,
which then scatter into
the final state of a particle-hole pair
with $(\omega_k+\omega_p, \vec k+\vec p)$ and 
$(\omega_k,\vec k)$.
For convenience, we assume that $\vec k=0, \omega_k=0$
and the outgoing fermion is also on the Fermi surface, 
that is, $\eps_{k+p} = 0$.
Then the resulting vertex function is a function 
of $\omega_p$ and $p_y$.
In order to examine the interplay 
between UV and IR scales,
we assume that the largest momentum 
along the Fermi surface 
is given by a finite UV cut-off $\Lambda$.
As we will see below, the Fermi liquid states with
${\theta} < 1$
and the non-Fermi liquid states with ${\theta} > 1$
show distinct behaviour in terms of UV/IR mixing.
In this appendix, we will use the conventional
energy-momentum space representation.

First we consider the Fermi liquids with ${\theta} < 1$.
We assume that the Yukawa coupling $g$ is small
and use the one-loop dressed propagators given by
\eqn{
  G^{-1}(\omega, \vec k) &= i c_F \om + k_x + \gamma k_y^2, 
\\
  D^{-1}(\nu, \vec q) &= |q_y|^{\theta} + c_B 
\frac{\abs{\nu}}{\abs{q_y}}, 
\label{eq: propagators}
}
where $c_F, c_B$ are constants.
The one-loop vertex correction 
with $\vec k=\omega_k=0$ and $\eps_p=0$ 
is given by
\eqn{
  \Gamma_{\theta}(\om_p,p_y;\Lambda) &= g^3\int 
\frac{d \nu d^2 \vec q}{(2\pi)^3}
~ G(\nu, \vec q) ~ G(\omega_p+\nu, \vec p+ \vec q) ~ D(\nu, 
\vec q) \\
  &= \frac{g^3}{2\pi^2} \frac{\Lambda^{1 - {\theta}}}{c_F} ~
\Upsilon_{\theta}(\alpha,L),
 \label{eq: upsilon_defn}
}
where the cross-over function is 
\eqn{
  \Upsilon_{\theta}(\alpha,L) &= \frac{\alpha^2}{L^{1 - 
{\theta}}} 
\Int{0}{L} dy ~ \frac{y}{y^2 + \alpha^2} ~ \log{\lt(1 +
\frac{1}{y^{{\theta} + 1}} \rt)} \label{eq: upsilon}
}
with
\eqn{
  L &= \frac{\Lambda}{(c_B \abs{\om_p})^{1/({\theta} + 
1)}}, 
\label{eq:dimlss1} \\
  \alpha & = \frac{c_F ~ \sgn{\om_p}}{2\gamma ~ 
c_B^{1/({\theta} + 1)}} ~ 
~\frac{\abs{\om_p}^{{\theta}/({\theta} + 1)}}{p_y}.
  \label{eq:dimlss}
}
Here $L$ and $1/\alpha$ correspond to 
the UV cut-off and the external $y$-momentum 
scaled by the energy.

\begin{figure}[!ht]
\centering     
\includegraphics[scale=0.3]{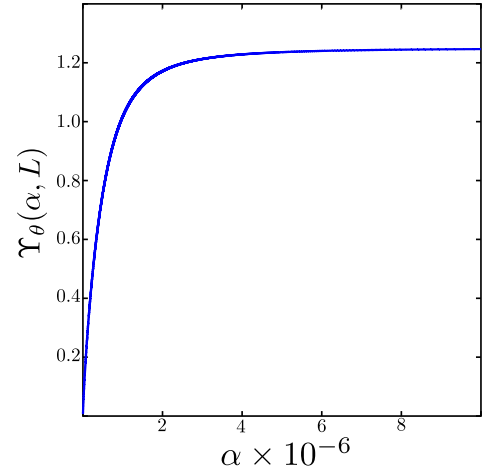} 
\caption{Plot of $\Upsilon_{\theta}(\alpha,L)$ 
as a function of $\alpha$ 
with $L=10^6$  and ${\theta} = 0.2$.
 }
\label{fig: upsilon}
\end{figure}

Now we examine the behavior of the vertex function
as a function of $p_y$.
Suppose that $\om_p$ and $\Lambda$ are fixed 
such that $L \gg 1$.
A typical shape of the cross-over function is 
shown in Fig. \ref{fig: upsilon} 
for a fixed value of $L$.
In the large $p_y$ limit with $L \gg 1 \gg |\alpha|$,
the vertex correction vanishes as
\begin{align}
  \Gamma_{\theta}(\om_p,p_y;\Lambda) & \sim \frac{c_F 
g^3}{c_B 
\gamma^2} ~ \frac{|\om_p|}{p_y^2} ~ \log^2\lt(\frac{c_F 
|\om_p|^{{\theta}/({\theta} + 1)}}{\gamma c_B^{1/({\theta}
+ 
1)} |p_y|} \rt).
\end{align}
As $p_y$ decreases,
$\alpha$ grows.
For an intermediate regime with $L \gg  |\alpha| \gg 1$, 
the vertex correction becomes
\begin{align}
  \Gamma_{\theta}(\om_p,p_y;\Lambda) &\sim \frac{g^3}{c_F} 
~ \lt[\frac{c_F}{\gamma}\rt]^{1 - {\theta}} ~ 
\abs{\frac{\om_p}{p_y}}^{1 - {\theta}}.
\label{eq: gamI} 
\end{align}
As $p_y$ decreases,
the vertex function tends to diverge 
as the number of virtual particle-hole pairs
that can be excited within the energy
provided by the boson increases.
When the energy of the boson is $\om_p$,
the range of $q_y$ that the virtual particle and hole
can take in the loop is given by
$ q_y \leq \frac{\om_p}{2\gamma p_y}$
as is shown in Fig. \ref{fig: finite_FS_big}.
This follows from the condition
$\epsilon_{q+p} - \epsilon_q \leq  \om_p$.
The volume of the phase space for the virtual excitations
increases as $p_y$ decreases.
However, $q_y$ is eventually bounded by $\Lambda$
in the presence of the UV cut-off,
and the vertex function saturates
to a constant as
\begin{align}
 & \Gamma_{\theta}(\om_p,p_y;\Lambda) 
\sim \frac{g^3}{c_F}~ \Lambda^{1 - {\theta}} 
\label{eq: gamII}
\end{align}
in the small $p_y$ limit 
with $\alpha \gg L \gg 1$.
In other words, the maximum value of $y$-momentum that
virtual particles can have is determined by the condition,
\bqa
q_y^{max} \sim \min \left( \frac{\om_p}{\gamma p_y}, 
\Lambda \right).
\label{twol}
\eqa
For $\frac{\om_p}{\gamma p_y} << \Lambda$, the phase space 
of the 
virtual particle-hole excitations is controlled by
the external momentum $p_y$.
In the opposite limit, 
the UV-cut off bounds
the phase space.
The two limits are 
illustrated in Fig. \ref{fig: finite_FS}.
From \eq{twol}, it is evident that 
$p_y \rightarrow 0$ and $\Lambda \rightarrow \infty$
limits do not commute.

It is interesting to note that the IR singularity
that is present in Eq. (\ref{eq: gamI})
is eventually cut-off by an IR scale 
$\tilde p_y \sim \frac{\om_p}{\gamma \Lambda}$
which is set by the inverse of the UV cut-off $\Lambda$.
This is a manifestation of UV/IR mixing where 
the IR behavior of the vertex function
depends on the UV structure in a singular manner.
The UV/IR mixing can be understood from a different 
perspective.
For fixed $\om_p$, 
the vertex function tends to diverges
as $\Lambda$ increases 
as far as $ \Lambda << \frac{\om_p}{\gamma p_y}$
as is shown in Eq. (\ref{eq: gamII}).
However, the UV divergence is cut-off 
by a scale $\tilde \Lambda \sim  \frac{\om_p}{\gamma p_y}$
which is set by the inverse of the IR scale $p_y$.
This non-trivial interplay between UV and IR scales
is a consequence of the fact 
that the low energy fermions near ${\vec k}=0$
`feel' the presence of other modes which 
carry large momenta.
This situation commonly arises in quantum field theories 
above the upper critical dimensions.
What is peculiar about the present case
is that modes with large momenta are 
not necessarily high energy modes
because gapless modes on the Fermi surface
can carry large momenta.
Therefore the modes at large momenta 
affect the low energy behaviour in a singular way.
This is the origin of the UV/IR mixing.

In the Fermi liquids, the UV/IR mixing 
is driven by the UV sensitive volume 
of the phase space for low energy particle-hole excitation 
available near the Fermi surface.
In the non-Fermi liquid state with ${\theta} > 1$, on the 
other hand,
the couplings that are non-local in momentum space are 
suppressed.
This is due to the fact that 
the energy of boson increases steeply at large momentum.
As a result, the intermediate states of particle-hole pairs 
with large momenta get dynamically suppressed 
although those states are equally available as in Fermi 
liquids.
This makes the vertex correction to be insensitive to the
virtual processes occurring at large momenta.
Therefore the UV cut-off is not important 
to the low energy processes for ${\theta} > 1$.
To see this explicitly,
we use the one-loop dressed fermion propagator,
\eqn{
G^{-1}(\omega,\vec k) = i c_F \sgn{\om} ~ 
\abs{\om}^{2/({\theta} + 1)} 
+ \eps_k
}
to compute the vertex correction,
\begin{align}
\Gamma_{\theta}(\om_p,p;\Lambda) &= \frac{g^3
c_B^{\frac{1 - {\theta}}{{\theta} + 1}}}{2\pi^2 ~ c_F }
\alpha^2 \int_0^1 dt \int_0^L
dy \frac{y}{y^{{\theta} + 1}+t} \nn
& \quad \times \frac{(1-t)^{\frac{2}{{\theta} + 1}} + 
t^{\frac{2}{{\theta} + 1}}}{y^2 + 
\alpha^2 \lt((1-t)^{\frac{2}{{\theta} + 1}} 
+ t^{\frac{2}{{\theta} + 1}}\rt)^2}
\label{B11}
\end{align}
with
\eqn{
 \alpha & = \frac{c_F ~ \sgn{\om_p}}{2\gamma ~ 
c_B^{1/({\theta} + 1)}} ~ ~\frac{\abs{\om_p}^{1/({\theta} + 
1)}}{p_y}
}
and $L$ is as defined in \eq{eq:dimlss1}.
In Eq. (\ref{B11}), the integration is convergent in the 
large $L$ limit
even when $p_y=0$
unlike the case in Fermi liquids.
For $L, \alpha >> 1$,
the vertex function saturates to a constant
\eqn{
\Gamma_{\theta}(\om_p,p) \sim \frac{1}{c_F ~ 
c_B^{\frac{{\theta} - 1}{{\theta} + 1}}} \label{eq: gam_nFL}
}
independent of the ratio  
$\frac{\om_p}{\gamma p_y \Lambda}$
in contrast to the non-trivial crossover
that is present in Fermi liquid state.
In this case there is no UV/IR mixing in the vertex 
function.
The insensitivity of the physics near $\vec k = 0$ 
to the gapless modes at large momenta 
is the reason 
for the emergent locality 
in the momentum space\cite{LEE2008}. 
As a result, one can use the patch description 
in non-Fermi liquid states.
On the contrary, 
all low energy modes remain coupled with each other
in the Fermi liquid state, 
and one has to keep the entire Fermi surface
in the low energy description.
Landau Fermi liquid theory indeed
includes Landau parameters associated with
the forward scattering across
the entire Fermi surface
as the low energy data.

\section{The RPA correction for the parabolic Fermi 
surface} 
\label{app: u2}

In this section we compute the RPA vertex 
correction,  the RPA self-energy, and derive the 
expressions for $c_B$ and $c_F$ in \eq{eq: calc_1B-2} and 
\eq{eq: c_F}, respectively.

\subsection{RPA vertex correction}
\label{app: u2_V}

The four-fermion vertex correction 
generated from the $L$-loop RPA diagram
shown in Fig. \ref{fig: RPA_4} is written as
\begin{widetext}
\begin{align}
  & \dl S_{4}^{(RPA,L)} = \int \frac{dk_1 ~d\om_1 ~dk_2 
~d\om_2 ~dq ~d\nu}{(2\pi)^6} ~dx_1 ~
e^{2i \gamma  (k_1 - k_2 ) q x_1} 
~ V_{ij;ln} ~ \chi_{\theta}(q) 
\sum_{\alpha =1}^{(L+1)!} \int_{\mc{C}_{\alpha}} 
\prod_{i = 1}^{L} dx_{i+1,i}\nn 
& \times  \lt(- \frac{g^2}{4\pi^2} 
\chi_{\theta}(q) \rt)^L ~~  \prod_{j=1}^L 
\lt[ \int d\om_j dp_j   
e^{2i \gamma x_{j+1,j} (k_1 - p_j) q} 
G_0(\om_j,- x_{j+1,j}) G_0(\om_j+\nu,x_{j+1,j}) \rt] \nn 
& \times 
\ccolon
\psi^*_{i,k_2}(\om_2,x_1) ~
\psi_{j,k_2+q}(\om_2+\nu,x_1)~
\psi^*_{l,k_1+q}(\om_1+\nu,x_{L+1})~ 
 \psi_{n,k_1}(\om_1,x_{L+1})  \ccolon.
\label{eq: vert_corr_L_exact_u2}
\end{align}
Here $x_i$ is the coordinate of 
the $i$-th vertex in the chain of RPA bubbles
and $x_{i+1,i} = x_{i+1} - x_i$.
$\mc{C}_{\alpha}$ represents 
non-overlapping sets of configurations 
of the $x$-coordinates of the vertices 
whose separations from their 
neighboring vertices are less than $X_0$.
For example, when there are three vertices at 
$x_1$, $x_2$ and $x_3$,  there exist six 
distinct sets of configurations given by
\bqa
\mc{C}_{1} & = & \{ (x_1,x_2,x_3) | x_1 > x_2 > x_3 \}, \nn
\mc{C}_{2} & = & \{ (x_1,x_2,x_3) | x_1 > x_3 > x_3 \}, \nn
& \vdots & \nn
\mc{C}_{6} & = & \{ (x_1,x_2,x_3) | x_3 > x_2 > x_1 \}.
\label{eq:sets}
\eqa
In the present case, only two out of the $(L+1)!$ 
sets contribute to the diagram
due to the chiral nature of the theory.
To see this, 
we first deduce the constraints 
on the relative coordinates of the vertices.
From the $\Theta$-functions in the propagators, 
we have
\eqn{
-\nu~ x_{j+1,j} > \om_j ~ x_{j+1,j} > 0. \label{eq: 
RPA_bound}
}
These inequalities not only put a bound on the internal 
frequencies $\{ \om_j \}$,
but also impose constraints on relative coordinates $\{ 
x_{ij} \}$  :
for any pair $\{ij\}$  
with $i>j$, $x_i$ and $x_j$ are strictly ordered depending 
on the sign of $\nu$, i.e. 
\begin{itemize}
 \item 
$x_{L+1} > x_L > \ldots > x_2 > x_1$ ~~for $\nu < 0$, 
\item 
$x_{L+1} < x_L < \ldots < x_2 < x_1$ ~~for $\nu > 0$.
\end{itemize}
The implication of the strict ordering is that 
for a fixed $\nu$ all $x_{ij}$ possess the 
same sign and $\abs{x_{ij}} \in (0,X_0]$. 
The integration over each $p_j$ leads to a 
$\dl$-function whose support is localized in the 
neighborhood of $x_{j+1}=x_j$. The width of the 
$\dl$-function goes to zero in the limit the UV cutoff of 
$p_j$'s is sent to $\infty$.
As a result, the RPA diagram generates a vertex
which is ultra-local in the $x$-direction,
\begin{align}
  \dl S_4^{(RPA, L)} =& \int \frac{dk_1 ~d\om_1 ~dk_2 
~d\om_2 ~dq ~d\nu}{(2\pi)^6} ~dx ~~ 
e^{2i \gamma  (k_1 - k_2 ) q x} 
~ V_{ij;ln} ~ \chi_{\theta}(q) ~ 
~ \Gamma_4^{(RPA,L)}(q,\nu,X_0) \nn 
& \qquad \times \ccolon \psi^*_{i,k_2}(\om_2,x) ~
\psi_{j,k_2+q}(\om_2+\nu,x)~
\psi^*_{l,k_1+q}(\om_1+\nu,x)~ 
\psi_{n,k_1}(\om_1,x)\ccolon, \label{eq: vert_corr_L_u2}
\end{align}
where
\begin{align}
\Gamma_4^{(RPA,L)}(q,\nu,X_0) &= \lt[ \lt(-
\frac{g^2}{4\pi^2} 
\chi_{\theta}(q) \rt)  \abs{\nu}
\int_{-\infty}^{\infty}
dp \int_0^{X_0} dx_R ~ e^{- 2i \gamma ~ sgn(\nu)~   
x_{R} (p - k_1 ) q}  ~ e^{-\eta \abs{x_R}| \nu|} \rt]^L
\nn
& \equiv \Bigl[- \Pi_2(q,\nu,X_0) ~ \chi_{\theta}(q) 
\Bigr]^L
\label{eq:B5}
\end{align}
with
\begin{align}
   \Pi_2(q,\nu,X_0) &= \lt(\frac{g^2}{4\pi^2} \rt) 
\abs{\nu} \int_0^{X_0} dx_R \int_{-\infty}^{\infty}
dp ~ e^{- 2i \gamma ~ sgn(\nu)~   
x_{R} p  q}  ~ e^{-\eta \abs{x_R}| \nu|} \\
&= \lt(\frac{g^2}{4\pi^2} 
\rt) \abs{\nu} \frac{\pi }{\gamma |q|} 
~ \times \half \int_{-X_0}^{X_0} dx_R ~ 
\dl(x_{R})  ~ e^{-\eta \abs{x_R}| \nu|} \\
&= c_B ~  \frac{|\nu|}{|q|} \label{eq: pi_u2_defn}
\end{align}
\end{widetext}
and
\eqn{
c_B = \frac{g^2}{8 \pi \gamma}.
}
It is noted that 
the dependence of $\Gamma_4^{(L)}$
on $k_1$ drops out as
$k_1$ is absorbed into the $p$
in \eq{eq:B5}.

\subsection{RPA self-energy}
\label{app: u2_SE}

The quantum correction generated from the $L$-loop RPA 
diagrams
shown in Fig. \ref{fig: RPA_2} is
\begin{widetext}
\begin{align}
  \dl S_2^{(RPA, L)} =& \int \frac{dk ~d\om}{(2\pi)^2} 
~dx_1 ~~  \Sigma_{ab}^{(RPA,L)}(k,\omega)  ~ \ccolon 
\psi^*_{a,k}(\om,x_1) ~
\psi_{b,k}(\om,x_1) \ccolon, \label{eq: self-energy_L_u2}
\end{align}
where
\begin{align}
 & \Sigma_{ab}^{(RPA,L)}(k,\omega) = 2^{L+1} ~ V_{al_1;j_1 
i_1} V_{i_1 j_1;j_2 i_2} \ldots V_{i_L j_L;l_1 b}  \int 
\frac{dq d\nu}{(2\pi)^2} \int_{\mc{C}} \prod_{l=1}^{L} 
\left[ dx_{l+1,l} ~ P_2(q,\nu,x_{l+1,l}) \right]
~ \lt[\chi_{\theta}(q)\rt]^{L+1} ~ G_0(\om + \nu, -
x_{L+1,1}) \label{eq: sig_L}
\end{align}
with
\begin{align}
 P_2(q,\nu,x_{l+1,l}) &=  \int \frac{dp_l 
d\nu_l}{(2\pi)^2} ~ e^{2i\gamma q(k-p_l)x_{l+1,l}} ~ 
G_0(\nu + \nu_l,  x_{l+1,l}) G_0(\nu_l, - x_{l+1,l}) \\
&= |\nu| ~\step{-\nu x_{l+1,l}} \int 
\frac{dp_l}{(2\pi)^2} ~ e^{2i\gamma q p_l x_{l+1,l}} 
~ e^{- \eta \abs{x_{l+1,l} \nu}}. \label{eq: pol_1}
\end{align}
\end{widetext}
The product of the $V$'s in \eq{eq: sig_L} yields
\begin{align}
 V_{al_1;j_1 i_1} \ldots V_{i_L j_L;l_1 b} = \dl_{ab} ~ 
\lt(-\frac{g^2 v}{2}\rt)~ \lt(-\frac{g^2}{2} \rt)^L, 
\label{eq: Vs}
\end{align}
where $v$ is defined in \eq{eq: v_defn}. 
Before integrating over $y$-momentum $p_l$ in 
\eq{eq: pol_1}, we use the constraints imposed 
by the $(L+1)$ $\Theta$-functions in \eq{eq: sig_L} 
to write
\begin{align}
 & \int_{\mc{C}} \prod_{l=1}^{L} 
\lt[ dx_{l+1,l} ~ \step{-x_{l+1,l} \nu} \rt]
~ \step{x_{L+1,1} (\nu + \om)} \nn
&= \step{\nu} \step{-\nu - \om} \int_{-X_0}^0 
\prod_{l=1}^{L}
dx_{l+1,l} \nn
& \quad + \step{-\nu} \step{\nu + \om} \int^{X_0}_0
\prod_{l=1}^{L} dx_{l+1,l}. 
\label{eq: split_rel}
\end{align}
Now we integrate over $p_l$ and use the property 
\begin{align}
 \int_0^{a} dx ~ \dl(x) = \half
\end{align}
along with results in 
Eqs.  (\ref{eq: pol_1}), (\ref{eq: Vs}) and (\ref{eq: 
split_rel}) 
to obtain
\begin{align}
\Sigma_{ab}^{(RPA,L)}(k,\om) =& i \dl_{ab}~ \sgn{\om}  
\frac{g^2 v}{(2\pi)^2} \int_{-\infty}^{\infty} 
dq \int_{0}^{|\om|} d\nu   \nn
& \quad \times \chi_{\theta}(q) \lt[ - c_B 
\frac{|\nu|}{|q|} \chi_{\theta}(q) \rt]^L.
\label{eq: sigma_L_fin}
\end{align}
The net contribution to the quadratic term from all 
RPA 
diagrams can be written as 
\begin{align}
 \dl S_2 =& \int \frac{dk ~d\om}{(2\pi)^2} 
~dx ~  \Sigma_{ab}^{(RPA)}(k,\omega) \ccolon 
\psi^*_{a,k}(\om,x) ~ \psi_{b,k}(\om,x) \ccolon,
\end{align}
where the RPA self-energy is
\begin{align}
 \Sigma_{ab}^{(RPA)}(k,\om) = \sum_{L=0}^{\infty} 
\Sigma_{ab}^{(RPA,L)}(k,\om). \label{eq: net_sig_1}
\end{align}
Here $\Sigma_{ab}^{(RPA,0)}(k,\om)$ is the Fock term 
in \eq{eq: fock}. 
After summing over all loops in \eq{eq: net_sig_1}
we obtain 
\begin{align}
 \Sigma_{ab}^{(RPA)}(k,\om) &= i \dl_{ab}~ \sgn{\om}
\frac{g^2 v}{(2\pi)^2} \nn
& \quad \times \int_{-\infty}^{\infty}  dq 
\int_{0}^{|\om|} d\nu ~ \chi_{\theta}^{(RPA)}(q,\nu)\nn
&= i \dl_{ab} ~ c_F ~ \sgn{\om}   |\om|^{2/(1+\theta)}, 
\label{eq: net_sig_fin}
\end{align}
where
$\chi_{\theta}^{(RPA)}(q,\nu)$ is given by \eq{eq:
RPA_dress_V} and 
\begin{align}
c_F = \frac{g^2 v}{2 \pi^2} ~
c_B^{\frac{1-\theta}{1+\theta}}  \int_0^{\infty} dy ~ y ~  
\log\lt(1 + \frac{1}{y^{\theta + 1}} \rt)
\end{align}
for $\mu=0$. 
As expected, 
the quantum corrections 
removes the spurious IR singularity at $\mu=0$.

\section{The RPA correction for general  shapes 
of the local Fermi surface} 
\label{app: general_n}

In this section we compute 
the RPA correction to the Wilsonian effective action
for the patch theory with a general shape 
of the local Fermi surface.

\subsection{RPA vertex correction}
\label{app: general_n_1}


The four-fermion vertex correction generated from the
$L$-loop RPA diagrams shown in Fig. \ref{fig: RPA_4} is 
\begin{widetext}
\begin{align}
  & \dl S_{4,u}^{(RPA,L)} = \int \frac{dk_1 ~d\om_1 ~dk_2 
~d\om_2 ~dq ~d\nu}{(2\pi)^6} ~dx_1 ~
\exp{\lt(i \gamma_u x_1  \sum_{m=1}^{u-1} \binom{u}{m}
(k_1^m - k_2^m ) q^{u-m} 
\rt)} ~ V_{ij;ln} ~ \chi_{\theta}(q) \sum_{\alpha
=1}^{(L+1)!} \int_{\mc{C}_{\alpha}} \prod_{i = 1}^{L}
dx_{i+1,i}\nn 
& \times  \lt(- \frac{g^2}{4\pi^2} 
\chi_{\theta}(q) \rt)^L \int \prod_{j=1}^L \lt[ d\om_j
 dp_j   \exp{\lt(i \gamma_u x_{j+1,j} \sum_{m=1}^{u-1}
\binom{u}{m} (k_1^m - p_j^m) q^{u-m} \rt)}  G_0(\om_j,-
x_{j+1,j}) G_0(\om_j+\nu,x_{j+1,j}) \rt] \nn 
& \times \sum_{m=0}^{\infty}
\frac{\lt(x_{(L+1)1}\rt)^{m}}{m!} ~\ccolon
\psi^*_{i,k_2}(\om_2,x_1) ~
\psi_{j,k_2+q}(\om_2+\nu,x_1)~
\partial_{x_1}^{m} \lt[\psi^*_{l,k_1+q}(\om_1+\nu,x_1)~ 
 \psi_{n,k_1}(\om_1,x_1) \rt] \ccolon,
\label{eq: vert_corr_L_exact}
\end{align}
where in the last line we have Taylor expanded the local
operator $\psi^*_{l,k_1+q}(\om_1+\nu,x_{L+1})~  
 \psi_{n,k_1}(\om_1,x_{L+1})$ around $x_1$. 
We first compute the leading order term in the Taylor 
expansion 
which renormalizes the marginal four-fermion vertex 
in the action (\eq{eq: mixed_action_n}).
Later we will comment on the sub-leading terms 
in the Taylor expansion.

\subsubsection{The leading order term}
The leading contribution to the four-fermion
vertex from the $L$-loop RPA diagram is
\begin{align}
  \dl S_{4,u}^{(RPA,L)} =& \int \frac{dk_1 ~d\om_1 ~dk_2 
~d\om_2 ~dq ~d\nu}{(2\pi)^6} ~dx ~~ \exp{\lt(i \gamma_u x 
\sum_{m=1}^{u-1} \binom{u}{m} (k_1^m - k_2^m ) q^{u-m} 
\rt)} ~ V_{ij;ln} ~ \chi_{\theta}(q) ~ 
~ \Gamma_{4,u}^{(RPA,L)}(k_1,q,\nu,X_0) \nn 
& \qquad \times \ccolon \psi^*_{i,k_2}(\om_2,x) ~
\psi_{j,k_2+q}(\om_2+\nu,x)~
\psi^*_{l,k_1+q}(\om_1+\nu,x)~ 
\psi_{n,k_1}(\om_1,x)\ccolon, \label{eq: vert_corr_L}
\end{align}
where 
\begin{align}
& \Gamma_{4,u}^{(RPA,L)}(k,q,\nu,X_0) = \lt(- 
\frac{g^2}{4\pi^2} \chi_{\theta}(q) \rt)^L ~~ \sum_{\alpha
=1}^{(L+1)!} \int_{\mc{C}_{\alpha}} \prod_{i = 1}^{L}
dx_{i+1,i}  \nn
& \qquad \times   \prod_{j=1}^L \lt[\int d\om_j dp_j ~
\exp{\lt(- i \gamma_u x_{j+1,j} \sum_{m=1}^{u-1}
\binom{u}{m} (p_j^m - k^m ) q^{u-m} \rt)} ~ G_0(\om_j,-
x_{j+1,j}) ~  G_0(\om_j+\nu,x_{j+1,j})\rt].
\label{eq: gamma_L}
\end{align}
Due to chirality, as discussed in Appendix \ref{app: u2}, 
the $x$-coordinates are strictly ordered. Hence, 
\eq{eq: gamma_L} factorizes as
\begin{align}
 \Gamma_{4,u}^{(RPA,L)}(k,q,\nu,X_0) & = \Bigl[-
\Pi_u(k,q,\nu,X_0) ~ \chi_{\theta}(q) 
\Bigr]^L,
\end{align}
where
\begin{align}
    \Pi_u(k,q,\nu,X_0) 
   &= \lt(\frac{g^2}{4\pi^2} \rt) 
\abs{\nu} \int_{-\infty}^{\infty}
dp \int_0^{X_0} dx_R ~ \exp{ \lt[i \gamma_u \sgn{\nu} x_R
\sum_{m=1}^{u-1} \binom{u}{m} \lt( p^m - k^m \rt)
q^{u-m} \rt]} ~ e^{-\eta \abs{x_R}| \nu|}.  
\label{eq: pi_gen_defn}
\end{align}
Unlike the case with $u=2$, 
$\Pi_u(k,q,\nu,X_0)$ in \eq{eq: pi_gen_defn}
for general $u>2$ depends not only on $q$
but also on one of the external $y$-momentum 
because $k^m$ can not be absorbed by $p^m$ in 
\eq{eq: pi_gen_defn}.
This is due to the fact that the inflection point
breaks the sliding symmetry along the Fermi surface.
By scaling 
\eqn{
p \mapsto \frac{y}{\abs{u \gamma_u x_R q}^{1/(u-1)}}, \quad 
\mbox{and } \quad   x_R \mapsto X_0 ~ x,
}
we rewrite \eq{eq: pi_gen_defn} as
\begin{align}
  \Pi_u(k,q,\nu,X_0) &= \lt(\frac{g^2}{4\pi^2} \rt) 
X_0^{\frac{u-2}{u-1}} ~  
 \frac{\abs{\nu}}{|u\gamma_u q|^{1/(u-1)}} 
\int_0^{1} 
 \frac{dx}{\abs{x}^{1/(u-1)}}  ~  e^{ -i ~ 
sgn(\nu) ~ \gamma_u X_0 x  \sum_{m=1}^{u-1} \binom{u}{m} 
k^m q^{u-m}} ~~ e^{-\eta X_0 \abs{x}| \nu|} \nn
& \quad \times h_u \Bigl(\abs{\gamma_u X_0  
q^u  x},\sgn{\gamma_u \nu},\sgn{q} \Bigr),
\end{align}
where
\begin{align}
  h_u(\alpha_u,s_{\gamma \nu},s_q ) = 
\int_{-\infty}^{\infty} dy ~ 
\exp{ \lt[ i~ s_{\gamma \nu} \sum_{m=1}^{u-1} 
\binom{u}{m} \frac{\alpha_u^{\frac{u-m-1}{u-1}} ~ 
s_q^{u-m}}{u^{m/(u-1)}} ~y^m \rt]}. \label{eq: f_u}
\end{align}
We can further simplify \eq{eq: f_u} by appealing to the 
parity of the integer $u$.
For even $u$, we have
\begin{align}
 & h_u(\alpha_u, s_{\gamma \nu}, s_q) = 2 
\int_{0}^{\infty} dy ~ 
\exp \lt[ i ~ s_{\gamma \nu} \sum_{m=1}^{(u-2)/2}
\binom{u}{2m}
\frac{\alpha_u^{\frac{u-2m-1}{u-1}}}{u^{2m/(u-1)}} ~y^{2m}
\rt]  \cos \lt( y^{u-1} +
\sum_{m=1}^{(u-2)/2}\binom{u}{2m-1}  
\frac{\alpha_u^{\frac{u-2m}{u-1}} }{u^{\frac{2m-1}{u-1}}}
~ y^{2m-1} \rt), \label{eq: f_u_even}
\end{align}
and for odd $u$,
\begin{align}
  & h_u(\alpha_u, s_{\gamma \nu}, s_q)  = 2 
\int_{0}^{\infty} dy ~ \exp \lt[ i ~ s_{\gamma \nu} 
s_q\sum_{m=1}^{(u-1)/2} \binom{u}{2m} 
\frac{\alpha_u^{\frac{u-2m-1}{u-1}}}{u^{2m/(u-1)}} ~y^{2m}
\rt]  \cos \lt( \sum_{m=1}^{(u-1)/2}\binom{u}{2m-1}  
\frac{\alpha_u^{\frac{u-2m}{u-1}}}{u^{\frac{2m-1}{u-1}}}
~ y^{2m-1} \rt). \label{eq: f_u_odd}
\end{align}

To the leading order in $X_0|q|^u << 1$,
$\Pi_u(k,q,\nu,X_0)$ takes the form,
\begin{align}
   & \Pi_u(k,q,\nu,X_0)  = \lt(\frac{g^2}{2\pi^2} \rt) 
X_0^{(u-2)/(u-1)} \frac{|\nu|}{|u \gamma_u q|^{1/(u-1)}} 
 \lt[ \xi_u(\sgn{\gamma_u \nu q}) + f_u 
\lt(X_0^{1/u} q, X_0^{1/u} k, \eta X_0 \abs{\nu} \rt) 
\rt], \label{eq: pi_u_leading}
\end{align}
where $\xi_u$ was defined in \eq{eq: xi}.  
The dimensionless function $f_u(s,t,v) \rtarw 0$ as 
$s \rtarw 0$ and is regular 
in the $t,v \rtarw 0$ limit. 
Therefore, to the leading order in $X_0^{1/u} q$,  
$X_0^{1/u} k$ and $\eta X_0 \abs{\nu}$,
\begin{align}
 & \Gamma_{4,u}^{(RPA,L)}(k,q,\nu,X_0) = \Biggl[- 
\lt(\frac{g^2}{2\pi^2} \rt) X_0^{\frac{u-2}{u-1}} 
\xi_u(\sgn{\nu q})   ~\frac{|\nu| ~\chi_{\theta}(q)}{|u 
\gamma_u q|^{1/(u-1)}} \Biggr]^L.
\end{align}
It is of note that 
$\Gamma_{4,u}^{(L)}$ is independent of $k$
to the leading order in $X_0$.

The infinite series of the RPA diagrams
combined with the bare four-fermion vertex,
\begin{align}
  S_4^{(u,{\theta})}  =& \int \frac{d  k_1 ~ dk_2 ~ dq ~ 
d\nu ~ d\om_1 ~ d\om_2}{(2\pi)^6}   \int dx ~
V_{ij;ln} ~ \chi_\theta(q)  ~ \exp{\lt(i \gamma_u 
x \sum_{m=1}^{u-1} 
\binom{u}{m} (k_1^m - k_2^m ) q^{u-m} \rt)} \nn
& \times \ccolon
\psi^*_{i,k_2}(\om_2,x)  
\psi_{j,k_2+q}(\om_2+\nu,x) ~
\psi^*_{l,k_1+q}(\om_1+\nu, x) ~ 
\psi_{n,k_1}(\om_1,x)\ccolon
\end{align}
gives the renormalized four-fermion vertex,
\begin{align}
  S_4^{(u,{\theta})} + \sum_{L=1}^{\infty} \dl
S_{4,u}^{(RPA,L)} =& 
\int \frac{dk_1 ~d\om_1 ~dk_2 ~d\om_2 ~dq ~d\nu}{(2\pi)^6} 
~dx ~~ \exp{\lt(i \gamma_u x \sum_{m=1}^{u-1}
\binom{u}{m} (k_1^m - k_2^m ) q^{u-m} \rt)} ~ V_{ij;ln} ~ 
\chi_{u,{\theta}}^{(RPA)}(q,\nu, X_0)  \nn 
& \qquad \times \ccolon \psi^*_{i,k_2}(\om_2,x) ~
\psi_{j,k_2+q}(\om_2+\nu,x)~
\psi^*_{l,k_1+q}(\om_1+\nu,x)~ 
\psi_{n,k_1}(\om_1,x)\ccolon, \label{eq: RPA_vertex}
\end{align}
where
 \begin{align}
  \chi^{(RPA)}_{u,{\theta}}(q,\nu,X_0)  &= \lt[
|q|^{\theta} 
+ c_B^{(u)}(X_0)~ 
\xi_u(\sgn{\gamma_u \nu q})~\frac{|\nu|}{|q|^{1/(u-1)}} 
\rt]^{-1} 
\label{eq: chi_n}
\end{align}
\end{widetext}
with
\eqn{
c_B^{(u)}(X_0) = \lt(\frac{g^2}{2\pi^2} \rt) 
\frac{X_0^{(u-2)/(u-1)}}{|u\gamma_u|^{1/(u-1)}} .
}

\subsubsection{The sub-leading terms}
\label{app: m!=0}

Now we consider the sub-leading terms with $m > 0$ in
\eq{eq: vert_corr_L_exact}.
As we discussed in the main text, 
for $u=2$ all the relative
coordinates between vertices 
in the RPA diagrams are fixed by the 
$\dl$-functions arising from the fermion loops.
As a result, the terms with $m > 0$ in \eq{eq: 
vert_corr_L_exact}
are absent for $u=2$.
For $u > 2$ the fermion loops do not produce 
$\dl$-functions. 
Consequently, one has to consider
the full gradient expansion in 
\eq{eq: vert_corr_L_exact}.
In this section, we show that the gradient expansion
is well defined when external momenta are small
with fixed $X_0$. 
This is less trivial than it naively looks 
because the two vertices at the end of the $L$-loop 
RPA chains can be as far as $L X_0$.
Since $L$ can be arbitrarily large,
one has to show that the contribution
from large $L$ is small.


The coefficient of the $m$ derivative term,
$\psi^* \psi ~\partial^m \lt[\psi^* \psi\rt]$ 
in \eq{eq: vert_corr_L_exact} is at most 
\eqn{
  \wtil{C}_{m,L} = \frac{(L \wtil{X}_0)^{m}}{m!} ~
\wtil{X}_0^{\frac{u-2}{u-1}  L}, \label{eq: C-tilde_defn}
}
where $\wtil{X}_0 \sim  X_0 |q|^{u},  X_0 q_x,  X_0 
\omega^{1/z}$
are small dimensionless parameter associated with external
momentum or frequency ($q, q_x, \omega$ )
measured in the unit of $X_0^{-1}$ ($q$ refers to 
$y$-momentum).
Here we used the fact that
$x_{(L+1)1} \leq L X_0$
and the fact that each fermion loop 
contributes a factor of $X_0^{(u-2)/(u-1)}$.
The question is how $\wtil{C}_{m,L}$ 
behaves in the large $m$ and $L$ limit
with a fixed value of $\wtil{X}_0 \ll 1$.

From \eq{eq: C-tilde_defn} it is obvious that at
fixed $m$, $\wtil{C}_{m,L}$ is
exponentially suppressed as a function of $L$ because of 
the exponential suppression in $\wtil{X}_0^{\frac{u-2}{u-1} 
 L}$.
In order to understand the behaviour of $\wtil{C}_{m,L}$ 
in the large $m$ limit, 
we consider
the logarithm of $\wtil{C}_{m,L}$,
\begin{widetext}
\begin{align}
\ln{\wtil{C}_{m,L}} &= - \ln{m!} - \lt(m + \frac{u-2}{u-1}
L\rt) \ln({\wtil{X}_0^{-1}}) + m\ln{L} \\
&\approx 
- L \lt[\frac{u-2}{u-1} \ln{(\wtil{X}_0^{-1})} 
 + \lt(\frac{m}{L}\rt) \lt\{ \lt(\ln{(\wtil{X}_0^{-1})} -
1\rt) + \ln{\lt(\frac{m}{L}\rt)} \rt\} \rt].
\label{eq: C-tilde_log}
\end{align}
\end{widetext}
Since $\wtil{X}_0^{-1} \gg 1$ and $x \ln{x} $ is bounded 
from below,
the expression within the square bracket
in \eq{eq: C-tilde_log} is positive definite for any
$(m/L) \in [0,\infty)$. Consequently, $\wtil{C}_{m,L}$ is
exponentially suppressed as a function of $m$ when $m \gg
1$. This shows that both in the large $m$ and
large $L$ limits $\wtil{C}_{m,L} \ll 1$. Therefore, the
derivative terms with $m > 0$ in the Taylor expansion in
\eq{eq: vert_corr_L_exact} are suppressed compared to 
the leading term at low momentum/frequency limit
with fixed $X_0$.

\subsection{RPA self-energy}
\label{app: general_n_2}

Here we compute the contributions 
of the RPA diagrams
to the quadratic action. 
The procedure is identical to
the one used in Appendix \ref{app: u2_SE}.

The self-energy generated from the $L$-loop RPA diagrams is
\begin{widetext}
\begin{align}
  \dl S_2^{(RPA, L)} =& \int \frac{dk ~d\om}{(2\pi)^2} 
~dx_1 ~~  \Sigma_{ab}^{(RPA,L)}(k,\omega,X_0)  ~ \ccolon 
\psi^*_{a,k}(\om,x_1) ~ \psi_{b,k}(\om,x_1) \ccolon, 
\label{eq: self-energy_L_u}
\end{align}
where
\begin{align}
  & \Sigma_{ab}^{(RPA,L)}(k,\omega,X_0) = 2^{L+1} ~ 
V_{al_1;j_1 i_1} V_{i_1 j_1;j_2 i_2} \ldots V_{i_L j_L;l_1 
b} \int \frac{dq d\nu}{(2\pi)^2} ~
\Bigl[\chi_{\theta}(q)\Bigr]^{L+1} \int_{\mc{C}}
\prod_{l=1}^{L} \lt[ dx_{l+1,l} \rt] 
G_0(\om + \nu, - x_{L+1,1})\nn
&\qquad \times \prod_{l=1}^{L} \lt[  \int \frac{dp_l 
d\nu_l}{(2\pi)^2} ~ G_0(\nu_l, -x_{l+1,l}) ~ G_0(\nu_l +
\nu, x_{l+1,l}) ~ e^{i\gamma_u x_{l+1,l} \sum_{m=1}^{u-1} 
\binom{u}{m} (k^m - p_l^m)q^{u-m}} \rt]. \label{eq:
sig_L_gen}
\end{align}
Note that \eq{eq: self-energy_L_u} is the leading term 
in the gradient expansion.
The terms with derivatives are dropped 
because they are irrelevant at low momentum
as discussed in Appendix \ref{app: m!=0}.
Integrating over $\nu_l$ in 
\eq{eq: sig_L_gen}, and using
the constraints imposed by 
the $\Theta$-functions in \eq{eq: sig_L_gen},
we obtain
\begin{align}
 & \Sigma_{ab}^{(RPA,L)}(k,\omega,X_0) = i ~\dl_{ab} 
~\frac{g^2 v}{(2\pi)^2} \int dq ~ 
\chi_{\theta}(q) \int d\nu \lt[ \step{-\nu} \step{\om + 
\nu} - \step{\nu} \step{-\om - \nu} \rt]
\lt[-\wtil{\Pi}_u(k,q,\om,\nu,X_0) \chi_{\theta}(q)\rt]^L \\
& = i ~ \dl_{ab} ~ \frac{g^2 v}{(2\pi)^2} \int dq ~ 
\chi_{\theta}(q) \int_0^{|\om|} d\nu \lt[ \step{\om} 
\lt[-\wtil{\Pi}_u(k,q,\om,-\nu,X_0) \chi_{\theta}(q)\rt]^L 
- \step{-\om} \lt[-\wtil{\Pi}_u(k,q,\om,\nu,X_0)
\chi_{\theta}(q)\rt]^L \rt],  \label{eq: sig_L_gen-1}
\end{align}
where  
\begin{align}
\wtil{\Pi}_u(k,q,\om,\nu,X_0)&= \lt(\frac{g^2}{4\pi^2} 
\rt) 
\abs{\nu} \int_{-\infty}^{\infty}
dp \int_0^{X_0} dx_R ~ \exp{ \lt[i \gamma_u \sgn{\nu} x_R
\sum_{m=1}^{u-1} \binom{u}{m} \lt( p^m - k^m \rt)
q^{u-m} \rt]} ~ e^{-\eta \abs{x_R}| \om|}.  
\label{eq: pi-tilde_defn}
\end{align}
Note that $\wtil{\Pi}_u$ differs from ${\Pi}_u$ defined in
\eq{eq: pi_gen_defn} because of the different
frequency dependence of the exponential damping factor.

We write the contribution from the Fock diagram in 
\eq{eq: fock}, which is valid for all $u$,  as
\begin{align}
 S_2^{'} = i \dl_{ab} \frac{g^2 v}{(2\pi)^2} \int
\frac{dk d\om}{(2\pi)^2} dx_1 \int dq ~
\chi_{\theta}(q) \int_0^{|\om|} d\nu \lt[ \step{\om} - 
\step{-\om} \rt] \ccolon 
\psi^*_{a,k}(\om,x_1) ~ \psi_{b,k}(\om,x_1) \ccolon.
\end{align}
To the Fock diagram we add the contributions 
$\dl S_2^{(RPA,L)}$ from $L=1$ to $L=\infty$ 
to obtain the RPA self-energy,
\begin{align}
 & \Sigma_{ab}^{(RPA)}(k,\omega,X_0) = i \dl_{ab} \frac{g^2 
v}{(2\pi)^2} \int dq  \int_0^{|\om|} d\nu 
\lt[ \step{\om} \wtil{\chi}_{u,\theta}(k,q,\om,-\nu,X_0) -  
\step{-\om} \wtil{\chi}_{u,\theta}(k,q,\om,\nu,X_0) \rt], 
\label{eq: RPA_SE_u_full} 
\end{align}
\end{widetext}
where
\begin{align}
 \wtil{\chi}_{u,\theta}(k,q,\om,\nu,X_0) = 
\frac{1}{|q|^{\theta} + \wtil{\Pi}_u(k,q,\om,\nu,X_0)}.
\end{align}
As in the case of $\Pi_u$ in \eq{eq: pi_u_leading}, to the
leading order in $X_0 |q|^{u}$, we have
\eqn{
\wtil{\Pi}_u(k,q,\om,\nu,X_0) = c_B^{(u)}(X_0)
~\xi(\sgn{\gamma_u q \nu}) \frac{|\nu|}{|q|^{1/(u-1)}}.
}
Accordingly, the renormalized vertex becomes
 \begin{align}
\chi^{(RPA)}_{u,\theta}(q, \nu,X_0)
= \frac{1}{|q|^{\theta}  + 
c_B^{u}(X_0) ~\xi(\sgn{\gamma_u q 
\nu}) \dsty{\frac{|\nu|}{|q|^{1/(u-1)}}} }
\label{chiRPAu}
 \end{align}
to the leading order.
Using \eq{chiRPAu} in 
\eq{eq: RPA_SE_u_full}, we obtain the
leading order contribution to the self-energy, 
\begin{align}
& \Sigma_{ab}^{(RPA)}(k,\omega,X_0) \nn
&= i \dl_{ab} \frac{g^2 v}{(2\pi)^2} \sgn{\om} \int dq  
\int_0^{|\om|} d\nu  ~ \chi^{(RPA)}_{u,\theta}(q,\nu,X_0) \\
&= i ~ \dl_{ab} ~ c_F^{(u)}(X_0) ~ \sgn{\om}
|\om|^{\frac{u}{u(\theta - 1) + 1}}.
\label{eq: RPA_SE_u_leading} 
\end{align}
Here $c_F^{(u)}(X_0)$ is a constant given by
\begin{align}
& c_F^{(u)}(X_0) = \frac{g^2 v}{4\pi^2}  
  \lt( c_B^{(u)}(X_0) \rt)^{\frac{(u-1)(1 - 
{\theta})}{{\theta}(u-1)+1}} 
\nn
  &  \times \int_{0}^1 dt \int_{-\infty}^{\infty} dy ~ 
\frac{|y|^{1/(u-1)}}{|y|^{\frac{{\theta}(u-1)+1}{u-1}} + 
\xi_u(\sgn{\gamma_u yt}) ~ |t|}. \label{eq: c_F^u}
\end{align}
Note that $c_F^{(u)}(X_0)$ depends on $X_0$ through 
$c_B^{(u)}(X_0)$. 
Since $c_B^{(u)}(X_0) \propto X_0^{(u-2)/(u-1)}$,
we have
\eqn{
  c_F^{(u)}(X_0) \propto X_0^{\frac{(u-2)(1 - 
{\theta})}{{\theta}(u-1)+1}}.
}

\subsection{Inversion symmetry along the 
\texorpdfstring{$y$}{y}-direction in 
the patch description }
\label{app: general_n_3}

\begin{figure}[!ht]
 \includegraphics[scale=0.5]{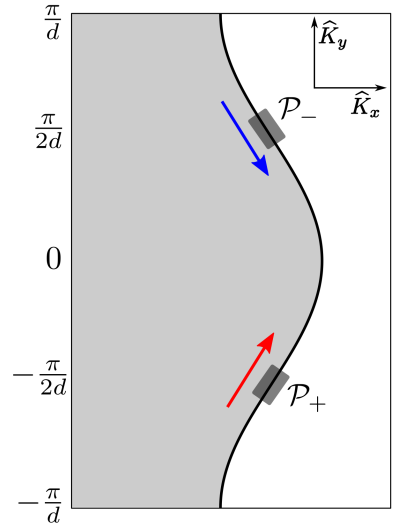}
\caption{
Under the inversion along the $y$ direction the 
two patches are exchanged.
Momentum defined away from the inversion 
points in each patch are also exchanged. 
For example, a particle-hole pair denoted as an arrow 
in patch $\mc{P}_+$ is mapped 
to the other arrow in patch $\mc{P}_-$.
}
\label{fig: inflectionFS-2}
\end{figure}

In this section  we show that 
the theory of the full Fermi surface
respects the inversion symmetry 
along the $y$-direction as it should be
although a single patch theory breaks the symmetry
for odd $u$.
The effective action for even $u$ in \eq{eq:
eff_act_general} 
is manifestly invariant under reversing $y$-momenta : 
$q \rightarrow - q$,  $k_i \rightarrow -k_i$.
For odd $u$, the symmetry is less obvious 
because the four-fermion vertex includes 
an imaginary components in $\xi_u$ 
which is odd under $q \rightarrow - q$ transformation
as is shown in \eq{eq: xi}.
In order to see the symmetry of the full theory,
one has to include the other patch 
connected by the inversion.
This is necessary even though the two patches are in general
decoupled at low energies because they
have different tangent vectors.

The inversion symmetry of the full Fermi surface 
guarantees that the inflection points with odd $u$
arise in pairs.
For example, there is a pair of inflection points
with $u=3$ in generic chiral Fermi surfaces
as is shown in \fig{fig: inflectionFS-2}. 
Suppose that the local dispersions near 
a pair of such inflection points are given by
\eqn{
  \eps_k^{(\sigma)} = k_x + \gamma_{\sigma} ~k_y^u,
}
where $\vec k$ measures the deviation of momenta away
from the inflection points,
$\sigma = +, -$ and
$\gamma_{+} = - \gamma_{-}$, where $\gamma_+ > 0$.
The fermion field in each local patch is related to the 
original field through
\begin{align}
 \psi^{(\sigma)}_i(\om,k_x,k_y) \equiv 
\psi_{i}\lt(\om,k_x,-\sigma K_y^* + 
k_y \rt), 
\end{align}
where we assume that the inflection points are at 
$(K_x,K_y)=(0, \pm K_y^*)$.
Under the inversion of $y$-component of momentum,
the original field transforms as
\begin{align}
 & \psi_{i}\lt(\om,k_x,-K_y^* + k_y \rt) 
\mapsto \psi_{i}
\lt(\om,k_x, K_y^*  - k_y \rt). 
\end{align}
Therefore, the inversion exchanges
$\psi^{(+)}$ and $\psi^{(-)}$ as
\begin{align}
&  \psi^{(\sigma)}_i(\om,k_x,k_y) \mapsto 
\psi^{(-\sigma)}_i(\om,k_x,-k_y).
\end{align}

In the mixed-space representation,
the effective action for the two patches is given by
\begin{align}
 & S^{(u,{\theta})}_{X_0} = \sum_{\sigma = \pm} \int dx 
\Bigg[
\int 
\frac{dk ~ d\om}{(2 \pi)^2} ~ 
\mc{L}^{(u,{\theta},\sigma)}_{2}(x,k,\om,X_0) \nn
& \qquad + \int \frac{dk_1 ~dk_2 ~dq ~d\om_1 
~d\om_2 ~d\nu}{(2 \pi)^6}  \nn 
& \quad \qquad \times 
\mc{L}^{(u,{\theta},\sigma)}_{4}(x,k_1,k_2,q,\om_1,\om_2, 
\nu,X_0) \Bigg], 
\end{align}
where 
\begin{align}
&\mc{L}^{(u,{\theta},\sigma)}_{2}(x,k,\om,X_0) =  
{\psi_{i,k}^{(\sigma)}}^*(\om,x) \nn
& \qquad \times \Bigl[  ic_F^{(u)}(X_0) ~ 
\sgn{\om} |\om|^{\frac{u}{{\theta}(u-1)+1}}  - i\partial_x 
\Bigr]
\psi_{i, k}^{(\sigma)}(\om,x),  \\
&\mc{L}^{(u,{\theta},\sigma)}_{4}(x,k_1,k_2,q,\om_1,\om_2, 
\nu,X_0)  =  V_{ij;ln} ~ 
\chi_{u,{\theta}}^{RPA(\sigma)}(q,\nu,X_0) \nn 
& \times  
\exp{\lt(i \gamma_\sigma x 
\sum_{m=1}^{u-1}
\binom{u}{m} (k_1^m - k_2^m ) q^{u-m} \rt)} 
~ {\psi^{(\sigma)~*}_{i, k_2}}(\om_2,x) \nn
& \times   {\psi^{(\sigma)}_{j, k_2+q}}(\om_2+\nu,x) 
{\psi_{l,k_1+q}^{(\sigma)~*}}(\om_1+\nu,x) ~ 
{\psi^{(\sigma)}_{n,k_1}}(\om_1,x) \label{eq: 
mixed_action_X}
\end{align}
with 
 \begin{align}
  & \chi^{RPA(\sigma)}_{u,{\theta}}(q,\nu,X_0) \nn
  &= \lt[ |q|^{\theta} + c_B^{(u)}(X_0)~ 
\xi_u(\sgn{\gamma_{\sigma} \nu 
q})~\frac{|\nu|}{|q|^{1/(u-1)}} 
\rt]^{-1}. \label{eq: chi_patch}
\end{align}
It is now easy to check that $S^{(u,{\theta})}_{X_0}$ 
is invariant under inversion of $y$-component of 
momenta because 
\begin{align}
\chi^{RPA(-\sigma)}_{u,{\theta}}(- q,\nu,X_0) & =  
\chi^{RPA(\sigma)}_{u,{\theta}}(q,\nu,X_0).
\end{align}


\end{document}